\documentclass[%
 reprint,
showpacs, %added to the original sample
amsmath,amssymb,
prb,
]{revtex4-2}

\usepackage{graphicx}% Include figure files
\usepackage{here}
\usepackage[abs]{overpic}

\usepackage{dcolumn}% Align table columns on decimal point
\usepackage{bm}% bold math
\usepackage{hyperref}% add hypertext capabilities
\usepackage{mathrsfs} % for \mathscr
\usepackage{braket} % for \bra, \ket, \braket, \set, \Bra, \Ket, \Braket, \Set

\usepackage{color} % temporal for coloring the text of NOTEs

\begin{document}

\preprint{ASTSSSLM}

\title{Analytical solutions of topological surface states in a series of lattice models}% Force line breaks with \\

\author{Masaru Onoda}
%\email{onoda@gipc.akita-u.ac.jp}
\affiliation{%
Mathematical Science Course,
%Department of Mathematical Science and Electrical-Electronic-Computer Engineering, \\
Graduate School of Engineering Science,
Akita University, Akita 010-8502, Japan
}%


\date{\today}% It is always \today, today,

\begin{abstract}
We derive the analytical solutions of surface states in a series of lattice models for three-dimensional topological insulators and their nontopological counterparts based on an ansatz.
A restriction on the spin-flip matrices in nearest-neighbor hopping characterizes the series.
This restriction affords the ansatz and favors analytical solvability of surface-state eigenvectors.
Despite the restriction, the series retains sufficient designability to describe various types of surface states.
We also describe how it can serve as a tractable tool for elucidating unique phenomena on topological surfaces.
\end{abstract}

%73.20.-r :  Electron states at surfaces and interfaces
%73.20.At : surface-states, band structure, electron density of states
%73.43.-f : Quantum Hall effects
%73.43.Cd : Theory and modeling
%72.25.-b : Spin polarized transport

\maketitle

%\tableofcontents

\section{\label{sec:Introduction}Introduction}
Topological insulators are electronic materials that insulate in the bulk but conduct along the edges in two-dimensional systems or over surfaces in three-dimensional systems~\cite{Moore2010, Hasan2010a, Qi2011, Shen2012, Ando2013, Bernevig2013, Franz2013, Ortmann2015, Chiu2016}.
This property is attributed to gapless boundary states within a bulk energy gap at the Fermi level of each topological insulator.
These boundary states reflect the topological nature of the electronic band structure under strong spin-orbit interactions.
They show characteristic spin polarization locked by momentum, leading to novel spin-dependent transport phenomena, and are named helical edge/surface states.
The gapless nature of such states is protected by time-reversal symmetry~\cite{Kane2005a, Kane2005}, which in some cases is accompanied by other discrete symmetries such as crystalline symmetries~\cite{Fu2011}.
The concept of topological order~\cite{Thouless1982} fostered in the investigation of quantum Hall systems~\cite{Klitzing1980} plays a crucial role in classifying various phases of topological insulators.
In a naive picture~\cite{Onoda2005}, a two-dimensional topological insulator can be interpreted as a superposition of a time-reversal pair of spontaneous integer quantum Hall systems~\cite{Haldane1988}.
More precisely, the parity of the number of Kramers' pairs of helical edge states is critical and corresponds to the topological invariant called the $\mathbb{Z}_{2}$ index~\cite{Kane2005a, Kane2005}.
Soon after the prediction of realistic two-dimensional topological insulator candidates~\cite{Bernevig2006}, novel phenomena caused by helical edge states were experimentally confirmed~\cite{Konig2007, Roth2009, Knez2011}.

Higher-dimensional extensions highlight the essential difference between quantum Hall systems and topological insulators.
A three-dimensional topological insulator is not necessarily interpreted as stacked layers of two-dimensional topological insulators.
Three-dimensional topological insulators are classified into two of the most prominent classes of topological materials, strong and weak topological insulators, by four $\mathbb{Z}_{2}$ topological invariants~\cite{Fu2007, Moore2007, Fu2007a, Roy2009}.
The helical surface states of a three-dimensional topological insulator appear as spin-polarized two-dimensional gapless Dirac fermions.
Thus, topological insulators can be interpreted as axion media~\cite{Peccei1977a, Peccei1977} characterized by an axion term with a special coefficient~\cite{Qi2008}.
Axion media realize the Witten effect~\cite{Witten1979}, in which a unit magnetic monopole~\cite{Dirac1948} will bind a fractional charge, e.g., a half-odd integer charge in the case of topological insulators. This effect can also be used to measure the topological nature of a given material~\cite{Rosenberg2010}.
Stimulated by the predictions of realistic candidate materials~\cite{Fu2007a,Zhang2009}, novel properties of gapless helical surface states were experimentally observed~\cite{Hsieh2008,Hsieh2009,Xia2009,Chen2009,Chen2010,Jiang2012}
Experimental studies of weak topological insulators have been rare despite early theoretical predictions~\cite{Hasan2010a}.
Very recently, the topological states of the candidate material ZrTe$_5$ were visualized on its side surface by spin- and angle-resolved photoemission spectroscopy~\cite{Zhang2021}.
The bulk bandgap of the material was shown to be controllable by external strain.
Enhanced confinement of side surface states of weak topological insulators could be advantageous for applications of dissipationless currents relative to strong topological insulators.

The concept of topological insulators has been combined with that of topological superconductors and extended by utilizing symmetry classification~\cite{Zirnbauer1996, Altland1997} to reveal five topological phases in each generic dimension~\cite{Schnyder2008, Kitaev2009, Ryu2010, Chiu2016}.
This extension has induced the idea of symmetry-protected topological phases, which are realizable across a comprehensive spectrum of physical systems, e.g., photonic systems~\cite{Yannopapas2011a, Khanikaev2013, Chen2014, Lu2014, Wu2015, Lu2016, Slobozhanyuk2017, Ozawa2019}, phononic systems~\cite{Prodan2009, Zhang2010, Berg2011, Li2012, Kane2014, Huber2016, Liu2017}, ultracold atomic systems~\cite{Goldman2016, Zhang2018, Cooper2019}, and superconducting circuits~\cite{Cai2019}.
This expanding spectrum of the research field will allow virtual implementation of topological states in artificial networks.
The quest for exotic phenomena in idealistic model systems is just as important as the quest for actual topological materials.

This paper presents a series of lattice models that can be a tractable tool for investigating topological and nontopological surfaces on an equal footing.
The series overcomes the restriction of the known family of topological lattice models with exactly solvable surface states studied in Ref.~\cite{Kunst2019} and can handle exactly solvable surface states on different types of surfaces and with different types of spin textures.
The organization of the paper is as follows.
Section~\ref{sec:ModelHamiltonian} defines the series and discusses its phase structure based on the four $\mathbb{Z}_{2}$ topological invariants.
Section~\ref{sec:AnalyticalSolutions} is the central part of this paper and derives the analytical solutions of surface states based on an ansatz.
These solutions can be exact at the semi-infinite limit of the system thickness.
Section~\ref{sec:Comparison} evaluates the quality of each analytical solution via comparison with a corresponding numerical solution obtained at finite thickness.
Section~\ref{sec:Discussions}  discusses the outlook for future studies.

\section{\label{sec:ModelHamiltonian}Model Hamiltonian}
We construct a series of lattice models on a simple cubic lattice with a conventional orthogonal basis $\{\bm{a}_{\mu}\}_{ \mu = 1,2,3}$ ($\vert\bm{a}_{\mu}\vert =a$).
Lattice sites $\bm{r}=\sum_{\mu}n_{\mu}\bm{a}_{\mu}$ ($n_{\mu}\in \mathbb{Z} $) are classified into the $\mathrm{A \Vert B}$ sublattice as follows: $\sum_{\mu} n_{\mu} = \mathrm{even \Vert odd} \Longrightarrow \bm{r} \in \mathrm{A \Vert B}$.
This classification introduces the symbol $\left(-1\right)^{\bm{r}}=\left(-1\right)^{\sum_{\mu}n_{\mu}}$.
Each sublattice forms a face-centered cubic lattice.
The following Hamiltonian defines the series via three kinds of real-valued parameters $\{t_{\mu}\}$, $\{t_{\mu\nu}\}$ ($\mu\neq\nu, t_{\mu\nu}=t_{\nu\mu}$), and $v_{\mathrm{s}}$ and SU(2) matrices $\{U_{\mu}\}$,
\begin{align}
\hat{H}  =&  \hat{H} _{\mathrm{n}}+\hat{H} _{\mathrm{nn}}+\hat{H} _{\mathrm{s}},
\nonumber\\
\hat{H} _{\mathrm{n}}
=&  \sum_{\bm{r}}\sum_{\mu}\left(-1\right)^{\bm{r}}t_{\mu}\hat{c}_{\bm{r}+\bm{a}_{\mu}}^{\dagger}U_{\mu}^{\left(-1\right)^{\bm{r}}}\hat{c}_{\bm{r}}+\mathrm{H.c.},
\nonumber
 \\
\hat{H} _{\mathrm{nn}}
=&  \sum_{\bm{r}}\sum_{\mu<\nu}\left(-1\right)^{\bm{r}}t_{\mu\nu}
\left(\hat{c}_{\bm{r}+\bm{a}_{\mu}+\bm{a}_{\nu}}^{\dagger}\hat{c}_{\bm{r}}+\hat{c}_{\bm{r}-\bm{a}_{\mu}+\bm{a}_{\nu}}^{\dagger}\hat{c}_{\bm{r}}\right)
\nonumber\\
&+\mathrm{H.c.}
\nonumber
,\\
\hat{H} _{\mathrm{s}} =&  \sum_{\bm{r}}\left(-1\right)^{\bm{r}}v_{\mathrm{s}}\hat{c}_{\bm{r}}^{\dagger}\hat{c}_{\bm{r}},
\label{eq:H}
\end{align}
where $\hat{c}^{\left(\dagger\right)}_{\bm{r}}$ is the spinor of the annihilation (creation) operator at site $\bm{r}$ and each H.c. stands for the Hermitian conjugate of the preceding expression.
The expression $U_{\mu}^{\left(-1\right)^{\bm{r}}}$ means $U_{\mu} \Vert U^{-1}_{\mu}$ for $\bm{r} \in \mathrm{A \Vert B}$.

We further characterize the series by the restriction
\begin{align}
&U^{\dagger}_{\mu}  U_{\nu} + U^{\dagger}_{\nu} U_{\mu}
=U_{\mu} U^{\dagger}_{\nu} + U_{\nu} U^{\dagger}_{\mu}
= 2\delta_{\mu\nu}\sigma_{0}
\label{eq:U-cond-Solv}
,
\end{align}
where $\sigma_{0}$ is the $2\times 2$ identity matrix for the spin degrees of freedom.
For later convenience, we also introduce Pauli matrices $\{\sigma_{\mu}\}_{\mu=1,2,3}$ for the spin degrees of freedom in the standard representation.
The restriction Eq.~(\ref{eq:U-cond-Solv}) helps find an ansatz for surface-state eigenvectors.
It also suggests the existence of $\pi$ flux per plaquette, i.e., $U^{\dagger}_{\mu} U_{\nu} U^{\dagger}_{\mu} U_{\nu}  = -\sigma_{0}$ $\left(\mu\neq\nu\right)$.
Despite this restriction, Eq.~(\ref{eq:H}) can describe a variety of topological and nontopological insulators, as shown in the Supplemental Material~\cite{SM}.
We can generally parametrize $\{U_{\mu}\}$ by unit vectors $\{\bm{n}_{\mu}\}$ ($\left\vert\bm{n}_{\mu}\right\vert=1$) and angle variables $\{\phi_{\mu}\}$ as
\begin{align}
&U_{\mu}  = \exp\left(i\phi_{\mu}\bm{n}_{\mu}\cdot\bm{\sigma}\right)
=\sigma_{0}\cos\phi_{\mu} + i\bm{n}_{\mu}\cdot\bm{\sigma}\sin\phi_{\mu}.
\label{eq:U}
\end{align}
Combining Eqs.~(\ref{eq:U-cond-Solv}) and (\ref{eq:U}), we obtain the following expression and will utilize it implicitly hereafter:
\begin{align}
U^{\dagger}_{\mu}U_{\nu}
&=\delta_{\mu\nu}\sigma_{0}+i\{
-\sin\phi_{\mu}\cos\phi_{\nu}\bm{n}_{\mu}\cdot\bm{\sigma}
\nonumber\\
&\quad
+\cos\phi_{\mu}\sin\phi_{\nu}\bm{n}_{\nu}\cdot\bm{\sigma}
\nonumber\\
&\quad+\sin\phi_{\mu}\sin\phi_{\nu}\left(\bm{n}_{\mu}\times\bm{n}_{\nu}\right)\cdot\bm{\sigma}
\}
\label{eq:UdU}
.
\end{align}

$\hat{H}$ is invariant under the time-reversal ($\hat{\mathcal{T}}$) and modified space-inversion ($\hat{\mathcal{P}}_{\mathrm{m}}$) transformations below,
\begin{align}
&\hat{\mathcal{T}}
z^{\dagger}\hat{c}_{\bm{r}}
\hat{\mathcal{T}}^{-1}
=\left(\Theta z\right)^{\dagger}\hat{c}_{\bm{r}}
,\quad
\hat{\mathcal{P}}_{\mathrm{m}} \hat{c}_{\bm{r}} \hat{\mathcal{P}}^{-1}_{\mathrm{m}} = (-1)^{\bm{r}}\hat{c}_{-\bm{r}},
\label{eq:Pm}
\end{align}
where $\Theta = e^{-i\frac{\pi}{2}\sigma_{2}} K$,
$K$ transforms the $c$ numbers on its right into their complex conjugates,
and $z^{\dagger}=(\overline{z}_{\uparrow} \: \overline{z}_{\downarrow})$ is a $c$-number spinor.
$\hat{c}^{(\dagger)}_{\bm{r}} $ on the $\mathrm{A \Vert B}$ sublattice appears to have $s \Vert p$-wave-like parity regarding $\hat{\mathcal{P}}_{\mathrm{m}}$.
However, this interpretation is not static.
The parities are interchangeable by changing the inversion center from an A site to a B site.
Only relative parity has substantive meaning.

The four $\mathbb{Z}_{2}$ topological invariants $(\nu_{0};\nu_{1},\nu_{2},\nu_{3})$ can be used to extract topological information of quantum states $\{\Ket{\bm{k}n}\}$ from the Pfaffian of the matrix $\left[w\left(\bm{k}\right)\right]_{mn}=\Bra{-\bm{k}m}\hat{\mathcal{T}}\Ket{\bm{k}n}$ to classify three-dimensional topological insulators~\cite{Fu2007,Moore2007,Fu2007a,Roy2009}, where $\bm{k}$ and $\{m, n\}$ represent wavevector and band indices in a system without surfaces.
The Supplemental Material~\cite{SM} details the derivation of the invariants.
A system with $\nu_{0}=1$ is in the strong topological insulator phase.
The other invariants classify phases more finely and define weak topological insulators in the case of $\nu_{0}=0$.
We select two types of examples: (i) $t_{12}=t_{23}=t_{31}=t_{\mathrm{nn}}$ and (ii) $t_{\mu\nu}\neq 0$, $t_{\nu\lambda}=t_{\lambda\mu}=0$, where $[\mu,\nu,\lambda]=[1,2,3]\Vert[2,3,1]\Vert[3,1,2]$.
The strong topological insulator phase of type $(1;000)$ appears in case (i) under the condition $\left\vert v_{\mathrm{s}}+4t_{\mathrm{nn}}\right\vert<\left\vert 8t_{\mathrm{nn}}\right\vert$.
The weak topological insulator phase of type $(0;110)\Vert(0;011)\Vert(0;101)$ appears in case (ii) with $[\mu,\nu,\lambda]=[1,2,3]\Vert[2,3,1]\Vert[3,1,2]$ under the condition $\left\vert v_{\mathrm{s}}\right\vert<\left\vert 4t_{\mu\nu}\right\vert$.

The unique feature of the series of lattice models also emerges more prominently in the transition between $(1;000)$ and $(0;110)\Vert(0;011)\Vert(0;101)$. We can find at least three patterns of transitions: $(1;000) \leftrightarrow (0;110)\Vert(0;011)\Vert(0;101)$, $(1;000) \leftrightarrow (0;000) \leftrightarrow  (0;110)\Vert(0;011)\Vert(0;101)$, and $(1;000)  \leftrightarrow (1;110)\Vert(1;011)\Vert(1;101) \leftrightarrow (0;110)\Vert(0;011)\Vert(0;101)$.
The Supplemental Material~\cite{SM} details the alterations of the invariants around transition points in these processes.

\section{\label{sec:AnalyticalSolutions}Analytical Solutions of Surface States}
We represent a system with a pair of $(hkl)$ surfaces as $\mathrm{Sys}^{(hkl)}$ and its unconventional basis as $\{\bm{a}^{(hkl)}_{\mu}\}$.
Each $\mathrm{Sys}^{(hkl)}$ has periodic boundary conditions imposed along the directions parallel to the $(hkl)$ surfaces.
For convenience, we associate the three indices $[\alpha,\beta,\gamma]$ with $\mathrm{Sys}^{(hkl)}$ as below and use this association throughout the paper:
\begin{align}
&[\alpha,\beta,\gamma] =
\begin{cases}
[1,2,3] & \text{for $\mathrm{Sys}^{(100)}$},\\
[2,3,1] & \text{for $\mathrm{Sys}^{(010)}$},\\
[3,1,2] & \text{for $\mathrm{Sys}^{(001)}$}.
\end{cases}
\label{eq:axis}
\end{align}
The basis of $\mathrm{Sys}^{(100)\Vert(010)\Vert(001)}$ is given by
$\bm{a}^{(hkl)}_{\alpha} = \bm{a}_{\alpha}+\bm{a}_{\beta}$,
$\bm{a}^{(hkl)}_{\beta} = \bm{a}_{\beta}-\bm{a}_{\gamma}$, and
$\bm{a}^{(hkl)}_{\gamma} = \bm{a}_{\beta}+\bm{a}_{\gamma}$,
and the basis of $\mathrm{Sys}^{(111)}$ is given by
$\bm{a}^{(111)}_{1} = \bm{a}_{1}-\bm{a}_{3}$,
$\bm{a}^{(111)}_{2} = \bm{a}_{2}-\bm{a}_{3}$, and
$\bm{a}^{(111)}_{3} = \bm{a}_{1}+\bm{a}_{2}$.
The reciprocal lattice basis of $\mathrm{Sys}^{(hkl)}$, $\{\bm{b}^{(hkl)}_{\mu}\}$, is defined in a conventional manner from $\{\bm{a}^{(hkl)}_{\mu}\}$.
We introduce a two-dimensional wavevector space ($\tilde{\bm{k}}$ space) for each $\mathrm{Sys}^{(hkl)}$, which is generated by $\{\bm{b}^{(hkl)}_{\beta}, \bm{b}^{(hkl)}_{\gamma}\}$ for $\mathrm{Sys}^{(100)\Vert(010)\Vert(001)}$ and by $\{\bm{b}^{(111}_{1}, \bm{b}^{(111)}_{2}\}$ for $\mathrm{Sys}^{(111)}$.
The projected Brillouin zone of $\mathrm{Sys}^{(hkl)}$ is obtained by projecting the $\tilde{\bm{k}}$ space to the $(hkl)$ plane.
We also call the $\tilde{\bm{k}}$ space the projected Brillouin zone because of the one-to-one correspondence between them.

We partially Fourier transform the annihilation operators,
\begin{align}
& \hat{c}_{\mathrm{A}\tilde{\bm{k}}n} = \frac{1}{\sqrt{\tilde{N}}}\sum_{\tilde{\bm{R}}} e^{-i\tilde{\bm{k}}\cdot\tilde{\bm{R}}}\hat{c}_{\tilde{\bm{R}}+n\bm{a}^{(hkl)}_{\alpha}}
,
\nonumber\\
&
\hat{c}_{\mathrm{B}\tilde{\bm{k}}n} = \frac{1}{\sqrt{\tilde{N}}}\sum_{\tilde{\bm{R}}} e^{-i\tilde{\bm{k}}\cdot\tilde{\bm{R}}}\hat{c}_{\tilde{\bm{R}}+\tilde{\bm{\Delta}}^{(hkl)}+n\bm{a}^{(hkl)}_{\alpha}}
,
\nonumber\\
&\tilde{\bm{\Delta}}^{(hkl)} =
\begin{cases}
\bm{a}_{\beta} & \text{for $\mathrm{Sys}^{(100)\Vert(010)\Vert(001)}$},\\
\bm{a}_{3} & \text{for $\mathrm{Sys}^{(111)}$},
\end{cases}
\label{eq:Delta_tilde}
\end{align}
where $\tilde{\bm{R}}$ takes $\mathrm{A}$ sites within each layer, the association Eq.~(\ref{eq:axis}) is used for $\mathrm{Sys}^{(100)\Vert(010)\Vert(001)}$, and $\alpha=3$ for $\mathrm{Sys}^{(111)}$.
$\tilde{N}$ is the number of unit cells in each layer.
The index $n\in\mathbb{Z}$ labels each layer for $\mathrm{Sys}^{(100)\Vert(010)\Vert(001)}$ and each pair of layers for $\mathrm{Sys}^{(111)}$.
Note that a single layer in $\mathrm{Sys}^{(111)}$ contains only A or B sublattice sites.
Let $N_{\perp}$ be the number of layers for $\mathrm{Sys}^{(100)\Vert(010)\Vert(001)}$ and the number of pairs of layers for $\mathrm{Sys}^{(111)}$, and the layer index $n$ takes values from $\left\lfloor -\frac{N_{\perp}-1}{2} \right\rfloor$ to $\left\lfloor \frac{N_{\perp}-1}{2} \right\rfloor$, where $\left\lfloor x \right\rfloor$ is the floor function of $x$.
The partially Fourier-transformed Hamiltonian takes the following form,
\begin{align}
&\hat{H}=
\sum_{\tilde{\bm{k}}\in\mathrm{pBZ}}
\begin{pmatrix}
\hat{\Phi}^{\dagger}_{\mathrm{A}\tilde{\bm{k}}} & \hat{\Phi}^{\dagger}_{\mathrm{B}\tilde{\bm{k}}}
\end{pmatrix}
\mathcal{H}^{(hkl)}_{\tilde{\bm{k}}}
\begin{pmatrix}
\hat{\Phi}_{\mathrm{A}\tilde{\bm{k}}} \\
\hat{\Phi}_{\mathrm{B}\tilde{\bm{k}}}
\end{pmatrix}
,\nonumber\\
&\mathcal{H}^{(hkl)}_{\tilde{\bm{k}}}
=
\begin{pmatrix}
\mathcal{M}^{(hkl)}_{\tilde{\bm{k}}}  &e^{-i\tilde{\bm{k}}\cdot\tilde{\bm{\Delta}}^{(hkl)}}\mathcal{T}^{(hkl)\dagger}_{\tilde{\bm{k}}} \\
e^{i\tilde{\bm{k}}\cdot\tilde{\bm{\Delta}}^{(hkl)}}\mathcal{T}^{(hkl)}_{\tilde{\bm{k}}}  & -\mathcal{M}^{(hkl)}_{\tilde{\bm{k}}}
\end{pmatrix}
,\nonumber\\
&\hat{\Phi}_{X\tilde{\bm{k}}}
=
\begin{pmatrix}
\hat{c}_{X\tilde{\bm{k}}\left\lfloor -\frac{N_{\perp}-1}{2} \right\rfloor}
& \cdots
& \hat{c}_{X\tilde{\bm{k}}\left\lfloor \frac{N_{\perp}-1}{2} \right\rfloor}
\end{pmatrix}
^{\mathrm{T}}
,
\end{align}
where $\mathrm{pBZ}$ means the projected first Brillouin zone and $X =\mathrm{A},\mathrm{B}$.
The matrices $\mathcal{M}^{(hkl)}_{\tilde{\bm{k}}}$ and $\mathcal{T}^{(hkl)}_{\tilde{\bm{k}}}$ are given in the Supplemental Material~\cite{SM}.
For later convenience, we introduce the symbols $\{\mu^{(hkl)}_{\tilde{\bm{k}}}\}$ to represent mass-gap functions related to the off-diagonal part of $\mathcal{M}^{(hkl)}_{\tilde{\bm{k}}}$.

Here, we bring in the ansatz for the surface-state eigenvectors of the matrix $\mathcal{H}^{(hkl)}_{\tilde{\bm{k}}}$ by assuming that each surface-state eigenvector layer-dependent component is decoupled from its spin- and sublattice-dependent components,
$u^{(hkl)}_{Y\tilde{\bm{k}}\pm}=v^{(hkl)}_{Y\tilde{\bm{k}}\pm} \otimes w^{(hkl)}_{Y\tilde{\bm{k}}\pm}$ ($Y=\mathrm{U}, \mathrm{L}$).
$u^{(hkl)}_{\mathrm{U}\tilde{\bm{k}}\pm}$ and $u^{(hkl)}_{\mathrm{L}\tilde{\bm{k}}\pm}$ represent the eigenvectors of surface states localized around the upper and lower surfaces, respectively.
$\{v^{(hkl)}_{Y\tilde{\bm{k}}\pm}\}_{Y=\mathrm{U}, \mathrm{L}}$ are their spin- and sublattice-dependent components, and $\{w^{(hkl)}_{Y\tilde{\bm{k}}\pm}\}_{Y=\mathrm{U},\mathrm{L}}$ are their layer-dependent components.

Let $\chi^{(hkl)}_{\tilde{\bm{k}}\pm}$ be $c$-number spinors.
The ansatz for $\{v^{(100)\Vert(010)\Vert(001)}_{Y\tilde{\bm{k}}\pm}\}_{Y=\mathrm{U}, \mathrm{L}}$ is expressed as
\begin{align}
v^{(hkl)}_{\mathrm{U}\tilde{\bm{k}}\pm}
&\propto
\begin{pmatrix}
\chi^{(hkl)}_{\tilde{\bm{k}}\pm}\\
-e^{i\tilde{\bm{k}}\cdot\tilde{\bm{\Delta}}^{(hkl)}}
\mathrm{sgn}\left(t_{\alpha}\mu^{(hkl)}_{\tilde{\bm{k}}} \right)
U_{\alpha}\chi^{(hkl)}_{\tilde{\bm{k}}\pm}
\end{pmatrix}
,
\nonumber\\
v^{(hkl)}_{\mathrm{L}\tilde{\bm{k}}\pm}
&\propto
\begin{pmatrix}
\chi^{(hkl)}_{\tilde{\bm{k}}\pm} \\
e^{i\tilde{\bm{k}}\cdot\tilde{\bm{\Delta}}^{(hkl)}}
\mathrm{sgn}\left(t_{\alpha}\mu^{(hkl)}_{\tilde{\bm{k}}} \right)
U_{\alpha}\chi^{(hkl)}_{\tilde{\bm{k}}\pm}
\end{pmatrix}
,
\end{align}
and the ansatz for $\{v^{(111)}_{Y\tilde{\bm{k}}\pm}\}_{Y=\mathrm{U}, \mathrm{L}}$ is
\begin{align}
v^{(111)}_{\mathrm{U}\tilde{\bm{k}}\pm}
&\propto
\begin{pmatrix}
-\gamma_{\tilde{\bm{k}}\pm}
\chi^{(111)}_{\tilde{\bm{k}}\pm}
\\
\frac{\mu^{(111)}_{\tilde{\bm{k}}}}{\left\vert \mu^{(111)}_{\tilde{\bm{k}}} \right\vert}
\frac{e^{i\tilde{\bm{k}}\cdot\tilde{\bm{\Delta}}^{(111)}}}{\left\vert \epsilon_{\tilde{\bm{k}}\pm}\right\vert}
\tilde{T}^{(111)}_{-\tilde{\bm{k}}}
\chi^{(hkl)}_{\tilde{\bm{k}}\pm}
\end{pmatrix}
,
\nonumber\\
v^{(111)}_{\mathrm{L}\tilde{\bm{k}}\pm}
&\propto
\begin{pmatrix}
\chi^{(111)}_{\tilde{\bm{k}}\pm} \\
\gamma_{\tilde{\bm{k}}\pm}
\frac{\mu^{(111)}_{\tilde{\bm{k}}}}{\left\vert \mu^{(111)}_{\tilde{\bm{k}}} \right\vert}
\frac{e^{i\tilde{\bm{k}}\cdot\tilde{\bm{\Delta}}^{(111)}}}{\left\vert \epsilon_{\tilde{\bm{k}}\pm}\right\vert}
\tilde{T}^{(111)}_{-\tilde{\bm{k}}}\chi^{(111)}_{\tilde{\bm{k}}\pm}
\end{pmatrix}
,
\nonumber\\
\tilde{T}^{(111)}_{\tilde{\bm{k}}}
&=  t_{1}e^{-i\tilde{\bm{k}}\cdot\bm{a}_{1}}U_{1}
+t_{2}e^{-i\tilde{\bm{k}}\cdot\bm{a}_{2}}U_{2}
+t_{3}e^{-i\tilde{\bm{k}}\cdot\bm{a}_{3}}U_{3}
.
\end{align}
The spinors $\chi^{(hkl)}_{\tilde{\bm{k}}\pm}$ are given as eigenvectors of simple $2\times2$ Hermitian matrices.
The scalar-valued parameters $\gamma_{\tilde{\bm{k}}\pm}$ are determined by preliminary adjustments that simplify the action of $\mathcal{H}^{(111)}_{\tilde{\bm{k}}}$ on $\{u^{(111)}_{Y\tilde{\bm{k}}\pm}\}_{Y=\mathrm{U},\mathrm{L}}$.
The Supplemental Material~\cite{SM} provides further details.

The action of $\mathcal{H}^{(hkl)}_{\tilde{\bm{k}}}$ on $\{u^{(hkl)}_{Y\tilde{\bm{k}}\pm}\}_{Y=\mathrm{U},\mathrm{L}}$ takes the form
\begin{align}
\mathcal{H}^{(hkl)}_{\tilde{\bm{k}}} u^{(hkl)}_{Y\tilde{\bm{k}}\pm}
&=\left(E^{(hkl)}_{Y\tilde{\bm{k}}\pm}
+\Gamma^{(hkl)}_{Y\tilde{\bm{k}}\pm}
\otimes\Omega^{(hkl)}_{Y\tilde{\bm{k}}\pm}\right) u^{(hkl)}_{Y\tilde{\bm{k}}\pm}
\label{eq:Hhkl-eigeneq}
,
\end{align}
where $\{E^{(hkl)}_{Y\tilde{\bm{k}}\pm}\}_{Y=\mathrm{U},\mathrm{L}}$ are the candidates of the energy eigenvalues of surface states.
$\{\Gamma^{(hkl)}_{Y\tilde{\bm{k}}\pm}\}_{Y=\mathrm{U},\mathrm{L}}$ and $\{\Omega^{(hkl)}_{Y\tilde{\bm{k}}\pm}\}_{Y=\mathrm{U},\mathrm{L}}$ represent $2\times 2$ matrices with sublattice indices and $N_{\perp}\times N_{\perp}$ tridiagonal matrices with layer indices.
Equation~(\ref{eq:Hhkl-eigeneq}) suggests that $\{u^{(hkl)}_{Y\tilde{\bm{k}}\pm}\}_{Y=\mathrm{U},\mathrm{L}}$ can be the eigenvectors of $\mathcal{H}^{(hkl)}_{\tilde{\bm{k}}}$ with eigenvalues $\{E^{(hkl)}_{Y\tilde{\bm{k}}\pm}\}_{Y=\mathrm{U},\mathrm{L}}$ when $\Omega^{(hkl)}_{Y\tilde{\bm{k}}\pm} w^{(hkl)}_{Y\tilde{\bm{k}}\pm} =0$ ($Y=\mathrm{U},\mathrm{L}$).
Since $\{\Omega^{(hkl)}_{Y\tilde{\bm{k}}\pm}\}_{Y=\mathrm{U},\mathrm{L}}$ are tridiagonal, their eigenvalue problems are tractable and formally solvable except for the boundary condition issue.
We adopt the strategy of taking the semi-infinite limit $N_{\perp}\to\infty$ instead of strictly treating the boundary conditions of a finite-thickness system.
Every system under consideration is assumed to be sufficiently thicker than any penetration depth of its surface states.
We introduce, for convenience, other layer indices $\{n_{Y}\}_{Y=\mathrm{U},\mathrm{L}}$ related to the original index $n$ ($\left\lfloor -\frac{N_{\perp}-1}{2} \right\rfloor\leq n \leq \left\lfloor \frac{N_{\perp}-1}{2} \right\rfloor$) as
$
n_{\mathrm{U}} = \left\lfloor\frac{N_{\perp}+1}{2}\right\rfloor -n
$
and
$
n_{\mathrm{L}} = \left\lceil\frac{N_{\perp}+1}{2}\right\rceil +n
$
($1\le n_{\mathrm{U},\mathrm{L}} \le N_{\perp}$)
,
where $\left\lceil x \right\rceil$ is the ceiling function of $x$.
Taking into account the boundary conditions immediately outside the surfaces,
$\left[w^{(hkl)}_{Y\tilde{\bm{k}}\pm}\right]_{n} = 0$ at $n_{Y}=0$ ($Y=\mathrm{U}, \mathrm{L}$),
the solutions are constructed as
\begin{align}
\left[w^{(hkl)}_{\mathrm{U}\tilde{\bm{k}}S}\right]_{n}
&\propto
\frac{1}{2i}\left(
\overline{\varphi}^{(hkl)n_{\mathrm{U}}}_{\tilde{\bm{k}}S+}-\overline{\varphi}^{(hkl)n_{\mathrm{U}}}_{\tilde{\bm{k}}S-}
\right)
,
\nonumber\\
\left[w^{(hkl)}_{\mathrm{L}\tilde{\bm{k}}S}\right]_{n}
&\propto
\frac{1}{2i}\left(
\varphi^{(hkl)n_{\mathrm{L}}}_{\tilde{\bm{k}}S+}-\varphi^{(hkl)n_{\mathrm{L}}}_{\tilde{\bm{k}}S-}
\right),
\end{align}
where $\{\varphi^{(hkl)}_{\tilde{\bm{k}}S\pm}\}_{S=\pm}$ are given in the Supplemental Material~\cite{SM}.
The boundary conditions on the decaying side of surface states complicate the problem.
For the semi-infinite limit, requiring that $\left\vert w^{(hkl)}_{Y\tilde{\bm{k}}\pm}\right\vert$ decay as the layer index $n_{Y}$ increases, where $Y=\mathrm{U}, \mathrm{L}$, is sufficient.
The explicit expression of the attenuation condition for each $\mathrm{Sys}^{(hkl)}$ appears in the Supplemental Material~\cite{SM}.

\section{\label{sec:Comparison}Comparison with Numerical Solutions}
To evaluate the quality of the surface-state analytical solutions obtained above in the semi-infinite-thick limit, we compare them with corresponding numerical solutions at finite thickness.
Taking $t_{\mathrm{n}}$ as an energy unit, we select models with $t_{\mu}=t_{\mathrm{n}}$, $\phi_{\mu}=\pi/3$ and $\bm{n}_{\mu} =\left(-\bm{a}_{\mu}+\bm{a}_{\nu}+\bm{a}_{\lambda}\right)/\left(\sqrt{3}a\right)$, where $[\mu,\nu,\lambda]=[1,2,3], [2,3,1], [3,1,2]$.
Anisotropy is controlled through $\{t_{\mu\nu}\}$.
The evaluations are performed for an isotropic strong topological insulator (TI) and an anisotropic weak topological insulator with the parameter sets TI-$t_{\mathrm{nn}}$ and TI-$t_{12}$ listed in Table~\ref{table:Parameter_Sets}.
All numerical results given below are obtained with thicknesses of 60 pairs of layers for $\mathrm{Sys}^{(111)}$ and 60 layers for $\mathrm{Sys}^{(100)\Vert(001)}$.
Each layer contains $360\times 360$ unit cells.

\begin{table}[h]
 \caption{Parameter sets}
 \label{table:Parameter_Sets}
 \centering
  \begin{tabular}{l c rrr c r c l}
    \hline \hline
    Set name&& $ t_{12}/t_{\mathrm{n}}$ & $t_{23}/t_{\mathrm{n}}$ & $t_{31}/t_{\mathrm{n}}$ && $v_{\mathrm{s}}/t_{\mathrm{n}}$&  &Model system\\
    \hline
    TI-$t_{\mathrm{nn}}$       && $-0.150$ & $-0.150$ & $-0.150$ && $ 0.800$ && Strong TI\\
    TI-$t_{12}$      && $-0.300$ & $ 0.000$ & $ 0.000$ && $ 0.200$ && Weak TI\\
    \hline \hline
  \end{tabular}
\end{table}

A few notes are necessary before proceeding with specific examples.
An eigenvector has complete information about its eigenvalue, while the eigenvalue has only limited information about the eigenvector.
The attenuation condition is sufficient for a given analytical solution, i.e., an eigenvector candidate, to be a surface-state solution.
Thus, when the condition holds, the eigenvalue of the solution matches its corresponding numerical result by necessity.
However, the coincidence between an eigenvalue candidate and a numerical result is a necessary condition.
An eigenvalue candidate can approximately or accidentally coincide with a bulk-state eigenvalue in the proximity of bulk dispersions.
Such a coincidence does not guarantee that an eigenvector candidate with such an eigenvalue represents a surface state.

\begin{figure}[htb]
\vspace{5mm}
\begin{tabular}{ccc}
\begin{overpic}[scale=0.25, clip]{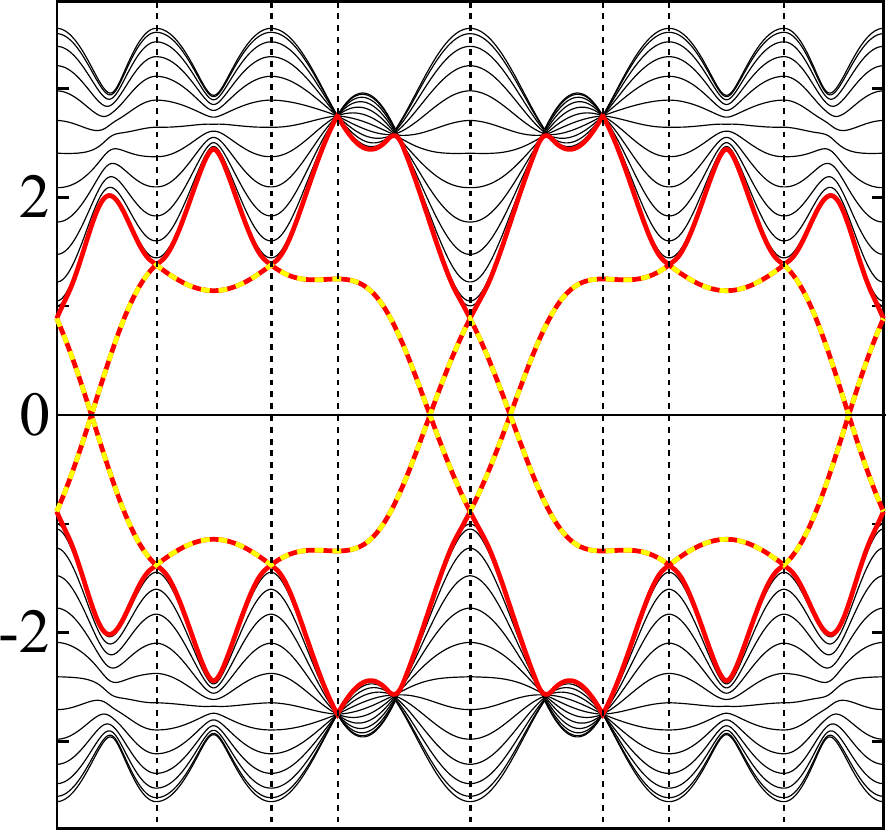}
\thicklines
\put( -6, 102){(a)}
\put( -8, 43){\rotatebox{90}{$\epsilon/t_{\mathrm{n}}$}}
\put(   4,    -8){$\Gamma$}
\put(  14, 102){$\mathrm{M}_{2}$}
\put(  28, 102){$\mathrm{M}_{3}$}
\put(  37,    -8){$\mathrm{K}'$}
\put(  54,    -8){$\Gamma$}
\put( 69,    -8){$\mathrm{K}$}
\put( 76, 102){$\mathrm{M}_{2}$}
\put( 90, 102){$\mathrm{M}_{1}$}
\put(103,   -8){$\Gamma$}
\end{overpic}
&
\hspace{2mm}
&
\begin{overpic}[scale=0.25, clip]{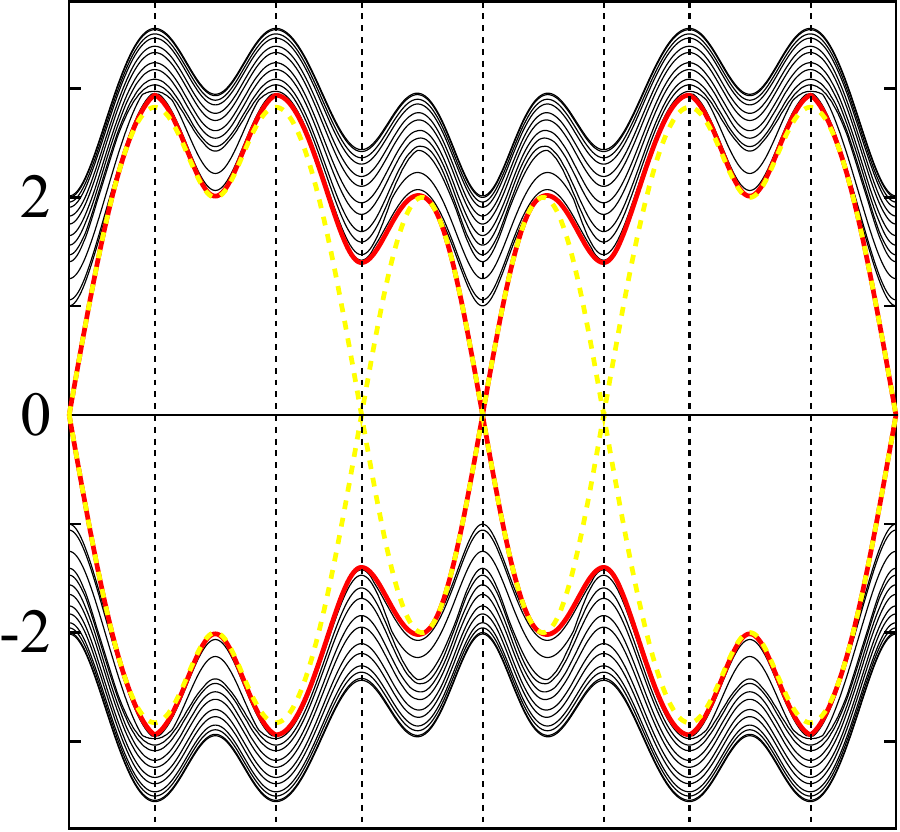}
\thicklines
\put( -6, 102){(b)}
\put(    5,  -8){$\Gamma$}
\put(  15, 102){$\mathrm{X}'$}
\put(  30, 102){$\overline{\mathrm{X}}$}
\put(  39,  -8){$\mathrm{M}'$}
\put(  55,  -8){$\Gamma$}
\put(  68,  -8){$\mathrm{M}$}
\put(  80, 102){$\mathrm{X}'$}
\put(  95, 102){$\mathrm{X}$}
\put(105,  -8){$\Gamma$}
\end{overpic}
\end{tabular}
\caption{
Projected band diagrams of a strong topological insulator ($t_{\mathrm{nn}}/t_{\mathrm{n}}=-0.15, v_{\mathrm{s}}/t_{\mathrm{n}}=0.8$) in a slab shape with (a) $(111)$ surfaces and (b) $(100)$ surfaces.
}
\label{fig:band-diagrams_strong-TI_tnn}
\end{figure}

\begin{figure}[t]
\vspace{5mm}
\begin{tabular}{ccc}
\begin{overpic}[scale=0.25, clip]{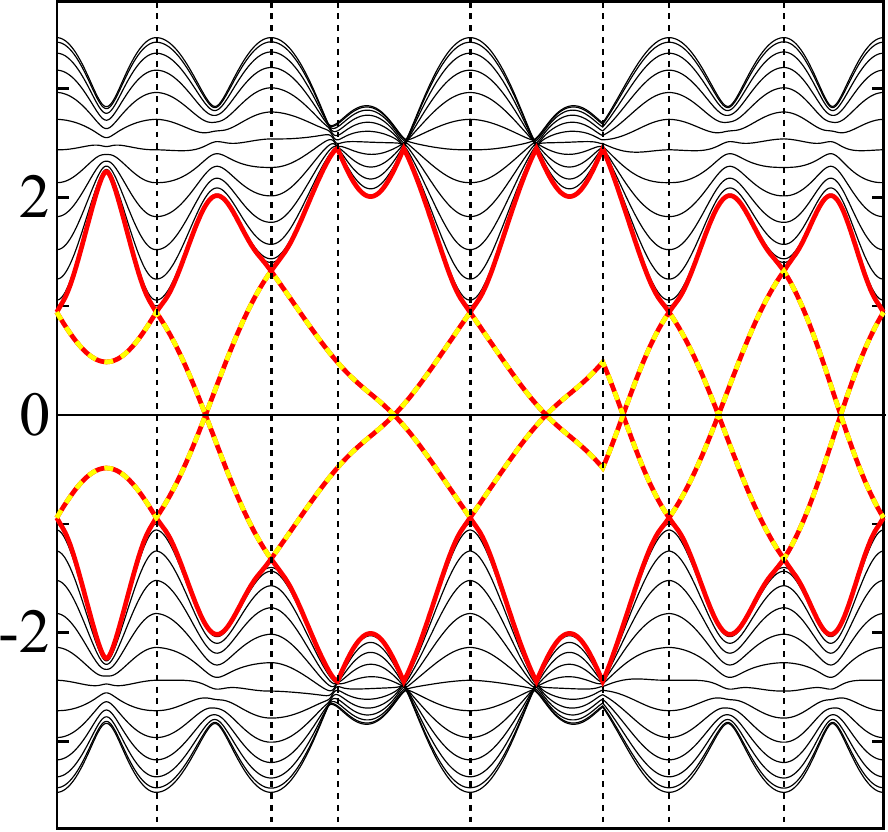}
\thicklines
\put( -6, 102){(a)}
\put( -8, 43){\rotatebox{90}{$\epsilon/t_{\mathrm{n}}$}}
\put(   4,    -8){$\Gamma$}
\put(  14, 102){$\mathrm{M}_{2}$}
\put(  28, 102){$\mathrm{M}_{3}$}
\put(  37,    -8){$\mathrm{K}'$}
\put(  54,    -8){$\Gamma$}
\put( 69,    -8){$\mathrm{K}$}
\put( 76, 102){$\mathrm{M}_{2}$}
\put( 90, 102){$\mathrm{M}_{1}$}
\put(103,   -8){$\Gamma$}
\end{overpic}
&
\hspace{2mm}
&
\begin{overpic}[scale=0.25, clip]{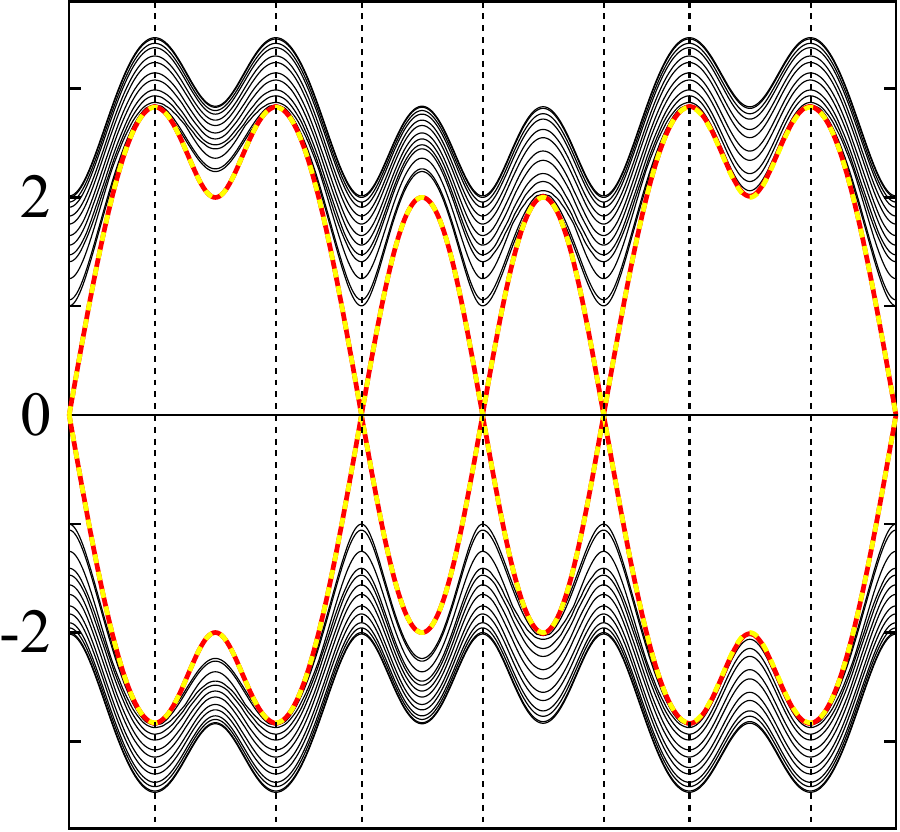}
\thicklines
\put( -6, 102){(b)}
\put(    5,  -8){$\Gamma$}
\put(  15, 102){$\mathrm{X}'$}
\put(  30, 102){$\overline{\mathrm{X}}$}
\put(  39,  -8){$\mathrm{M}'$}
\put(  55,  -8){$\Gamma$}
\put(  68,  -8){$\mathrm{M}$}
\put(  80, 102){$\mathrm{X}'$}
\put(  95, 102){$\mathrm{X}$}
\put(105,  -8){$\Gamma$}
\end{overpic}
\end{tabular}
\vspace{0mm}
\caption{
Projected band diagrams of a weak topological insulator ($t_{12}/t_{\mathrm{n}}=-0.3, v_{\mathrm{s}}/t_{\mathrm{n}}=0.2$) in a slab shape with (a) $(111)$ surfaces and (b) $(100)$ surfaces.
}
\label{fig:band-diagrams_weak-TI_t12}
\end{figure}
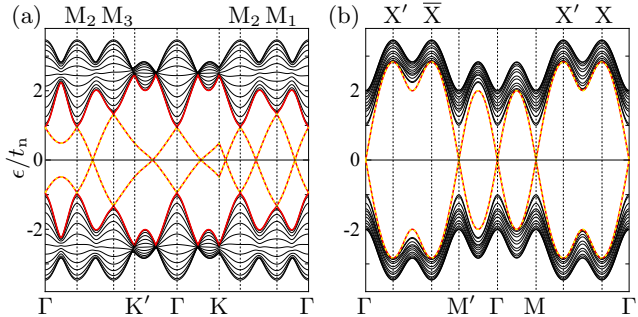

Figures~\ref{fig:band-diagrams_strong-TI_tnn} and \ref{fig:band-diagrams_weak-TI_t12} show the projected band diagrams of TI-$t_{\mathrm{nn}}$ and TI-$t_{12}$, respectively.
Panels (a) and (b) of each figure correspond to $\mathrm{Sys}^{(111)}$ and $\mathrm{Sys}^{(100)}$.
The definitions of symmetry points appear in the Supplemental Material~\cite{SM}.
The band dispersions are plotted at every ten indices from the top and bottom with thin black lines, and the middle four bands are plotted with thick red lines.
Dashed yellow lines represent the candidates of energy eigenvalues of surface states, while only the middle two of the four appear in Figs.~\ref{fig:band-diagrams_strong-TI_tnn}(a) and \ref{fig:band-diagrams_weak-TI_t12}(a).
When the analytical solutions satisfy the attenuation condition, the corresponding parts of the dashed yellow lines precisely overlap the red lines, forming vivid two-tone-colored dashed lines.
The faint dashed yellow lines suggest that the corresponding candidates violate the attenuation condition and are no longer appropriate solutions.

%%%%%%%%%%%%%%%%%%%%%%%%%%%%%%%%%%%%%%%%%%%%%%%%%%
% Figs of Band Diagrams of topological insulators
%%%%%%%%%%%%%%%%%%%%%%%%%%%%%%%%%%%%%%%%%%%%%%%%%%T

%%%%%%%%%%%%%%%%%%%%%%%%%%%%%%%%%%%%%%%%%%%%%%%%%%

The yellow-red two-tone-colored lines in Figs.~\ref{fig:band-diagrams_strong-TI_tnn}(a) and \ref{fig:band-diagrams_weak-TI_t12}(a) correspond to the lower half of the massless Dirac-type surface bands at the upper $(111)$ surface and the upper half of the massless Dirac-type surface bands at the lower $(111)$ surface.
The eigenvectors of these bands satisfy the attenuation condition over most of the symmetry lines of the projected Brillouin zone.
However, for the other halves, the attenuation condition only holds within small domains around the time-reversal-invariant symmetry points.
Such bands (not shown) do not overlap with the bands of corresponding numerical solutions (the lowermost and uppermost red lines) except for in these small domains.

In contrast to the $(111)$ surface bands, the $(100)$ surface bands at the upper and lower surfaces are degenerate.
For the $(100)$ surface states of TI-$t_{\mathrm{nn}}$, the attenuation condition holds around $\Gamma$ and breaks down around $M$ and $M'$.
The numerical results confirm that there are no well-isolated surface states except for around $\Gamma$.
Although the attenuation condition breaks down also along $X'\overline{X}$ and $X' X$, in the proximity of bulk dispersions, the analytical and numerical dispersions can appear to overlap.
For the $(100)$ surface states of TI-$t_{12}$, the attenuation condition holds over a large portion except for around the midpoint of $\Gamma M$ and on $X'X$.
The situation for the $(001)$ surface states (not shown) depends more drastically on the model systems TI-$t_{\mathrm{nn}}$ or TI-$t_{12}$.
For TI-$t_{\mathrm{nn}}$, the $(001)$ surface states are almost identical to the $(100)$ surface states.
For TI-$t_{12}$, the $(001)$ surface-state candidates satisfy the attenuation condition nowhere.
Correspondingly, the weak topological insulator of TI-$t_{12}$ has no $(001)$ surface states.

As discussed at the end of Sec.~\ref{sec:ModelHamiltonian}, the transitions between the phases $(1;000)$ and $(0;110)\Vert(0;011)\Vert(0;101)$ also bring out the unique feature of the series of lattice models. For example, between TI-$t_{\mathrm{nn}}$ and TI-$t_{12}$, a direct transition of type $(1;000) \leftrightarrow (0;110)$ emerges.
The Supplemental Material~\cite{SM} presents the band diagrams of a semimetal at the transition point, which suggests that a discussion similar to the above also holds even for semimetals at transition points.

%%%%%%%%%%%%%%%%%%%%%%%%%%%%%%%%%%%%%%%%%%%%%%%%%%
% Figs of Charge and Spin Density Distributions
%%%%%%%%%%%%%%%%%%%%%%%%%%%%%%%%%%%%%%%%%%%%%%%%%%
\begin{figure}[htb]
\vspace{5mm}
\begin{tabular}{cc}
\begin{overpic}[scale=0.25, clip]{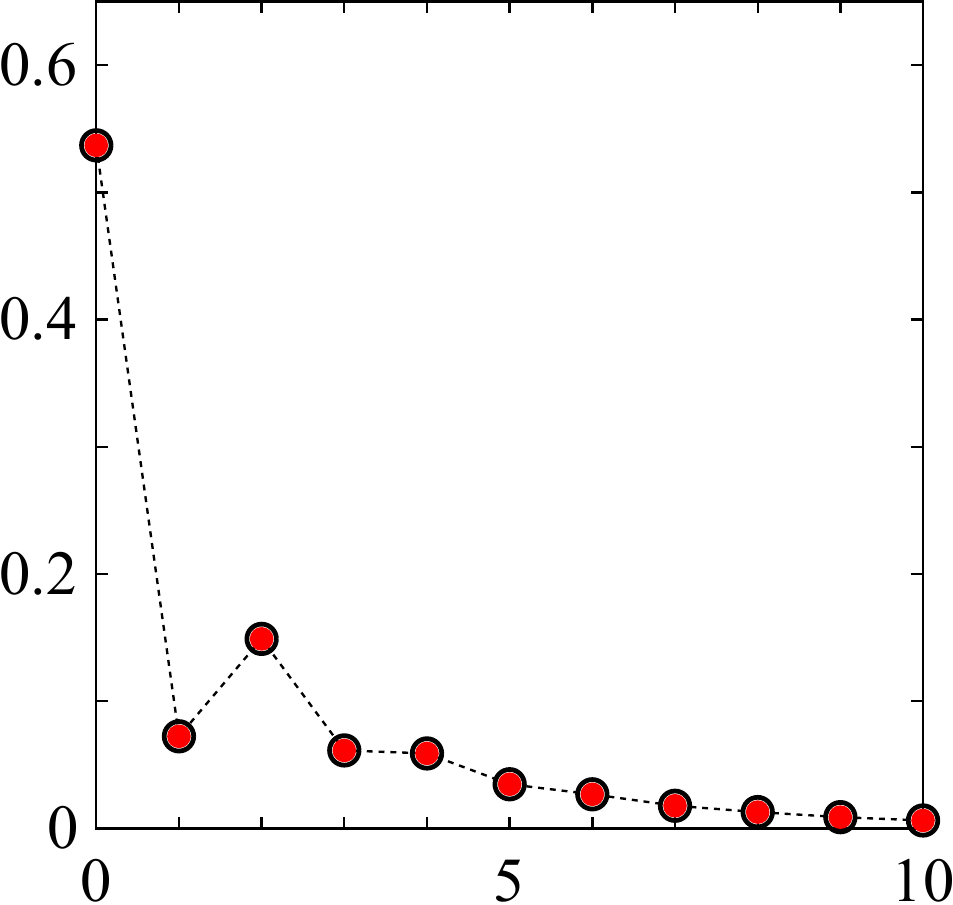}
\thicklines
\put( -2, 110){(a)}
\put( 58, -9){$n_{\mathrm{L}}$}
\end{overpic}
&
\begin{overpic}[scale=0.25, clip]{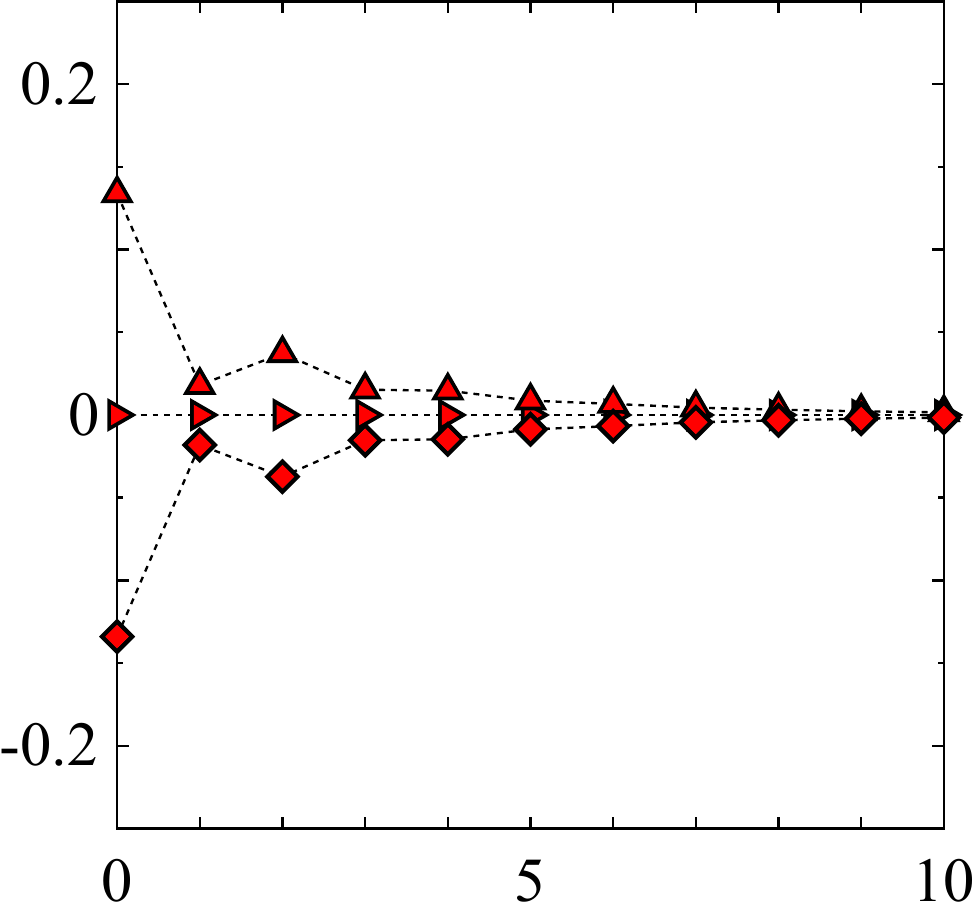}
\thicklines
\put( 0, 110){(b)}
\put( 80,  49){ \rotatebox{-90}{$\blacktriangle$}}
\put( 90,  44){ : $S_{1}$}
\put( 82,  32){$\blacktriangle$}
\put( 90,  32){ : $S_{2}$}
\put( 82,  19){\rotatebox{45}{\tiny$\blacksquare$}}
\put( 90,  20){ : $S_{3}$}
\put( 61, -9){$n_{\mathrm{L}}$}
\end{overpic}
\end{tabular}
\caption{
(a) Charge and (b) spin density distributions of a surface state
of the strong topological insulator (TI-$t_{\mathrm{nn}}$)
near the lower $(001)$ surface.
}
\label{fig:charge&spin_strong-TI_tnn}
\end{figure}
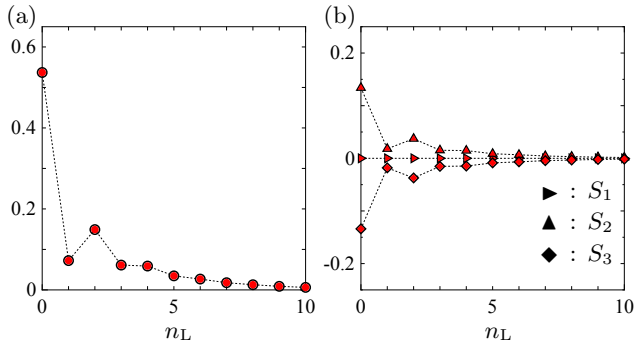
%%%%%%%%%%%%%%%%%%%%%%%%%%%%%%%%%%%%%%%%%%%%%%%%%%

We also examine surface-state eigenvectors through their charge and spin density distributions.
Figures~\ref{fig:charge&spin_strong-TI_tnn}(a) and \ref{fig:charge&spin_strong-TI_tnn}(b) show the charge and spin density distributions of a surface state of TI-$t_{\mathrm{nn}}$ near the lower $(001)$ surface.
The wavevector of the surface state is $\tilde{\bm{k}}=7/40\left(-\bm{b}^{(001)}_{1}+\bm{b}^{(001)}_{2}\right)$.
Solid red and open black marks represent the analytical and numerical results, respectively.
Circles represent the charge density. Laterally facing triangles, upward triangles, and diamonds represent the first, second, and third components of the spin density, respectively.
The analytical solution coincides nicely with the numerical solution in the charge and spin density distributions.
We also find good coincidence between the analytical and numerical results in other cases, including surface-state solutions of ordinary insulators, as long as the attenuation condition holds for such solutions.
Verifying the bulk-boundary correspondence in the context of Ref.~\cite{Rhim2018}, which treats topological and nontopological phases on an equal footing, would be worthy of research in the future.

\section{\label{sec:Discussions}Discussion}
We derived analytical solutions of the surface states in the series of lattice models defined by Eq.~(\ref{eq:H}).
The surface-state solvability comes from the restriction on spin-flip matrices in nearest-neighbor hopping.
However, the models with this restriction can still describe various topological surfaces.
Finally, we discuss some future directions below.

The analytical solutions of surface states will reduce the computational costs in the numerical analysis of phenomena on topological surfaces beyond the limitation of complete modeling, including bulk states.
We can implement them in real-time simulations of the transport dynamics of helical surface states.
Extending them to interface-state solutions would aid the study of dynamics near the boundary between a topological insulator and a conductor.
This would reveal the conversion process from a homogeneously spin-polarized wave packet to a wave packet with toroidal spin polarization by the reflection at the topological interface~\cite{Onoda2012, Ohmura2012, Onoda2019}.
The analytical solutions can also serve as a basis for studying the effects of interactions between surface states.
Building them into self-consistent equations for magnetic and excitonic instabilities on topological surfaces as a complementary approach to the effective model of a pair of topological surfaces~\cite{Marchand2012} is worthwhile.
Beyond mean-field analysis, the analytical solutions can be building blocks for constructing effective models of interacting surface states to reveal the Berezinski\u{\i}-Kosterlitz-Thouless-type transitions~\cite{Berezinskii1971, Berezinskii1972, Kosterlitz1973, Kosterlitz1974} on topological surfaces.

Topological interfaces are now emerging in various artificial systems, e.g., photonic and phononic lattices, resonator networks, and cold atoms in optical lattices.
Ideal situations are more likely to be realized in such systems than in electronic systems.
The simplicity and designability of our series can be an advantage in designing novel functionalities of such artificial topological surfaces.

\begin{acknowledgments}
JSPS KAKENHI Grant No. JP16K05467 supported this work in its early stages.
\end{acknowledgments}

%\nocite{*}

%


\begin{thebibliography}{70}%
\makeatletter
\providecommand \@ifxundefined [1]{%
 \@ifx{#1\undefined}
}%
\providecommand \@ifnum [1]{%
 \ifnum #1\expandafter \@firstoftwo
 \else \expandafter \@secondoftwo
 \fi
}%
\providecommand \@ifx [1]{%
 \ifx #1\expandafter \@firstoftwo
 \else \expandafter \@secondoftwo
 \fi
}%
\providecommand \natexlab [1]{#1}%
\providecommand \enquote  [1]{``#1''}%
\providecommand \bibnamefont  [1]{#1}%
\providecommand \bibfnamefont [1]{#1}%
\providecommand \citenamefont [1]{#1}%
\providecommand \href@noop [0]{\@secondoftwo}%
\providecommand \href [0]{\begingroup \@sanitize@url \@href}%
\providecommand \@href[1]{\@@startlink{#1}\@@href}%
\providecommand \@@href[1]{\endgroup#1\@@endlink}%
\providecommand \@sanitize@url [0]{\catcode `\\12\catcode `\$12\catcode
  `\&12\catcode `\#12\catcode `\^12\catcode `\_12\catcode `\%12\relax}%
\providecommand \@@startlink[1]{}%
\providecommand \@@endlink[0]{}%
\providecommand \url  [0]{\begingroup\@sanitize@url \@url }%
\providecommand \@url [1]{\endgroup\@href {#1}{\urlprefix }}%
\providecommand \urlprefix  [0]{URL }%
\providecommand \Eprint [0]{\href }%
\providecommand \doibase [0]{https://doi.org/}%
\providecommand \selectlanguage [0]{\@gobble}%
\providecommand \bibinfo  [0]{\@secondoftwo}%
\providecommand \bibfield  [0]{\@secondoftwo}%
\providecommand \translation [1]{[#1]}%
\providecommand \BibitemOpen [0]{}%
\providecommand \bibitemStop [0]{}%
\providecommand \bibitemNoStop [0]{.\EOS\space}%
\providecommand \EOS [0]{\spacefactor3000\relax}%
\providecommand \BibitemShut  [1]{\csname bibitem#1\endcsname}%
\let\auto@bib@innerbib\@empty
%</preamble>
\bibitem [{\citenamefont {Moore}(2010)}]{Moore2010}%
  \BibitemOpen
  \bibfield  {author} {\bibinfo {author} {\bibfnamefont {J.~E.}\ \bibnamefont
  {Moore}},\ }\href {https://doi.org/10.1038/nature08916} {\bibfield  {journal}
  {\bibinfo  {journal} {Nature}\ }\textbf {\bibinfo {volume} {464}},\ \bibinfo
  {pages} {194} (\bibinfo {year} {2010})}\BibitemShut {NoStop}%
\bibitem [{\citenamefont {Hasan}\ and\ \citenamefont
  {Kane}(2010)}]{Hasan2010a}%
  \BibitemOpen
  \bibfield  {author} {\bibinfo {author} {\bibfnamefont {M.~Z.}\ \bibnamefont
  {Hasan}}\ and\ \bibinfo {author} {\bibfnamefont {C.~L.}\ \bibnamefont
  {Kane}},\ }\href {https://doi.org/10.1103/RevModPhys.82.3045} {\bibfield
  {journal} {\bibinfo  {journal} {Rev. Mod. Phys.}\ }\textbf {\bibinfo {volume}
  {82}},\ \bibinfo {pages} {3045} (\bibinfo {year} {2010})}\BibitemShut
  {NoStop}%
\bibitem [{\citenamefont {Qi}\ and\ \citenamefont {Zhang}(2011)}]{Qi2011}%
  \BibitemOpen
  \bibfield  {author} {\bibinfo {author} {\bibfnamefont {X.-L.}\ \bibnamefont
  {Qi}}\ and\ \bibinfo {author} {\bibfnamefont {S.-C.}\ \bibnamefont {Zhang}},\
  }\href {https://doi.org/10.1103/RevModPhys.83.1057} {\bibfield  {journal}
  {\bibinfo  {journal} {Rev. Mod. Phys.}\ }\textbf {\bibinfo {volume} {83}},\
  \bibinfo {pages} {1057} (\bibinfo {year} {2011})}\BibitemShut {NoStop}%
\bibitem [{\citenamefont {Shen}(2012)}]{Shen2012}%
  \BibitemOpen
  \bibfield  {author} {\bibinfo {author} {\bibfnamefont {S.-Q.}\ \bibnamefont
  {Shen}},\ }\href {https://doi.org/10.1007/978-3-642-32858-9} {\emph {\bibinfo
  {title} {{Topological Insulators}}}},\ \bibinfo {edition} {2nd}\ ed.,\
  \bibinfo {series} {Springer Series in Solid-State Sciences}, Vol.\ \bibinfo
  {volume} {174}\ (\bibinfo  {publisher} {Springer, Berlin},\ \bibinfo {year}
  {2012})\BibitemShut {NoStop}%
\bibitem [{\citenamefont {Ando}(2013)}]{Ando2013}%
  \BibitemOpen
  \bibfield  {author} {\bibinfo {author} {\bibfnamefont {Y.}~\bibnamefont
  {Ando}},\ }\href {https://doi.org/10.7566/JPSJ.82.102001} {\bibfield
  {journal} {\bibinfo  {journal} {J. Phys. Soc. Japan}\ }\textbf {\bibinfo
  {volume} {82}},\ \bibinfo {pages} {102001} (\bibinfo {year}
  {2013})}\BibitemShut {NoStop}%
\bibitem [{\citenamefont {Bernevig}(2013)}]{Bernevig2013}%
  \BibitemOpen
  \bibfield  {author} {\bibinfo {author} {\bibfnamefont {B.~A.}\ \bibnamefont
  {Bernevig}},\ }\href
  {https://press.princeton.edu/books/hardcover/9780691151755/topological-insulators-and-topological-superconductors}
  {\emph {\bibinfo {title} {{Topological Insulators and Topological
  Superconductors}}}}\ (\bibinfo  {publisher} {Princeton University Press, Princeton, NJ},\
  \bibinfo {year} {2013})\BibitemShut {NoStop}%
\bibitem [{\citenamefont {Franz}\ and\ \citenamefont
  {Molenkamp}(2013)}]{Franz2013}%
  \BibitemOpen
  \href{https://www.elsevier.com/books/topological-insulators/franz/978-0-444-63314-9}
  {\emph {\bibinfo {title} {{Topological Insulators}}}},\ \bibinfo {edition}{1st}\ ed.,\ 
  edited by\ \bibinfo {editor} {\bibfnamefont {M.}~\bibnamefont {Franz}}\ and\ 
  \bibinfo{editor} {\bibfnamefont {L.}~\bibnamefont {Molenkamp}},\ 
  \bibinfo {series} {Contemporary Concepts in Condensed Matter Science}
  Vol.~\bibinfo {volume} {6}\ (\bibinfo  {publisher} {Elsevier, Amsterdam},\ 
  \bibinfo{year} {2013})\BibitemShut {NoStop}%
\bibitem [{\citenamefont {Ortmann}\ \emph {et~al.}(2015)\citenamefont
  {Ortmann}, \citenamefont {Roche},\ and\ \citenamefont
  {Valenzuela}}]{Ortmann2015}%
  \BibitemOpen
  \href{https://doi.org/10.1002/9783527681594} 
  {\emph {\bibinfo {title} {Topological Insulators: Fundamentals and Perspectives}}},\ 
  edited by\ \bibinfo {editor} {\bibfnamefont {F.}~\bibnamefont {Ortmann}}, 
  \bibinfo{editor} {\bibfnamefont {S.}~\bibnamefont {Roche}},\ and\ 
  \bibinfo {editor}{\bibfnamefont {S.~O.}\ \bibnamefont {Valenzuela}}\ 
  (\bibinfo  {publisher}{Wiley-VCH, Weinheim},\ \bibinfo {year} {2015})\BibitemShut
  {NoStop}%
\bibitem [{\citenamefont {Chiu}\ \emph {et~al.}(2016)\citenamefont {Chiu},
  \citenamefont {Teo}, \citenamefont {Schnyder},\ and\ \citenamefont
  {Ryu}}]{Chiu2016}%
  \BibitemOpen
  \bibfield  {author} {\bibinfo {author} {\bibfnamefont {C.-K.}\ \bibnamefont
  {Chiu}}, \bibinfo {author} {\bibfnamefont {J.~C.~Y.}\ \bibnamefont {Teo}},
  \bibinfo {author} {\bibfnamefont {A.~P.}\ \bibnamefont {Schnyder}},\ and\
  \bibinfo {author} {\bibfnamefont {S.}~\bibnamefont {Ryu}},\ }\href
  {https://doi.org/10.1103/RevModPhys.88.035005} {\bibfield  {journal}
  {\bibinfo  {journal} {Rev. Mod. Phys.}\ }\textbf {\bibinfo {volume} {88}},\
  \bibinfo {pages} {035005} (\bibinfo {year} {2016})}\BibitemShut {NoStop}%
\bibitem [{\citenamefont {Kane}\ and\ \citenamefont
  {Mele}(2005{\natexlab{a}})}]{Kane2005a}%
  \BibitemOpen
  \bibfield  {author} {\bibinfo {author} {\bibfnamefont {C.~L.}\ \bibnamefont
  {Kane}}\ and\ \bibinfo {author} {\bibfnamefont {E.~J.}\ \bibnamefont
  {Mele}},\ }\href {https://doi.org/10.1103/PhysRevLett.95.146802} {\bibfield
  {journal} {\bibinfo  {journal} {Phys. Rev. Lett.}\ }\textbf {\bibinfo
  {volume} {95}},\ \bibinfo {pages} {146802} (\bibinfo {year}
  {2005}{\natexlab{a}})}\BibitemShut {NoStop}%
\bibitem [{\citenamefont {Kane}\ and\ \citenamefont
  {Mele}(2005{\natexlab{b}})}]{Kane2005}%
  \BibitemOpen
  \bibfield  {author} {\bibinfo {author} {\bibfnamefont {C.~L.}\ \bibnamefont
  {Kane}}\ and\ \bibinfo {author} {\bibfnamefont {E.~J.}\ \bibnamefont
  {Mele}},\ }\href {https://doi.org/10.1103/PhysRevLett.95.226801} {\bibfield
  {journal} {\bibinfo  {journal} {Phys. Rev. Lett.}\ }\textbf {\bibinfo
  {volume} {95}},\ \bibinfo {pages} {226801} (\bibinfo {year}
  {2005}{\natexlab{b}})}\BibitemShut {NoStop}%
\bibitem [{\citenamefont {Fu}(2011)}]{Fu2011}%
  \BibitemOpen
  \bibfield  {author} {\bibinfo {author} {\bibfnamefont {L.}~\bibnamefont
  {Fu}},\ }\href {https://doi.org/10.1103/PhysRevLett.106.106802} {\bibfield
  {journal} {\bibinfo  {journal} {Phys. Rev. Lett.}\ }\textbf {\bibinfo
  {volume} {106}},\ \bibinfo {pages} {106802} (\bibinfo {year}
  {2011})}\BibitemShut {NoStop}%
\bibitem [{\citenamefont {Thouless}\ \emph {et~al.}(1982)\citenamefont
  {Thouless}, \citenamefont {Kohmoto}, \citenamefont {Nightingale},\ and\
  \citenamefont {den Nijs}}]{Thouless1982}%
  \BibitemOpen
  \bibfield  {author} {\bibinfo {author} {\bibfnamefont {D.~J.}\ \bibnamefont
  {Thouless}}, \bibinfo {author} {\bibfnamefont {M.}~\bibnamefont {Kohmoto}},
  \bibinfo {author} {\bibfnamefont {M.~P.}\ \bibnamefont {Nightingale}},\ and\
  \bibinfo {author} {\bibfnamefont {M.}~\bibnamefont {den Nijs}},\ }\href
  {https://doi.org/10.1103/PhysRevLett.49.405} {\bibfield  {journal} {\bibinfo
  {journal} {Phys. Rev. Lett.}\ }\textbf {\bibinfo {volume} {49}},\ \bibinfo
  {pages} {405} (\bibinfo {year} {1982})}\BibitemShut {NoStop}%
\bibitem [{\citenamefont {Klitzing}\ \emph {et~al.}(1980)\citenamefont
  {Klitzing}, \citenamefont {Dorda},\ and\ \citenamefont
  {Pepper}}]{Klitzing1980}%
  \BibitemOpen
  \bibfield  {author} {\bibinfo {author} {\bibfnamefont {K.}\ \bibnamefont
  {v.~Klitzing}}, \bibinfo {author} {\bibfnamefont {G.}~\bibnamefont {Dorda}},\
  and\ \bibinfo {author} {\bibfnamefont {M.}~\bibnamefont {Pepper}},\ }\href
  {https://doi.org/10.1103/PhysRevLett.45.494} {\bibfield  {journal} {\bibinfo
  {journal} {Phys. Rev. Lett.}\ }\textbf {\bibinfo {volume} {45}},\ \bibinfo
  {pages} {494} (\bibinfo {year} {1980})}\BibitemShut {NoStop}%
\bibitem [{\citenamefont {Onoda}\ and\ \citenamefont
  {Nagaosa}(2005)}]{Onoda2005}%
  \BibitemOpen
  \bibfield  {author} {\bibinfo {author} {\bibfnamefont {M.}~\bibnamefont
  {Onoda}}\ and\ \bibinfo {author} {\bibfnamefont {N.}~\bibnamefont
  {Nagaosa}},\ }\href {https://doi.org/10.1103/PhysRevLett.95.106601}
  {\bibfield  {journal} {\bibinfo  {journal} {Phys. Rev. Lett.}\ }\textbf
  {\bibinfo {volume} {95}},\ \bibinfo {pages} {106601} (\bibinfo {year}
  {2005})}\BibitemShut {NoStop}%
\bibitem [{\citenamefont {Haldane}(1988)}]{Haldane1988}%
  \BibitemOpen
  \bibfield  {author} {\bibinfo {author} {\bibfnamefont {F.~D.~M.}\
  \bibnamefont {Haldane}},\ }\href
  {https://doi.org/10.1103/PhysRevLett.61.2015} {\bibfield  {journal} {\bibinfo
   {journal} {Phys. Rev. Lett.}\ }\textbf {\bibinfo {volume} {61}},\ \bibinfo
  {pages} {2015} (\bibinfo {year} {1988})}\BibitemShut {NoStop}%
\bibitem [{\citenamefont {Bernevig}\ \emph {et~al.}(2006)\citenamefont
  {Bernevig}, \citenamefont {Hughes},\ and\ \citenamefont
  {Zhang}}]{Bernevig2006}%
  \BibitemOpen
  \bibfield  {author} {\bibinfo {author} {\bibfnamefont {B.~A.}\ \bibnamefont
  {Bernevig}}, \bibinfo {author} {\bibfnamefont {T.~L.}\ \bibnamefont
  {Hughes}},\ and\ \bibinfo {author} {\bibfnamefont {S.-C.}\ \bibnamefont
  {Zhang}},\ }\href {https://doi.org/10.1126/science.1133734} {\bibfield
  {journal} {\bibinfo  {journal} {Science}\ }\textbf {\bibinfo {volume}
  {314}},\ \bibinfo {pages} {1757} (\bibinfo {year} {2006})}\BibitemShut
  {NoStop}%
\bibitem [{\citenamefont {Konig}\ \emph {et~al.}(2007)\citenamefont {Konig},
  \citenamefont {Wiedmann}, \citenamefont {Brune}, \citenamefont {Roth},
  \citenamefont {Buhmann}, \citenamefont {Molenkamp}, \citenamefont {Qi},\ and\
  \citenamefont {Zhang}}]{Konig2007}%
  \BibitemOpen
  \bibfield  {author} {\bibinfo {author} {\bibfnamefont {M.}~\bibnamefont
  {K\"{o}nig}}, \bibinfo {author} {\bibfnamefont {S.}~\bibnamefont {Wiedmann}},
  \bibinfo {author} {\bibfnamefont {C.}~\bibnamefont {Brune}}, \bibinfo
  {author} {\bibfnamefont {A.}~\bibnamefont {Roth}}, \bibinfo {author}
  {\bibfnamefont {H.}~\bibnamefont {Buhmann}}, \bibinfo {author} {\bibfnamefont
  {L.~W.}\ \bibnamefont {Molenkamp}}, \bibinfo {author} {\bibfnamefont {X.-L.}\
  \bibnamefont {Qi}},\ and\ \bibinfo {author} {\bibfnamefont {S.-C.}\
  \bibnamefont {Zhang}},\ }\href {https://doi.org/10.1126/science.1148047}
  {\bibfield  {journal} {\bibinfo  {journal} {Science}\ }\textbf {\bibinfo
  {volume} {318}},\ \bibinfo {pages} {766} (\bibinfo {year}
  {2007})}\BibitemShut {NoStop}%
\bibitem [{\citenamefont {Roth}\ \emph {et~al.}(2009)\citenamefont {Roth},
  \citenamefont {Brune}, \citenamefont {Buhmann}, \citenamefont {Molenkamp},
  \citenamefont {Maciejko}, \citenamefont {Qi},\ and\ \citenamefont
  {Zhang}}]{Roth2009}%
  \BibitemOpen
  \bibfield  {author} {\bibinfo {author} {\bibfnamefont {A.}~\bibnamefont
  {Roth}}, \bibinfo {author} {\bibfnamefont {C.}~\bibnamefont {Brune}},
  \bibinfo {author} {\bibfnamefont {H.}~\bibnamefont {Buhmann}}, \bibinfo
  {author} {\bibfnamefont {L.~W.}\ \bibnamefont {Molenkamp}}, \bibinfo {author}
  {\bibfnamefont {J.}~\bibnamefont {Maciejko}}, \bibinfo {author}
  {\bibfnamefont {X.-L.}\ \bibnamefont {Qi}},\ and\ \bibinfo {author}
  {\bibfnamefont {S.-C.}\ \bibnamefont {Zhang}},\ }\href
  {https://doi.org/10.1126/science.1174736} {\bibfield  {journal} {\bibinfo
  {journal} {Science}\ }\textbf {\bibinfo {volume} {325}},\ \bibinfo {pages}
  {294} (\bibinfo {year} {2009})}\BibitemShut {NoStop}%
\bibitem [{\citenamefont {Knez}\ \emph {et~al.}(2011)\citenamefont {Knez},
  \citenamefont {Du},\ and\ \citenamefont {Sullivan}}]{Knez2011}%
  \BibitemOpen
  \bibfield  {author} {\bibinfo {author} {\bibfnamefont {I.}~\bibnamefont
  {Knez}}, \bibinfo {author} {\bibfnamefont {R.-R.}\ \bibnamefont {Du}},\ and\
  \bibinfo {author} {\bibfnamefont {G.}~\bibnamefont {Sullivan}},\ }\href
  {https://doi.org/10.1103/PhysRevLett.107.136603} {\bibfield  {journal}
  {\bibinfo  {journal} {Phys. Rev. Lett.}\ }\textbf {\bibinfo {volume} {107}},\
  \bibinfo {pages} {136603} (\bibinfo {year} {2011})}
  \BibitemShut {NoStop}%
\bibitem [{\citenamefont {Fu}\ \emph {et~al.}(2007)\citenamefont {Fu},
  \citenamefont {Kane},\ and\ \citenamefont {Mele}}]{Fu2007}%
  \BibitemOpen
  \bibfield  {author} {\bibinfo {author} {\bibfnamefont {L.}~\bibnamefont
  {Fu}}, \bibinfo {author} {\bibfnamefont {C.~L.}\ \bibnamefont {Kane}},\ and\
  \bibinfo {author} {\bibfnamefont {E.~J.}\ \bibnamefont {Mele}},\ }\href
  {https://doi.org/10.1103/PhysRevLett.98.106803} {\bibfield  {journal}
  {\bibinfo  {journal} {Phys. Rev. Lett.}\ }\textbf {\bibinfo {volume} {98}},\
  \bibinfo {pages} {106803} (\bibinfo {year} {2007})}\BibitemShut {NoStop}%
\bibitem [{\citenamefont {Moore}\ and\ \citenamefont
  {Balents}(2007)}]{Moore2007}%
  \BibitemOpen
  \bibfield  {author} {\bibinfo {author} {\bibfnamefont {J.~E.}\ \bibnamefont
  {Moore}}\ and\ \bibinfo {author} {\bibfnamefont {L.}~\bibnamefont
  {Balents}},\ }\href {https://doi.org/10.1103/PhysRevB.75.121306} {\bibfield
  {journal} {\bibinfo  {journal} {Phys. Rev. B}\ }\textbf {\bibinfo {volume}
  {75}},\ \bibinfo {pages} {121306} (\bibinfo {year} {2007})}\BibitemShut
  {NoStop}%
\bibitem [{\citenamefont {Fu}\ and\ \citenamefont {Kane}(2007)}]{Fu2007a}%
  \BibitemOpen
  \bibfield  {author} {\bibinfo {author} {\bibfnamefont {L.}~\bibnamefont
  {Fu}}\ and\ \bibinfo {author} {\bibfnamefont {C.~L.}\ \bibnamefont {Kane}},\
  }\href {https://doi.org/10.1103/PhysRevB.76.045302} {\bibfield  {journal}
  {\bibinfo  {journal} {Phys. Rev. B}\ }\textbf {\bibinfo {volume} {76}},\
  \bibinfo {pages} {045302} (\bibinfo {year} {2007})}\BibitemShut {NoStop}%
\bibitem [{\citenamefont {Roy}(2009)}]{Roy2009}%
  \BibitemOpen
  \bibfield  {author} {\bibinfo {author} {\bibfnamefont {R.}~\bibnamefont
  {Roy}},\ }\href {https://doi.org/10.1103/PhysRevB.79.195322} {\bibfield
  {journal} {\bibinfo  {journal} {Phys. Rev. B}\ }\textbf {\bibinfo {volume}
  {79}},\ \bibinfo {pages} {195322} (\bibinfo {year} {2009})}\BibitemShut
  {NoStop}%
\bibitem [{\citenamefont {Peccei}\ and\ \citenamefont
  {Quinn}(1977{\natexlab{a}})}]{Peccei1977a}%
  \BibitemOpen
  \bibfield  {author} {\bibinfo {author} {\bibfnamefont {R.~D.}\ \bibnamefont
  {Peccei}}\ and\ \bibinfo {author} {\bibfnamefont {H.~R.}\ \bibnamefont
  {Quinn}},\ }\href {https://doi.org/10.1103/PhysRevLett.38.1440} {\bibfield
  {journal} {\bibinfo  {journal} {Phys. Rev. Lett.}\ }\textbf {\bibinfo
  {volume} {38}},\ \bibinfo {pages} {1440} (\bibinfo {year}
  {1977}{\natexlab{a}})}\BibitemShut {NoStop}%
\bibitem [{\citenamefont {Peccei}\ and\ \citenamefont
  {Quinn}(1977{\natexlab{b}})}]{Peccei1977}%
  \BibitemOpen
  \bibfield  {author} {\bibinfo {author} {\bibfnamefont {R.~D.}\ \bibnamefont
  {Peccei}}\ and\ \bibinfo {author} {\bibfnamefont {H.~R.}\ \bibnamefont
  {Quinn}},\ }\href {https://doi.org/10.1103/PhysRevD.16.1791} {\bibfield
  {journal} {\bibinfo  {journal} {Phys. Rev. D}\ }\textbf {\bibinfo {volume}
  {16}},\ \bibinfo {pages} {1791} (\bibinfo {year}
  {1977}{\natexlab{b}})}\BibitemShut {NoStop}%
\bibitem [{\citenamefont {Qi}\ \emph {et~al.}(2008)\citenamefont {Qi},
  \citenamefont {Hughes},\ and\ \citenamefont {Zhang}}]{Qi2008}%
  \BibitemOpen
  \bibfield  {author} {\bibinfo {author} {\bibfnamefont {X.-L.}\ \bibnamefont
  {Qi}}, \bibinfo {author} {\bibfnamefont {T.~L.}\ \bibnamefont {Hughes}},\
  and\ \bibinfo {author} {\bibfnamefont {S.-C.}\ \bibnamefont {Zhang}},\ }\href
  {https://doi.org/10.1103/PhysRevB.78.195424} {\bibfield  {journal} {\bibinfo
  {journal} {Phys. Rev. B}\ }\textbf {\bibinfo {volume} {78}},\ \bibinfo
  {pages} {195424} (\bibinfo {year} {2008})}\BibitemShut {NoStop}%
\bibitem [{\citenamefont {Witten}(1979)}]{Witten1979}%
  \BibitemOpen
  \bibfield  {author} {\bibinfo {author} {\bibfnamefont {E.}~\bibnamefont
  {Witten}},\ }\href {https://doi.org/10.1016/0370-2693(79)90838-4} {\bibfield
  {journal} {\bibinfo  {journal} {Phys. Lett. B}\ }\textbf {\bibinfo {volume}
  {86}},\ \bibinfo {pages} {283} (\bibinfo {year} {1979})}\BibitemShut
  {NoStop}%
\bibitem [{\citenamefont {Dirac}(1948)}]{Dirac1948}%
  \BibitemOpen
  \bibfield  {author} {\bibinfo {author} {\bibfnamefont {P.~A.~M.}\
  \bibnamefont {Dirac}},\ }\href {https://doi.org/10.1103/PhysRev.74.817}
  {\bibfield  {journal} {\bibinfo  {journal} {Phys. Rev.}\ }\textbf {\bibinfo
  {volume} {74}},\ \bibinfo {pages} {817} (\bibinfo {year} {1948})}\BibitemShut
  {NoStop}%
\bibitem [{\citenamefont {Rosenberg}\ and\ \citenamefont
  {Franz}(2010)}]{Rosenberg2010}%
  \BibitemOpen
  \bibfield  {author} {\bibinfo {author} {\bibfnamefont {G.}~\bibnamefont
  {Rosenberg}}\ and\ \bibinfo {author} {\bibfnamefont {M.}~\bibnamefont
  {Franz}},\ }\href {https://doi.org/10.1103/PhysRevB.82.035105} {\bibfield
  {journal} {\bibinfo  {journal} {Phys. Rev. B}\
  }\textbf {\bibinfo {volume} {82}},\ \bibinfo {pages} {035105} (\bibinfo {year}
  {2010})}\BibitemShut {NoStop}%
\bibitem [{\citenamefont {Zhang}\ \emph {et~al.}(2009)\citenamefont {Zhang},
  \citenamefont {Liu}, \citenamefont {Qi}, \citenamefont {Dai}, \citenamefont
  {Fang},\ and\ \citenamefont {Zhang}}]{Zhang2009}%
  \BibitemOpen
  \bibfield  {author} {\bibinfo {author} {\bibfnamefont {H.}~\bibnamefont
  {Zhang}}, \bibinfo {author} {\bibfnamefont {C.-X.}\ \bibnamefont {Liu}},
  \bibinfo {author} {\bibfnamefont {X.-L.}\ \bibnamefont {Qi}}, \bibinfo
  {author} {\bibfnamefont {X.}~\bibnamefont {Dai}}, \bibinfo {author}
  {\bibfnamefont {Z.}~\bibnamefont {Fang}},\ and\ \bibinfo {author}
  {\bibfnamefont {S.-C.}\ \bibnamefont {Zhang}},\ }\href
  {https://doi.org/10.1038/nphys1270} {\bibfield  {journal} {\bibinfo
  {journal} {Nat. Phys.}\ }\textbf {\bibinfo {volume} {5}},\ \bibinfo {pages}
  {438} (\bibinfo {year} {2009})}\BibitemShut {NoStop}%
\bibitem [{\citenamefont {Hsieh}\ \emph {et~al.}(2008)\citenamefont {Hsieh},
  \citenamefont {Qian}, \citenamefont {Wray}, \citenamefont {Xia},
  \citenamefont {Hor}, \citenamefont {Cava},\ and\ \citenamefont
  {Hasan}}]{Hsieh2008}%
  \BibitemOpen
  \bibfield  {author} {\bibinfo {author} {\bibfnamefont {D.}~\bibnamefont
  {Hsieh}}, \bibinfo {author} {\bibfnamefont {D.}~\bibnamefont {Qian}},
  \bibinfo {author} {\bibfnamefont {L.}~\bibnamefont {Wray}}, \bibinfo {author}
  {\bibfnamefont {Y.}~\bibnamefont {Xia}}, \bibinfo {author} {\bibfnamefont
  {Y.~S.}\ \bibnamefont {Hor}}, \bibinfo {author} {\bibfnamefont {R.~J.}\
  \bibnamefont {Cava}},\ and\ \bibinfo {author} {\bibfnamefont {M.~Z.}\
  \bibnamefont {Hasan}},\ }\href {https://doi.org/10.1038/nature06843}
  {\bibfield  {journal} {\bibinfo  {journal} {Nature}\ }\textbf {\bibinfo
  {volume} {452}},\ \bibinfo {pages} {970} (\bibinfo {year}
  {2008})}\BibitemShut {NoStop}%
\bibitem [{\citenamefont {Hsieh}\ \emph {et~al.}(2009)\citenamefont {Hsieh},
  \citenamefont {Xia}, \citenamefont {Wray}, \citenamefont {Qian},
  \citenamefont {Pal}, \citenamefont {Dil}, \citenamefont {Osterwalder},
  \citenamefont {Meier}, \citenamefont {Bihlmayer}, \citenamefont {Kane},
  \citenamefont {Hor}, \citenamefont {Cava},\ and\ \citenamefont
  {Hasan}}]{Hsieh2009}%
  \BibitemOpen
  \bibfield  {author} {\bibinfo {author} {\bibfnamefont {D.}~\bibnamefont
  {Hsieh}}, \bibinfo {author} {\bibfnamefont {Y.}~\bibnamefont {Xia}}, \bibinfo
  {author} {\bibfnamefont {L.}~\bibnamefont {Wray}}, \bibinfo {author}
  {\bibfnamefont {D.}~\bibnamefont {Qian}}, \bibinfo {author} {\bibfnamefont
  {A.}~\bibnamefont {Pal}}, \bibinfo {author} {\bibfnamefont {J.~H.}\
  \bibnamefont {Dil}}, \bibinfo {author} {\bibfnamefont {J.}~\bibnamefont
  {Osterwalder}}, \bibinfo {author} {\bibfnamefont {F.}~\bibnamefont {Meier}},
  \bibinfo {author} {\bibfnamefont {G.}~\bibnamefont {Bihlmayer}}, \bibinfo
  {author} {\bibfnamefont {C.~L.}\ \bibnamefont {Kane}}, \bibinfo {author}
  {\bibfnamefont {Y.~S.}\ \bibnamefont {Hor}}, \bibinfo {author} {\bibfnamefont
  {R.~J.}\ \bibnamefont {Cava}},\ and\ \bibinfo {author} {\bibfnamefont
  {M.~Z.}\ \bibnamefont {Hasan}},\ }\href
  {https://doi.org/10.1126/science.1167733} {\bibfield  {journal} {\bibinfo
  {journal} {Science}\ }\textbf {\bibinfo {volume} {323}},\ \bibinfo {pages}
  {919} (\bibinfo {year} {2009})}\BibitemShut {NoStop}%
\bibitem [{\citenamefont {Xia}\ \emph {et~al.}(2009)\citenamefont {Xia},
  \citenamefont {Qian}, \citenamefont {Hsieh}, \citenamefont {Wray},
  \citenamefont {Pal}, \citenamefont {Lin}, \citenamefont {Bansil},
  \citenamefont {Grauer}, \citenamefont {Hor}, \citenamefont {Cava},\ and\
  \citenamefont {Hasan}}]{Xia2009}%
  \BibitemOpen
  \bibfield  {author} {\bibinfo {author} {\bibfnamefont {Y.}~\bibnamefont
  {Xia}}, \bibinfo {author} {\bibfnamefont {D.}~\bibnamefont {Qian}}, \bibinfo
  {author} {\bibfnamefont {D.}~\bibnamefont {Hsieh}}, \bibinfo {author}
  {\bibfnamefont {L.}~\bibnamefont {Wray}}, \bibinfo {author} {\bibfnamefont
  {A.}~\bibnamefont {Pal}}, \bibinfo {author} {\bibfnamefont {H.}~\bibnamefont
  {Lin}}, \bibinfo {author} {\bibfnamefont {A.}~\bibnamefont {Bansil}},
  \bibinfo {author} {\bibfnamefont {D.}~\bibnamefont {Grauer}}, \bibinfo
  {author} {\bibfnamefont {Y.~S.}\ \bibnamefont {Hor}}, \bibinfo {author}
  {\bibfnamefont {R.~J.}\ \bibnamefont {Cava}},\ and\ \bibinfo {author}
  {\bibfnamefont {M.~Z.}\ \bibnamefont {Hasan}},\ }\href
  {https://doi.org/10.1038/nphys1274} {\bibfield  {journal} {\bibinfo
  {journal} {Nat. Phys.}\ }\textbf {\bibinfo {volume} {5}},\ \bibinfo {pages}
  {398} (\bibinfo {year} {2009})}\BibitemShut {NoStop}%
\bibitem [{\citenamefont {Chen}\ \emph {et~al.}(2009)\citenamefont {Chen},
  \citenamefont {Analytis}, \citenamefont {Chu}, \citenamefont {Liu},
  \citenamefont {Mo}, \citenamefont {Qi}, \citenamefont {Zhang}, \citenamefont
  {Lu}, \citenamefont {Dai}, \citenamefont {Fang}, \citenamefont {Zhang},
  \citenamefont {Fisher}, \citenamefont {Hussain},\ and\ \citenamefont
  {Shen}}]{Chen2009}%
  \BibitemOpen
  \bibfield  {author} {\bibinfo {author} {\bibfnamefont {Y.~L.}\ \bibnamefont
  {Chen}}, \bibinfo {author} {\bibfnamefont {J.~G.}\ \bibnamefont {Analytis}},
  \bibinfo {author} {\bibfnamefont {J.-H.}\ \bibnamefont {Chu}}, \bibinfo
  {author} {\bibfnamefont {Z.~K.}\ \bibnamefont {Liu}}, \bibinfo {author}
  {\bibfnamefont {S.-K.}\ \bibnamefont {Mo}}, \bibinfo {author} {\bibfnamefont
  {X.~L.}\ \bibnamefont {Qi}}, \bibinfo {author} {\bibfnamefont {H.~J.}\
  \bibnamefont {Zhang}}, \bibinfo {author} {\bibfnamefont {D.~H.}\ \bibnamefont
  {Lu}}, \bibinfo {author} {\bibfnamefont {X.}~\bibnamefont {Dai}}, \bibinfo
  {author} {\bibfnamefont {Z.}~\bibnamefont {Fang}}, \bibinfo {author}
  {\bibfnamefont {S.~C.}\ \bibnamefont {Zhang}}, \bibinfo {author}
  {\bibfnamefont {I.~R.}\ \bibnamefont {Fisher}}, \bibinfo {author}
  {\bibfnamefont {Z.}~\bibnamefont {Hussain}},\ and\ \bibinfo {author}
  {\bibfnamefont {Z.-X.}\ \bibnamefont {Shen}},\ }\href
  {https://doi.org/10.1126/science.1173034} {\bibfield  {journal} {\bibinfo
  {journal} {Science}\ }\textbf {\bibinfo {volume} {325}},\ \bibinfo {pages}
  {178} (\bibinfo {year} {2009})}\BibitemShut {NoStop}%
\bibitem [{\citenamefont {Chen}\ \emph {et~al.}(2010)\citenamefont {Chen},
  \citenamefont {Liu}, \citenamefont {Analytis}, \citenamefont {Chu},
  \citenamefont {Zhang}, \citenamefont {Yan}, \citenamefont {Mo}, \citenamefont
  {Moore}, \citenamefont {Lu}, \citenamefont {Fisher}, \citenamefont {Zhang},
  \citenamefont {Hussain},\ and\ \citenamefont {Shen}}]{Chen2010}%
  \BibitemOpen
  \bibfield  {author} {\bibinfo {author} {\bibfnamefont {Y.~L.}\ \bibnamefont
  {Chen}}, \bibinfo {author} {\bibfnamefont {Z.~K.}\ \bibnamefont {Liu}},
  \bibinfo {author} {\bibfnamefont {J.~G.}\ \bibnamefont {Analytis}}, \bibinfo
  {author} {\bibfnamefont {J.-H.}\ \bibnamefont {Chu}}, \bibinfo {author}
  {\bibfnamefont {H.~J.}\ \bibnamefont {Zhang}}, \bibinfo {author}
  {\bibfnamefont {B.~H.}\ \bibnamefont {Yan}}, \bibinfo {author} {\bibfnamefont
  {S.-K.}\ \bibnamefont {Mo}}, \bibinfo {author} {\bibfnamefont {R.~G.}\
  \bibnamefont {Moore}}, \bibinfo {author} {\bibfnamefont {D.~H.}\ \bibnamefont
  {Lu}}, \bibinfo {author} {\bibfnamefont {I.~R.}\ \bibnamefont {Fisher}},
  \bibinfo {author} {\bibfnamefont {S.~C.}\ \bibnamefont {Zhang}}, \bibinfo
  {author} {\bibfnamefont {Z.}~\bibnamefont {Hussain}},\ and\ \bibinfo {author}
  {\bibfnamefont {Z.-X.}\ \bibnamefont {Shen}},\ }\href
  {https://doi.org/10.1103/PhysRevLett.105.266401} {\bibfield  {journal}
  {\bibinfo  {journal} {Phys. Rev. Lett.}\ }\textbf {\bibinfo {volume} {105}},\
  \bibinfo {pages} {266401} (\bibinfo {year} {2010})}\BibitemShut {NoStop}%
\bibitem [{\citenamefont {Jiang}\ \emph {et~al.}(2012)\citenamefont {Jiang},
  \citenamefont {Wang}, \citenamefont {Chen}, \citenamefont {Li}, \citenamefont
  {Song}, \citenamefont {He}, \citenamefont {Wang}, \citenamefont {Chen},
  \citenamefont {Ma},\ and\ \citenamefont {Xue}}]{Jiang2012}%
  \BibitemOpen
  \bibfield  {author} {\bibinfo {author} {\bibfnamefont {Y.}~\bibnamefont
  {Jiang}}, \bibinfo {author} {\bibfnamefont {Y.}~\bibnamefont {Wang}},
  \bibinfo {author} {\bibfnamefont {M.}~\bibnamefont {Chen}}, \bibinfo {author}
  {\bibfnamefont {Z.}~\bibnamefont {Li}}, \bibinfo {author} {\bibfnamefont
  {C.}~\bibnamefont {Song}}, \bibinfo {author} {\bibfnamefont {K.}~\bibnamefont
  {He}}, \bibinfo {author} {\bibfnamefont {L.}~\bibnamefont {Wang}}, \bibinfo
  {author} {\bibfnamefont {X.}~\bibnamefont {Chen}}, \bibinfo {author}
  {\bibfnamefont {X.}~\bibnamefont {Ma}},\ and\ \bibinfo {author}
  {\bibfnamefont {Q.-K.}\ \bibnamefont {Xue}},\ }\href
  {https://doi.org/10.1103/PhysRevLett.108.016401} {\bibfield  {journal}
  {\bibinfo  {journal} {Phys. Rev. Lett.}\ }\textbf {\bibinfo {volume} {108}},\
  \bibinfo {pages} {016401} (\bibinfo {year} {2012})}\BibitemShut {NoStop}%
\bibitem [{\citenamefont {Zhang}\ \emph {et~al.}(2021)\citenamefont {Zhang},
  \citenamefont {Noguchi}, \citenamefont {Kuroda}, \citenamefont {Lin},
  \citenamefont {Kawaguchi}, \citenamefont {Yaji}, \citenamefont {Harasawa},
  \citenamefont {Lippmaa}, \citenamefont {Nie}, \citenamefont {Weng},
  \citenamefont {Kandyba}, \citenamefont {Giampietri}, \citenamefont {Barinov},
  \citenamefont {Li}, \citenamefont {Gu}, \citenamefont {Shin},\ and\
  \citenamefont {Kondo}}]{Zhang2021}%
  \BibitemOpen
  \bibfield  {author} {\bibinfo {author} {\bibfnamefont {P.}~\bibnamefont
  {Zhang}}, \bibinfo {author} {\bibfnamefont {R.}~\bibnamefont {Noguchi}},
  \bibinfo {author} {\bibfnamefont {K.}~\bibnamefont {Kuroda}}, \bibinfo
  {author} {\bibfnamefont {C.}~\bibnamefont {Lin}}, \bibinfo {author}
  {\bibfnamefont {K.}~\bibnamefont {Kawaguchi}}, \bibinfo {author}
  {\bibfnamefont {K.}~\bibnamefont {Yaji}}, \bibinfo {author} {\bibfnamefont
  {A.}~\bibnamefont {Harasawa}}, \bibinfo {author} {\bibfnamefont
  {M.}~\bibnamefont {Lippmaa}}, \bibinfo {author} {\bibfnamefont
  {S.}~\bibnamefont {Nie}}, \bibinfo {author} {\bibfnamefont {H.}~\bibnamefont
  {Weng}}, \bibinfo {author} {\bibfnamefont {V.}~\bibnamefont {Kandyba}},
  \bibinfo {author} {\bibfnamefont {A.}~\bibnamefont {Giampietri}}, \bibinfo
  {author} {\bibfnamefont {A.}~\bibnamefont {Barinov}}, \bibinfo {author}
  {\bibfnamefont {Q.}~\bibnamefont {Li}}, \bibinfo {author} {\bibfnamefont
  {G.~D.}\ \bibnamefont {Gu}}, \bibinfo {author} {\bibfnamefont
  {S.}~\bibnamefont {Shin}},\ and\ \bibinfo {author} {\bibfnamefont
  {T.}~\bibnamefont {Kondo}},\ }\href
  {https://doi.org/10.1038/s41467-020-20564-8} {\bibfield  {journal} {\bibinfo
  {journal} {Nat. Commun.}\ }\textbf {\bibinfo {volume} {12}},\ \bibinfo
  {pages} {406} (\bibinfo {year} {2021})}\BibitemShut {NoStop}%
\bibitem [{\citenamefont {Zirnbauer}(1996)}]{Zirnbauer1996}%
  \BibitemOpen
  \bibfield  {author} {\bibinfo {author} {\bibfnamefont {M.~R.}\ \bibnamefont
  {Zirnbauer}},\ }\href {https://doi.org/10.1063/1.531675} {\bibfield
  {journal} {\bibinfo  {journal} {J. Math. Phys.}\ }\textbf {\bibinfo {volume}
  {37}},\ \bibinfo {pages} {4986} (\bibinfo {year} {1996})}\BibitemShut
  {NoStop}%
\bibitem [{\citenamefont {Altland}\ and\ \citenamefont
  {Zirnbauer}(1997)}]{Altland1997}%
  \BibitemOpen
  \bibfield  {author} {\bibinfo {author} {\bibfnamefont {A.}~\bibnamefont
  {Altland}}\ and\ \bibinfo {author} {\bibfnamefont {M.~R.}\ \bibnamefont
  {Zirnbauer}},\ }\href {https://doi.org/10.1103/PhysRevB.55.1142} {\bibfield
  {journal} {\bibinfo  {journal} {Phys. Rev. B}\ }\textbf {\bibinfo {volume}
  {55}},\ \bibinfo {pages} {1142} (\bibinfo {year} {1997})}\BibitemShut
  {NoStop}%
\bibitem [{\citenamefont {Schnyder}\ \emph {et~al.}(2008)\citenamefont
  {Schnyder}, \citenamefont {Ryu}, \citenamefont {Furusaki},\ and\
  \citenamefont {Ludwig}}]{Schnyder2008}%
  \BibitemOpen
  \bibfield  {author} {\bibinfo {author} {\bibfnamefont {A.~P.}\ \bibnamefont
  {Schnyder}}, \bibinfo {author} {\bibfnamefont {S.}~\bibnamefont {Ryu}},
  \bibinfo {author} {\bibfnamefont {A.}~\bibnamefont {Furusaki}},\ and\
  \bibinfo {author} {\bibfnamefont {A.~W.~W.}\ \bibnamefont {Ludwig}},\ }\href
  {https://doi.org/10.1103/PhysRevB.78.195125} {\bibfield  {journal} {\bibinfo
  {journal} {Phys. Rev. B}\ }\textbf {\bibinfo {volume} {78}},\ \bibinfo
  {pages} {195125} (\bibinfo {year} {2008})}\BibitemShut {NoStop}%
\bibitem [{\citenamefont {Kitaev}\ \emph {et~al.}(2009)\citenamefont {Kitaev},
  \citenamefont {Lebedev},\ and\ \citenamefont {Feigel'man}}]{Kitaev2009}%
  \BibitemOpen
  \bibfield  {author} {\bibinfo {author} {\bibfnamefont {A.}~\bibnamefont {Y.}~\bibnamefont {Kitaev}}, \ }in\
  \href {https://doi.org/10.1063/1.3149495} 
  {\emph {\bibinfo {booktitle} {Advances in Theoretical Physics: Landau Memorial Conference}}},\ 
  edited by\ \bibinfo {editor} {\bibfnamefont {V.}~\bibnamefont {Lebedev}}\
  and\ \bibinfo {editor} {\bibfnamefont {M.}~\bibnamefont {Feigel'man}},\ AIP
  Conf. Proc. Vol.\ \bibinfo {volume} {1134}\ (\bibinfo  {publisher}
  {AIP, Melville, NY},\ \bibinfo {year} {2009})\ pp.\ \bibinfo {pages} {22--30}
  \BibitemShut {NoStop}%
\bibitem [{\citenamefont {Ryu}\ \emph {et~al.}(2010)\citenamefont {Ryu},
  \citenamefont {Schnyder}, \citenamefont {Furusaki},\ and\ \citenamefont
  {Ludwig}}]{Ryu2010}%
  \BibitemOpen
  \bibfield  {author} {\bibinfo {author} {\bibfnamefont {S.}~\bibnamefont
  {Ryu}}, \bibinfo {author} {\bibfnamefont {A.~P.}\ \bibnamefont {Schnyder}},
  \bibinfo {author} {\bibfnamefont {A.}~\bibnamefont {Furusaki}},\ and\
  \bibinfo {author} {\bibfnamefont {A.~W.~W.}\ \bibnamefont {Ludwig}},\ }\href
  {https://doi.org/10.1088/1367-2630/12/6/065010} {\bibfield  {journal}
  {\bibinfo  {journal} {New J. Phys.}\ }\textbf {\bibinfo {volume} {12}},\
  \bibinfo {pages} {065010} (\bibinfo {year} {2010})}\BibitemShut {NoStop}%
\bibitem [{\citenamefont {Yannopapas}(2011)}]{Yannopapas2011a}%
  \BibitemOpen
  \bibfield  {author} {\bibinfo {author} {\bibfnamefont {V.}~\bibnamefont
  {Yannopapas}},\ }\href {https://doi.org/10.1103/PhysRevB.84.195126}
  {\bibfield  {journal} {\bibinfo  {journal} {Phys. Rev. B}\ }\textbf {\bibinfo
  {volume} {84}},\ \bibinfo {pages} {195126} (\bibinfo {year}
  {2011})}\BibitemShut {NoStop}%
\bibitem [{\citenamefont {Khanikaev}\ \emph {et~al.}(2013)\citenamefont
  {Khanikaev}, \citenamefont {{Hossein Mousavi}}, \citenamefont {Tse},
  \citenamefont {Kargarian}, \citenamefont {MacDonald}, \citenamefont {Shvets},
  \citenamefont {Mousavi}, \citenamefont {Tse}, \citenamefont {Kargarian},
  \citenamefont {MacDonald},\ and\ \citenamefont {Shvets}}]{Khanikaev2013}%
  \BibitemOpen
  \bibfield  {author} {\bibinfo {author} {\bibfnamefont {A.~B.}\ \bibnamefont
  {Khanikaev}}, \bibinfo {author} {\bibfnamefont {S.}~\bibnamefont {{Hossein
  Mousavi}}}, \bibinfo {author} {\bibfnamefont {W.-k.~K.}\ \bibnamefont {Tse}},
  \bibinfo {author} {\bibfnamefont {M.}~\bibnamefont {Kargarian}}, \bibinfo
  {author} {\bibfnamefont {A.~H.}\ \bibnamefont {MacDonald}}, \bibinfo {author}
  {\bibfnamefont {G.}~\bibnamefont {Shvets}}, \bibinfo {author} {\bibfnamefont
  {S.~H.}\ \bibnamefont {Mousavi}}, \bibinfo {author} {\bibfnamefont
  {W.-k.~K.}\ \bibnamefont {Tse}}, \bibinfo {author} {\bibfnamefont
  {M.}~\bibnamefont {Kargarian}}, \bibinfo {author} {\bibfnamefont {A.~H.}\
  \bibnamefont {MacDonald}},\ and\ \bibinfo {author} {\bibfnamefont
  {G.}~\bibnamefont {Shvets}},\ }\href {https://doi.org/10.1038/nmat3520}
  {\bibfield  {journal} {\bibinfo  {journal} {Nat. Mater.}\ }\textbf {\bibinfo
  {volume} {12}},\ \bibinfo {pages} {233} (\bibinfo {year} {2013})}\BibitemShut
  {NoStop}%
\bibitem [{\citenamefont {Chen}\ \emph {et~al.}(2014)\citenamefont {Chen},
  \citenamefont {Jiang}, \citenamefont {Chen}, \citenamefont {Zhu},
  \citenamefont {Zhou}, \citenamefont {Dong},\ and\ \citenamefont
  {Chan}}]{Chen2014}%
  \BibitemOpen
  \bibfield  {author} {\bibinfo {author} {\bibfnamefont {W.-J.}\ \bibnamefont
  {Chen}}, \bibinfo {author} {\bibfnamefont {S.-J.}\ \bibnamefont {Jiang}},
  \bibinfo {author} {\bibfnamefont {X.-D.}\ \bibnamefont {Chen}}, \bibinfo
  {author} {\bibfnamefont {B.}~\bibnamefont {Zhu}}, \bibinfo {author}
  {\bibfnamefont {L.}~\bibnamefont {Zhou}}, \bibinfo {author} {\bibfnamefont
  {J.-W.}\ \bibnamefont {Dong}},\ and\ \bibinfo {author} {\bibfnamefont
  {C.~T.}\ \bibnamefont {Chan}},\ }\href {https://doi.org/10.1038/ncomms6782}
  {\bibfield  {journal} {\bibinfo  {journal} {Nat. Commun.}\ }\textbf {\bibinfo
  {volume} {5}},\ \bibinfo {pages} {5782} (\bibinfo {year} {2014})}\BibitemShut
  {NoStop}%
\bibitem [{\citenamefont {Lu}\ \emph {et~al.}(2014)\citenamefont {Lu},
  \citenamefont {Joannopoulos},\ and\ \citenamefont
  {Solja{\v{c}}i{\'{c}}}}]{Lu2014}%
  \BibitemOpen
  \bibfield  {author} {\bibinfo {author} {\bibfnamefont {L.}~\bibnamefont
  {Lu}}, \bibinfo {author} {\bibfnamefont {J.~D.}\ \bibnamefont
  {Joannopoulos}},\ and\ \bibinfo {author} {\bibfnamefont {M.}~\bibnamefont
  {Solja{\v{c}}i{\'{c}}}},\ }\href {https://doi.org/10.1038/nphoton.2014.248}
  {\bibfield  {journal} {\bibinfo {journal} {Nat. Photonics}\ }\textbf {\bibinfo {volume} {8}},\ 
  \bibinfo {pages}{821} (\bibinfo {year} {2014})}
  \BibitemShut {NoStop}%
\bibitem [{\citenamefont {Wu}\ and\ \citenamefont {Hu}(2015)}]{Wu2015}%
  \BibitemOpen
  \bibfield  {author} {\bibinfo {author} {\bibfnamefont {L.-H.}\ \bibnamefont
  {Wu}}\ and\ \bibinfo {author} {\bibfnamefont {X.}~\bibnamefont {Hu}},\ }\href
  {https://doi.org/10.1103/PhysRevLett.114.223901} {\bibfield  {journal}
  {\bibinfo  {journal} {Phys. Rev. Lett.}\ }\textbf {\bibinfo {volume} {114}},\
  \bibinfo {pages} {223901} (\bibinfo {year} {2015})}\BibitemShut {NoStop}%
\bibitem [{\citenamefont {Lu}\ \emph {et~al.}(2016)\citenamefont {Lu},
  \citenamefont {Fang}, \citenamefont {Fu}, \citenamefont {Johnson},
  \citenamefont {Joannopoulos},\ and\ \citenamefont
  {Solja{\v{c}}i{\'{c}}}}]{Lu2016}%
  \BibitemOpen
  \bibfield  {author} {\bibinfo {author} {\bibfnamefont {L.}~\bibnamefont
  {Lu}}, \bibinfo {author} {\bibfnamefont {C.}~\bibnamefont {Fang}}, \bibinfo
  {author} {\bibfnamefont {L.}~\bibnamefont {Fu}}, \bibinfo {author}
  {\bibfnamefont {S.~G.}\ \bibnamefont {Johnson}}, \bibinfo {author}
  {\bibfnamefont {J.~D.}\ \bibnamefont {Joannopoulos}},\ and\ \bibinfo {author}
  {\bibfnamefont {M.}~\bibnamefont {Solja{\v{c}}i{\'{c}}}},\ }\href
  {https://doi.org/10.1038/nphys3611} {\bibfield  {journal} {\bibinfo
  {journal} {Nat. Phys.}\ }\textbf {\bibinfo {volume} {12}},\ \bibinfo {pages}
  {337} (\bibinfo {year} {2016})}\BibitemShut {NoStop}%
\bibitem [{\citenamefont {Slobozhanyuk}\ \emph {et~al.}(2017)\citenamefont
  {Slobozhanyuk}, \citenamefont {Mousavi}, \citenamefont {Ni}, \citenamefont
  {Smirnova}, \citenamefont {Kivshar},\ and\ \citenamefont
  {Khanikaev}}]{Slobozhanyuk2017}%
  \BibitemOpen
  \bibfield  {author} {\bibinfo {author} {\bibfnamefont {A.}~\bibnamefont
  {Slobozhanyuk}}, \bibinfo {author} {\bibfnamefont {S.~H.}\ \bibnamefont
  {Mousavi}}, \bibinfo {author} {\bibfnamefont {X.}~\bibnamefont {Ni}},
  \bibinfo {author} {\bibfnamefont {D.}~\bibnamefont {Smirnova}}, \bibinfo
  {author} {\bibfnamefont {Y.~S.}\ \bibnamefont {Kivshar}},\ and\ \bibinfo
  {author} {\bibfnamefont {A.~B.}\ \bibnamefont {Khanikaev}},\ }\href
  {https://doi.org/10.1038/nphoton.2016.253} {\bibfield  {journal} {\bibinfo
  {journal} {Nat. Photonics}\ }\textbf {\bibinfo {volume} {11}},\ \bibinfo
  {pages} {130} (\bibinfo {year} {2017})}\BibitemShut {NoStop}%
\bibitem [{\citenamefont {Ozawa}\ \emph {et~al.}(2019)\citenamefont {Ozawa},
  \citenamefont {Price}, \citenamefont {Amo}, \citenamefont {Goldman},
  \citenamefont {Hafezi}, \citenamefont {Lu}, \citenamefont {Rechtsman},
  \citenamefont {Schuster}, \citenamefont {Simon}, \citenamefont {Zilberberg},\
  and\ \citenamefont {Carusotto}}]{Ozawa2019}%
  \BibitemOpen
  \bibfield  {author} {\bibinfo {author} {\bibfnamefont {T.}~\bibnamefont
  {Ozawa}}, \bibinfo {author} {\bibfnamefont {H.~M.}\ \bibnamefont {Price}},
  \bibinfo {author} {\bibfnamefont {A.}~\bibnamefont {Amo}}, \bibinfo {author}
  {\bibfnamefont {N.}~\bibnamefont {Goldman}}, \bibinfo {author} {\bibfnamefont
  {M.}~\bibnamefont {Hafezi}}, \bibinfo {author} {\bibfnamefont
  {L.}~\bibnamefont {Lu}}, \bibinfo {author} {\bibfnamefont {M.~C.}\
  \bibnamefont {Rechtsman}}, \bibinfo {author} {\bibfnamefont {D.}~\bibnamefont
  {Schuster}}, \bibinfo {author} {\bibfnamefont {J.}~\bibnamefont {Simon}},
  \bibinfo {author} {\bibfnamefont {O.}~\bibnamefont {Zilberberg}},\ and\
  \bibinfo {author} {\bibfnamefont {I.}~\bibnamefont {Carusotto}},\ }\href
  {https://doi.org/10.1103/RevModPhys.91.015006} {\bibfield  {journal}
  {\bibinfo  {journal} {Rev. Mod. Phys.}\ }\textbf {\bibinfo {volume} {91}},\
  \bibinfo {pages} {15006} (\bibinfo {year} {2019})}\BibitemShut {NoStop}%
\bibitem [{\citenamefont {Prodan}\ and\ \citenamefont
  {Prodan}(2009)}]{Prodan2009}%
  \BibitemOpen
  \bibfield  {author} {\bibinfo {author} {\bibfnamefont {E.}~\bibnamefont
  {Prodan}}\ and\ \bibinfo {author} {\bibfnamefont {C.}~\bibnamefont
  {Prodan}},\ }\href {https://doi.org/10.1103/PhysRevLett.103.248101}
  {\bibfield  {journal} {\bibinfo  {journal} {Phys. Rev. Lett.}\ }\textbf
  {\bibinfo {volume} {103}},\ \bibinfo {pages} {248101} (\bibinfo {year}
  {2009})}\BibitemShut {NoStop}%
\bibitem [{\citenamefont {Zhang}\ \emph {et~al.}(2010)\citenamefont {Zhang},
  \citenamefont {Ren}, \citenamefont {Wang},\ and\ \citenamefont
  {Li}}]{Zhang2010}%
  \BibitemOpen
  \bibfield  {author} {\bibinfo {author} {\bibfnamefont {L.}~\bibnamefont
  {Zhang}}, \bibinfo {author} {\bibfnamefont {J.}~\bibnamefont {Ren}}, \bibinfo
  {author} {\bibfnamefont {J.-S.}\ \bibnamefont {Wang}},\ and\ \bibinfo
  {author} {\bibfnamefont {B.}~\bibnamefont {Li}},\ }\href
  {https://doi.org/10.1103/PhysRevLett.105.225901} {\bibfield  {journal}
  {\bibinfo  {journal} {Phys. Rev. Lett.}\ }\textbf {\bibinfo {volume} {105}},\
  \bibinfo {pages} {225901} (\bibinfo {year} {2010})}\BibitemShut {NoStop}%
\bibitem [{\citenamefont {Berg}\ \emph {et~al.}(2011)\citenamefont {Berg},
  \citenamefont {Joel}, \citenamefont {Koolyk},\ and\ \citenamefont
  {Prodan}}]{Berg2011}%
  \BibitemOpen
  \bibfield  {author} {\bibinfo {author} {\bibfnamefont {N.}~\bibnamefont
  {Berg}}, \bibinfo {author} {\bibfnamefont {K.}~\bibnamefont {Joel}}, \bibinfo
  {author} {\bibfnamefont {M.}~\bibnamefont {Koolyk}},\ and\ \bibinfo {author}
  {\bibfnamefont {E.}~\bibnamefont {Prodan}},\ }\href
  {https://doi.org/10.1103/PhysRevE.83.021913} {\bibfield  {journal} {\bibinfo
  {journal} {Phys. Rev. E}\ }\textbf {\bibinfo {volume} {83}},\ \bibinfo
  {pages} {021913} (\bibinfo {year} {2011})}\BibitemShut {NoStop}%
\bibitem [{\citenamefont {Li}\ \emph {et~al.}(2012)\citenamefont {Li},
  \citenamefont {Ren}, \citenamefont {Wang}, \citenamefont {Zhang},
  \citenamefont {H{\"{a}}nggi},\ and\ \citenamefont {Li}}]{Li2012}%
  \BibitemOpen
  \bibfield  {author} {\bibinfo {author} {\bibfnamefont {N.}~\bibnamefont
  {Li}}, \bibinfo {author} {\bibfnamefont {J.}~\bibnamefont {Ren}}, \bibinfo
  {author} {\bibfnamefont {L.}~\bibnamefont {Wang}}, \bibinfo {author}
  {\bibfnamefont {G.}~\bibnamefont {Zhang}}, \bibinfo {author} {\bibfnamefont
  {P.}~\bibnamefont {H{\"{a}}nggi}},\ and\ \bibinfo {author} {\bibfnamefont
  {B.}~\bibnamefont {Li}},\ }\href {https://doi.org/10.1103/RevModPhys.84.1045}
  {\bibfield  {journal} {\bibinfo  {journal} {Rev. Mod. Phys.}\ }\textbf
  {\bibinfo {volume} {84}},\ \bibinfo {pages} {1045} (\bibinfo {year}
  {2012})}\BibitemShut {NoStop}%
\bibitem [{\citenamefont {Kane}\ and\ \citenamefont
  {Lubensky}(2013)}]{Kane2014}%
  \BibitemOpen
  \bibfield  {author} {\bibinfo {author} {\bibfnamefont {C.~L.}\ \bibnamefont
  {Kane}}\ and\ \bibinfo {author} {\bibfnamefont {T.~C.}\ \bibnamefont
  {Lubensky}},\ }\href {https://doi.org/10.1038/nphys2835} {\bibfield
  {journal} {\bibinfo  {journal} {Nat. Phys.}\ }\textbf {\bibinfo {volume}
  {10}},\ \bibinfo {pages} {39} (\bibinfo {year} {2014})}\BibitemShut {NoStop}%
\bibitem [{\citenamefont {Huber}(2016)}]{Huber2016}%
  \BibitemOpen
  \bibfield  {author} {\bibinfo {author} {\bibfnamefont {S.~D.}\ \bibnamefont
  {Huber}},\ }\href {https://doi.org/10.1038/nphys3801} {\bibfield  {journal}
  {\bibinfo  {journal} {Nat. Phys.}\ }\textbf {\bibinfo {volume} {12}},\
  \bibinfo {pages} {621} (\bibinfo {year} {2016})}\BibitemShut {NoStop}%
\bibitem [{\citenamefont {Liu}\ \emph {et~al.}(2017)\citenamefont {Liu},
  \citenamefont {Xu}, \citenamefont {Zhang},\ and\ \citenamefont
  {Duan}}]{Liu2017}%
  \BibitemOpen
  \bibfield  {author} {\bibinfo {author} {\bibfnamefont {Y.}~\bibnamefont
  {Liu}}, \bibinfo {author} {\bibfnamefont {Y.}~\bibnamefont {Xu}}, \bibinfo
  {author} {\bibfnamefont {S.-C.}\ \bibnamefont {Zhang}},\ and\ \bibinfo
  {author} {\bibfnamefont {W.}~\bibnamefont {Duan}},\ }\href
  {https://doi.org/10.1103/PhysRevB.96.064106} {\bibfield  {journal} {\bibinfo
  {journal} {Phys. Rev. B}\ }\textbf {\bibinfo {volume} {96}},\ \bibinfo
  {pages} {064106} (\bibinfo {year} {2017})}\BibitemShut {NoStop}%
\bibitem [{\citenamefont {Goldman}\ \emph {et~al.}(2016)\citenamefont
  {Goldman}, \citenamefont {Budich},\ and\ \citenamefont
  {Zoller}}]{Goldman2016}%
  \BibitemOpen
  \bibfield  {author} {\bibinfo {author} {\bibfnamefont {N.}~\bibnamefont
  {Goldman}}, \bibinfo {author} {\bibfnamefont {J.~C.}\ \bibnamefont
  {Budich}},\ and\ \bibinfo {author} {\bibfnamefont {P.}~\bibnamefont
  {Zoller}},\ }\href {https://doi.org/10.1038/nphys3803} {\bibfield  {journal}
  {\bibinfo  {journal} {Nat. Phys.}\ }\textbf {\bibinfo {volume} {12}},\
  \bibinfo {pages} {639} (\bibinfo {year} {2016})}\BibitemShut {NoStop}%
\bibitem [{\citenamefont {Zhang}\ \emph {et~al.}(2018)\citenamefont {Zhang},
  \citenamefont {Zhu}, \citenamefont {Zhao}, \citenamefont {Yan},\ and\
  \citenamefont {Zhu}}]{Zhang2018}%
  \BibitemOpen
  \bibfield  {author} {\bibinfo {author} {\bibfnamefont {D.~W.}\ \bibnamefont
  {Zhang}}, \bibinfo {author} {\bibfnamefont {Y.~Q.}\ \bibnamefont {Zhu}},
  \bibinfo {author} {\bibfnamefont {Y.~X.}\ \bibnamefont {Zhao}}, \bibinfo
  {author} {\bibfnamefont {H.}~\bibnamefont {Yan}},\ and\ \bibinfo {author}
  {\bibfnamefont {S.~L.}\ \bibnamefont {Zhu}},\ }\href
  {https://doi.org/10.1080/00018732.2019.1594094} {\bibfield  {journal}
  {\bibinfo  {journal} {Adv. Phys.}\ }\textbf {\bibinfo {volume} {67}},\
  \bibinfo {pages} {253} (\bibinfo {year} {2018})}\BibitemShut {NoStop}%
\bibitem [{\citenamefont {Cooper}\ \emph {et~al.}(2019)\citenamefont {Cooper},
  \citenamefont {Dalibard},\ and\ \citenamefont {Spielman}}]{Cooper2019}%
  \BibitemOpen
  \bibfield  {author} {\bibinfo {author} {\bibfnamefont {N.~R.}\ \bibnamefont
  {Cooper}}, \bibinfo {author} {\bibfnamefont {J.}~\bibnamefont {Dalibard}},\
  and\ \bibinfo {author} {\bibfnamefont {I.~B.}\ \bibnamefont {Spielman}},\
  }\href {https://doi.org/10.1103/RevModPhys.91.015005} {\bibfield  {journal}
  {\bibinfo  {journal} {Rev. Mod. Phys.}\ }\textbf {\bibinfo {volume} {91}},\
  \bibinfo {pages} {015005} (\bibinfo {year} {2019})}\BibitemShut {NoStop}%
\bibitem [{\citenamefont {Cai}\ \emph {et~al.}(2019)\citenamefont {Cai},
  \citenamefont {Han}, \citenamefont {Mei}, \citenamefont {Xu}, \citenamefont
  {Ma}, \citenamefont {Li}, \citenamefont {Wang}, \citenamefont {Song},
  \citenamefont {Xue}, \citenamefont {Yin}, \citenamefont {Jia},\ and\
  \citenamefont {Sun}}]{Cai2019}%  
  \BibitemOpen
  \bibfield  {author} {\bibinfo {author} {\bibfnamefont {W.}~\bibnamefont
  {Cai}}, \bibinfo {author} {\bibfnamefont {J.}~\bibnamefont {Han}}, \bibinfo
  {author} {\bibfnamefont {F.}~\bibnamefont {Mei}}, \bibinfo {author}
  {\bibfnamefont {Y.}~\bibnamefont {Xu}}, \bibinfo {author} {\bibfnamefont
  {Y.}~\bibnamefont {Ma}}, \bibinfo {author} {\bibfnamefont {X.}~\bibnamefont
  {Li}}, \bibinfo {author} {\bibfnamefont {H.}~\bibnamefont {Wang}}, \bibinfo
  {author} {\bibfnamefont {Y.~P.}\ \bibnamefont {Song}}, \bibinfo {author}
  {\bibfnamefont {Z.-Y.}\ \bibnamefont {Xue}}, \bibinfo {author} {\bibfnamefont
  {Z.-q.}\ \bibnamefont {Yin}}, \bibinfo {author} {\bibfnamefont
  {S.}~\bibnamefont {Jia}},\ and\ \bibinfo {author} {\bibfnamefont
  {L.}~\bibnamefont {Sun}},\ }\href
  {https://doi.org/10.1103/PhysRevLett.123.080501} {\bibfield  {journal}
  {\bibinfo  {journal} {Phys. Rev. Lett.}\ }\textbf {\bibinfo {volume} {123}},\
  \bibinfo {pages} {080501} (\bibinfo {year} {2019})}\BibitemShut {NoStop}%  
\bibitem [{\citenamefont { Kunst}\ \emph {et~al.}(2019)\citenamefont {  Kunst},
  \citenamefont {van Miert},\ and\ \citenamefont
  {Bergholtz}}]{Kunst2019}%
  \BibitemOpen
 \bibfield  {author} {\bibinfo {author} {\bibfnamefont {F.~K.}~\bibnamefont
  {Kunst}}, \bibinfo {author} {\bibfnamefont {G.}\ \bibnamefont
  {van~Miert}},\ and\ \bibinfo {author} {\bibfnamefont {E.~J.}~\bibnamefont
  {Bergholtz}},\ }\href {https://doi.org/10.1103/PhysRevB.99.085427} {\bibfield  {journal}
  {\bibinfo  {journal} {Phys. Rev. B}\ }\textbf {\bibinfo {volume} {99}},\
  \bibinfo {pages} {085427} (\bibinfo {year} {2019})}\BibitemShut {NoStop}%
\bibitem [{\citenamefont {Supplemental Material}(2022)}]{SM}%
  \BibitemOpen
   See Supplemental Material for the details about
   (1) the definition and concrete examples of the series of lattice models addressed in the main text, 
   (2) the analysis of systems with no surfaces to derive the four $\mathbb{Z}_{2}$ topological invariants, 
   (3) the analysis of systems with surfaces to derive the analytical solutions of surface-state eigenvectors, 
   and (4) the definition of symmetry points in band diagrams.
  \BibitemShut {NoStop}%
\bibitem [{\citenamefont {Rhim}\ \emph {et~al.}(2018)\citenamefont {Rhim},
  \citenamefont {Bardarson},\ and\ \citenamefont
  {Slager}}]{Rhim2018}%
  \BibitemOpen
 \bibfield  {author} {\bibinfo {author} {\bibfnamefont {J.-W.}~\bibnamefont
  {Rhim}}, \bibinfo {author} {\bibfnamefont {J.~H.}\ \bibnamefont
  {Bardarson}},\ and\ \bibinfo {author} {\bibfnamefont {R.-J.}~\bibnamefont
  {Slager}},\ }\href {https://doi.org/10.1103/PhysRevB.97.115143} {\bibfield  {journal}
  {\bibinfo  {journal} {Phys. Rev. B}\ }\textbf {\bibinfo {volume} {97}},\
  \bibinfo {pages} {115143} (\bibinfo {year} {2018})}\BibitemShut {NoStop}%
\bibitem [{\citenamefont {Onoda}(2012)}]{Onoda2012}%
  \BibitemOpen
  \bibfield  {author} {\bibinfo {author} {\bibfnamefont {M.}~\bibnamefont {Onoda}},\ }in\ 
  \href {http://ieeexplore.ieee.org/document/6318467/} 
  {\emph{\bibinfo {booktitle} {2012 Proceedings of SICE Annual Conference}}}\ 
  (\bibinfo  {publisher}{IEEE, New York},\ \bibinfo {year} {2012})\ 
  pp.\ \bibinfo {pages}{376--381}\BibitemShut {NoStop}%
\bibitem [{\citenamefont {Ohmura}\ \emph {et~al.}(2012)\citenamefont {Ohmura},
  \citenamefont {Shimokawa}, \citenamefont {Hosaka},\ and\ \citenamefont
  {Onoda}}]{Ohmura2012}%
  \BibitemOpen
  \bibfield  {author} {\bibinfo {author} {\bibfnamefont {Y.}~\bibnamefont {Ohmura}}, 
  \bibinfo {author} {\bibfnamefont {T.}~\bibnamefont {Shimokawa}},
  \bibinfo {author} {\bibfnamefont {S.}~\bibnamefont {Hosaka}},\ and\ 
  \bibinfo {author} {\bibfnamefont {M.}~\bibnamefont {Onoda}},\ }in\ 
  \href{http://ieeexplore.ieee.org/document/6318597/} 
  {\emph {\bibinfo {booktitle}{2012 Proceedings of SICE Annual Conference}}}\ 
  (\bibinfo  {publisher} {Ref.~\cite{Onoda2012}})\ 
  pp.\ \bibinfo {pages} {1044--1049}\BibitemShut {NoStop}%
\bibitem [{\citenamefont {Onoda}(2019)}]{Onoda2019}%
  \BibitemOpen
  \bibfield  {author} {\bibinfo {author} {\bibfnamefont {M.}~\bibnamefont
  {Onoda}},\ }\href {https://doi.org/10.3390/app9071468} {\bibfield  {journal}
  {\bibinfo  {journal} {Appl. Sci.}\ }\textbf {\bibinfo {volume} {9}},\
  \bibinfo {pages} {1468} (\bibinfo {year} {2019})}\BibitemShut {NoStop}%
\bibitem [{\citenamefont {Marchand}\ and\ \citenamefont
  {Franz}(2012)}]{Marchand2012}%
  \BibitemOpen
  \bibfield  {author} {\bibinfo {author} {\bibfnamefont {D.~J.~J.}\
  \bibnamefont {Marchand}}\ and\ \bibinfo {author} {\bibfnamefont
  {M.}~\bibnamefont {Franz}},\ }\href
  {https://doi.org/10.1103/PhysRevB.86.155146} {\bibfield  {journal} {\bibinfo
  {journal} {Phys. Rev. B}\ }\textbf {\bibinfo {volume} {86}},\ \bibinfo
  {pages} {155146} (\bibinfo {year} {2012})}\BibitemShut {NoStop}%
\bibitem [{\citenamefont {Berezinskii}(1971)}]{Berezinskii1971}%
  \BibitemOpen
  \bibfield  {author} {\bibinfo {author} {\bibfnamefont {V.~L.}\ \bibnamefont
  {Berezinski\u{\i}}},\ }\href {https://inspirehep.net/literature/61186} 
  {\bibfield {journal} {\bibinfo  {journal} {Zh. Eksp. Teor. Fiz.}\ }\textbf
  {\bibinfo {volume}{59}},\ \bibinfo {pages} {907} (\bibinfo {year} {1971})}
  [{\bibfield {journal} {\bibinfo  {journal} {Sov. Phys. JETP}\ }\textbf
  {\bibinfo {volume}{32}},\ \bibinfo {pages} {493} (\bibinfo {year} {1971})}]
  \BibitemShut {NoStop}%
\bibitem [{\citenamefont {Berezinskii}(1972)}]{Berezinskii1972}%
  \BibitemOpen
  \bibfield  {author} {\bibinfo {author} {\bibfnamefont {V.~L.}\ \bibnamefont
  {Berezinski\u{\i}}},\ }\href {https://inspirehep.net/literature/1716846}
  {\bibfield  {journal} {\bibinfo  {journal} {Zh. Eksp. Teor. Fiz.}\ }\textbf
  {\bibinfo {volume} {61}},\ \bibinfo {pages} {1144} (\bibinfo {year}{1972})}
  [{\bibfield  {journal} {\bibinfo  {journal} {Sov. Phys. JETP}\ }\textbf
  {\bibinfo {volume} {34}},\ \bibinfo {pages} {610} (\bibinfo {year}{1972})}]
  \BibitemShut {NoStop}%
\bibitem [{\citenamefont {Kosterlitz}\ and\ \citenamefont
  {Thouless}(1973)}]{Kosterlitz1973}%
  \BibitemOpen
  \bibfield  {author} {\bibinfo {author} {\bibfnamefont {J.~M.}\ \bibnamefont
  {Kosterlitz}}\ and\ \bibinfo {author} {\bibfnamefont {D.~J.}\ \bibnamefont
  {Thouless}},\ }\href {https://doi.org/10.1088/0022-3719/6/7/010} {\bibfield
  {journal} {\bibinfo  {journal} {J. Phys. C Solid State Phys.}\ }\textbf
  {\bibinfo {volume} {6}},\ \bibinfo {pages} {1181} (\bibinfo {year}
  {1973})}\BibitemShut {NoStop}%
\bibitem [{\citenamefont {Kosterlitz}(1974)}]{Kosterlitz1974}%
  \BibitemOpen
  \bibfield  {author} {\bibinfo {author} {\bibfnamefont {J.~M.}\ \bibnamefont
  {Kosterlitz}},\ }\href {https://doi.org/10.1088/0022-3719/7/6/005} {\bibfield
   {journal} {\bibinfo  {journal} {J. Phys. C Solid State Phys.}\ }\textbf
  {\bibinfo {volume} {7}},\ \bibinfo {pages} {1046} (\bibinfo {year}
  {1974})}\BibitemShut {NoStop}%
\end{thebibliography}
\end{document}

\preprint{ASTSSSLM_SM}

\title{Supplemental Material for \\"Analytical solutions of topological surface states in a series of lattice models"}% Force line breaks with \\

\author{Masaru Onoda}
%\email{onoda@gipc.akita-u.ac.jp}
\affiliation{%
Mathematical Science Course,
%Department of Mathematical Science and Electrical-Electronic-Computer Engineering, \\
Graduate School of Engineering Science,
Akita University, Akita 010-8502, Japan
}%



\begin{abstract}
This Supplemental Material provides the necessary details of the main article, including (1)~the definition and concrete examples of the series of lattice models addressed in the main article, (2)~the analysis of systems with no surfaces to derive the four $\mathbb{Z}_{2}$ topological invariants, (3)~the analysis of systems with surfaces to derive the analytical solutions of surface-state eigenvectors, and (4)~the definition of symmetry points in band diagrams. We also supplement the discussion of the band structures addressed in the main article. Some information described in the main article is shown again so that this Supplemental Material is as self-contained as possible.
\end{abstract}

\pacs{73.20.-r, 73.20.At, 73.43.-f, 73.43.Cd, 72.25.-b}
%73.20.-r :  Electron states at surfaces and interfaces
%73.20.At : surface-states, band structure, electron density of states
%73.43.-f : Quantum Hall effects
%73.43.Cd : Theory and modeling
%72.25.-b : Spin polarized transport

\maketitle

%\tableofcontents

\section{\label{sec:Series_of_Lattice_Models}Series of Lattice Models}
\subsection{\label{sec:Model_Hamiltonian}Model Hamiltonian and Symmetries}
The series of lattice models is constructed on a simple cubic lattice with a conventional orthogonal basis $\{\bm{a}_{\mu}\}_{\mu=1,2,3}$ ($\left\vert\bm{a}_{\mu}\right\vert =a$).
Lattice sites $\{\bm{r}\}$ are represented via the basis and classified into two face-centered cubic (FCC) sublattices as
\begin{align}
&\bm{r}=\sum_{\mu}n_{\mu}\bm{a}_{\mu}
\in
\begin{cases}
\text{$\mathrm{A}$ sublattice} & (\sum_{\mu} n_{\mu} = \mathrm{even}),
\\
\text{$\mathrm{B}$ sublattice}  & (\sum_{\mu} n_{\mu} = \mathrm{odd}),
\end{cases}
\end{align}
where $n_{\mu}\in \mathbb{Z}$.
The model Hamiltonian below defines the series of lattice models via three kinds of real-valued parameters $\{t_{\mu}\}$, $\{t_{\mu\nu}\}$ ($\mu\neq\nu, t_{\mu\nu}=t_{\nu\mu}$), and $v_{\mathrm{s}}$ and SU(2) matrices $\{U_{\mu}\}$,
\begin{align}
\hat{H}  =&  \hat{H} _{\mathrm{n}}+\hat{H} _{\mathrm{nn}}+\hat{H} _{\mathrm{s}},
\label{eq:H}
\\
\hat{H} _{\mathrm{n}}
=&  \sum_{\bm{r}}\sum_{\mu}\left(-1\right)^{\bm{r}}t_{\mu}\hat{c}_{\bm{r}+\bm{a}_{\mu}}^{\dagger}U_{\mu}^{\left(-1\right)^{\bm{r}}}\hat{c}_{\bm{r}}+\mathrm{H.c.},
\label{eq:Hn}
 \\
\hat{H} _{\mathrm{nn}}
=&  \sum_{\bm{r}}\sum_{\mu<\nu}\left(-1\right)^{\bm{r}}t_{\mu\nu}
\left(\hat{c}_{\bm{r}+\bm{a}_{\mu}+\bm{a}_{\nu}}^{\dagger}\hat{c}_{\bm{r}}+\hat{c}_{\bm{r}-\bm{a}_{\mu}+\bm{a}_{\nu}}^{\dagger}\hat{c}_{\bm{r}}\right)
\nonumber\\
&+\mathrm{H.c.}
,\\
\hat{H} _{\mathrm{s}} =&  \sum_{\bm{r}}\left(-1\right)^{\bm{r}}v_{\mathrm{s}}\hat{c}_{\bm{r}}^{\dagger}\hat{c}_{\bm{r}},
\end{align}
where $\hat{c}^{\dagger}_{\bm{r}}$ and $\hat{c}_{\bm{r}}$ are the spinors of creation and annihilation operators at site $\bm{r}$, i.e.,
\begin{align}
& \hat{c}^{\dagger}_{\bm{r}} =
\begin{pmatrix}
\hat{c}^{\dagger}_{\bm{r}\uparrow} & \hat{c}^{\dagger}_{\bm{r}\downarrow}\\
\end{pmatrix}
,
&\hat{c}_{\bm{r}}  =
\begin{pmatrix}
\hat{c}_{\bm{r}\uparrow} \\
\hat{c}_{\bm{r}\downarrow}\\
\end{pmatrix},
\end{align}
and
\begin{align}
&
\left(-1\right)^{\bm{r}}=\left(-1\right)^{\sum_{\mu}n_{\mu}}
,\quad U_{\mu}^{\left(-1\right)^{\bm{r}}}
=
\begin{cases}
U_{\mu} & \text{for $\bm{r} \in\mathrm{A}$},
\\
U^{\dagger}_{\mu} & \text{for $\bm{r} \in\mathrm{B}$}.
\end{cases}
\end{align}
Each H.c. stands for the Hermitian conjugate of the preceding expression.
$\{U_{\mu}\}$ satisfy the following relation:
\begin{align}
&U^{\dagger}_{\mu}  U_{\nu} + U^{\dagger}_{\nu} U_{\mu}
=U_{\mu} U^{\dagger}_{\nu} + U_{\nu} U^{\dagger}_{\mu}
= 2\delta_{\mu\nu}\sigma_{0}
\label{eq:U-cond-Solv}
,
\end{align}
where $\sigma_{0}$ is the $2\times 2$ identity matrix for the spin degrees of freedom.

We can generally parametrize $\{U_{\mu}\}$ by unit vectors $\{\bm{n}_{\mu}\}$ ($\left\vert\bm{n}_{\mu}\right\vert=1$) and angle variables $\{\phi_{\mu}\}$ as
\begin{align}
&U_{\mu}  = \exp\left(i\phi_{\mu}\bm{n}_{\mu}\cdot\bm{\sigma}\right)
=\sigma_{0}\cos\phi_{\mu} + i\bm{n}_{\mu}\cdot\bm{\sigma}\sin\phi_{\mu},
\label{eq:U}
\end{align}
where $\{\sigma_{\mu}\}_{\mu=1,2,3}$ are the Pauli matrices for the spin degrees of freedom in the standard representation.
This expression suggests that $U_{\mu}$ is a rotational transformation with the rotation angle $(-2\phi_{\mu})$ around the axis $\bm{n}_{\mu}$.
The condition Eq.~(\ref{eq:U-cond-Solv}) is rewritten as follows:
\begin{align}
\bm{n}_{\mu}\cdot\bm{n}_{\nu}
&=\frac{1}{\sin\phi_{\mu}\sin\phi_{\nu}}\left\{
\delta_{\mu\nu}-\cos\phi_{\mu}\cos\phi_{\nu}
\right\}
\nonumber
\\
&= \delta_{\mu\nu}\csc\phi_{\mu}\csc\phi_{\nu}-\cot\phi_{\mu}\cot\phi_{\nu}.
\label{eq:nn-cond}
\end{align}
The combination of Eqs.~(\ref{eq:U}) and (\ref{eq:nn-cond}) gives the following expression for the product $U^{\dagger}_{\mu}U_{\nu}$:
\begin{align}
U^{\dagger}_{\mu}U_{\nu}
&=\delta_{\mu\nu}\sigma_{0}+i\{
-\sin\phi_{\mu}\cos\phi_{\nu}\bm{n}_{\mu}\cdot\bm{\sigma}
\nonumber\\
&\quad
+\cos\phi_{\mu}\sin\phi_{\nu}\bm{n}_{\nu}\cdot\bm{\sigma}
\nonumber\\
&\quad+\sin\phi_{\mu}\sin\phi_{\nu}\left(\bm{n}_{\mu}\times\bm{n}_{\nu}\right)\cdot\bm{\sigma}
\}
\label{eq:UdU}
.
\end{align}

$\hat{H}$ is invariant under the time-reversal ($\hat{\mathcal{T}}$) and modified space-inversion ($\hat{\mathcal{P}}_{\mathrm{m}}$) transformations below,
\begin{align}
&\hat{\mathcal{T}}
z^{\dagger}\hat{c}_{\bm{r}}
\hat{\mathcal{T}}^{-1}
=\left(\Theta z\right)^{\dagger}\hat{c}_{\bm{r}}
,
\label{eq:T}
\\
&\hat{\mathcal{P}}_{\mathrm{m}} \hat{c}_{\bm{r}} \hat{\mathcal{P}}^{-1}_{\mathrm{m}} = (-1)^{\bm{r}}\hat{c}_{-\bm{r}},
\label{eq:Pm}
\end{align}
where $\Theta = e^{-i\frac{\pi}{2}\sigma_{2}} K$,
$K$ transforms the $c$ numbers on its right into their complex conjugates,
and
$z$ is a $c$-number spinor.

Despite the restriction of Eq.~(\ref{eq:U-cond-Solv}), the series of models can describe a variety of topological and nontopological insulators even with the additional restriction $\phi_{\mu}=\phi_{\mathrm{n}}$.
\begin{align}
&\text{Example I: } \phi_{\mathrm{n}}=\frac{\pi}{2},
\nonumber\\
&\hspace{2cm}
\bm{n}_{1} = \frac{\bm{a}_{1}}{a}
,\quad
\bm{n}_{2} = \frac{\bm{a}_{2}}{a}
,\quad
\bm{n}_{3} = \frac{\bm{a}_{3}}{a}
,
\label{eq:example-I}
\\
&\text{Example I\hspace{-1pt}I: } \phi_{\mathrm{n}}=\frac{\pi}{3},
\nonumber\\
&\hspace{2cm}
\bm{n}_{\mu} =
\frac{1}{\sqrt{3}a}
\left(
-\bm{a}_{\mu}+\bm{a}_{\nu}+\bm{a}_{\lambda}
\right)
,
\label{eq:example-II}
\\
&\text{Example I\hspace{-1pt}I\hspace{-1pt}I: } \sin\phi_{\mathrm{n}}=\sqrt{\frac{2}{3}}, \quad \cos\phi_{\mathrm{n}}=\frac{1}{\sqrt{3}},
\nonumber\\
&\hspace{2cm}
\bm{n}_{\mu} =
\frac{1}{\sqrt{2}a}
\left(
\bm{a}_{\nu}-\bm{a}_{\lambda}
\right)
,
\label{eq:example-III}
\end{align}
where $[\mu,\nu,\lambda]$ runs over $[1,2,3]$, $[2,3,1]$ and $[3,1,2]$.

\subsection{\label{sec:Basis_Vectors}Real and Reciprocal Lattice Basis Vectors}
This subsection introduces unconventional bases of lattice vectors for systems with a pair of surfaces.
As preparation, we start from the conventional basis $\{\bm{a}^{\mathrm{fcc}}_{\mu}\}$ of an FCC lattice,
\begin{align}
&\bm{a}^{\mathrm{fcc}}_{\mu} = \bm{a}_{\nu}+\bm{a}_{\lambda},
\end{align}
where $[\mu,\nu,\lambda]$ runs over $[1,2,3]$, $[2,3,1]$ and $[3,1,2]$.
The reciprocal lattice basis $\{\bm{b}^{\mathrm{fcc}}_{\mu}\}$ is defined from $\{\bm{a}^{\mathrm{fcc}}_{\mu}\}$ as
\begin{align}
&\bm{b}^{\mathrm{fcc}}_{\mu} =
\sum_{\nu,\lambda}
\frac{\pi \epsilon_{\mu\nu\lambda}\bm{a}^{\mathrm{fcc}}_{\nu}\times\bm{a}^{\mathrm{fcc}}_{\lambda}}{\bm{a}^{\mathrm{fcc}}_{1}\cdot\left(\bm{a}^{\mathrm{fcc}}_{2}\times\bm{a}^{\mathrm{fcc}}_{3}\right)}
.
\label{eq:bfcc}
\end{align}

We refer to a system with a pair of $(hkl)$ surfaces as $\mathrm{Sys}^{(hkl)}$ and represent its unconventional basis with $\{\bm{a}^{(hkl)}_{\mu}\}$.
We associate the three indices $[\alpha,\beta,\gamma]$ to $\mathrm{Sys}^{(hkl)}$ as follows:
\begin{align}
&[\alpha,\beta,\gamma] =
\begin{cases}
[1,2,3] & \text{for $\mathrm{Sys}^{(100)}$},\\
[2,3,1] & \text{for $\mathrm{Sys}^{(010)}$},\\
[3,1,2] & \text{for $\mathrm{Sys}^{(001)}$}.
\end{cases}
\label{eq:axis}
\end{align}
This Supplemental Material will continue to use the above associations in subsequent sections.
The basis of $\mathrm{Sys}^{(100)\Vert(010)\Vert(001)}$ is given by
\begin{align}
\bm{a}^{(hkl)}_{\alpha} = \bm{a}_{\alpha}+\bm{a}_{\beta},
\quad
\bm{a}^{(hkl)}_{\beta} = \bm{a}_{\beta}-\bm{a}_{\gamma},
\quad
\bm{a}^{(hkl)}_{\gamma}= \bm{a}_{\beta}+\bm{a}_{\gamma},
\end{align}
and the basis of $\mathrm{Sys}^{(111)}$ is given by
\begin{align}
\bm{a}^{(111)}_{1} = \bm{a}_{1}-\bm{a}_{3},
\quad
\bm{a}^{(111)}_{2} = \bm{a}_{2}-\bm{a}_{3},
\quad
\bm{a}^{(111)}_{3} = \bm{a}_{1}+\bm{a}_{2}.
\end{align}
The reciprocal lattice basis $\{\bm{b}^{(hkl)}_{\mu}\}$ of $\mathrm{Sys}^{(hkl)}$ is defined in the same manner as Eq.~(\ref{eq:bfcc}) by the replacement $\{\bm{a}^{\mathrm{fcc}}_{\mu}\} \to \{\bm{a}^{(hkl)}_{\mu}\}$,
\begin{align}
&\bm{b}^{(hkl)}_{\mu} =
\sum_{\nu,\lambda}
\frac{\pi \epsilon_{\mu\nu\lambda}\bm{a}^{(hkl)}_{\nu}\times\bm{a}^{(hkl)}_{\lambda}}{\bm{a}^{(hkl)}_{1}\cdot\left(\bm{a}^{(hkl)}_{2}\times\bm{a}^{(hkl)}_{3}\right)}
.
\label{eq:bhkl}
\end{align}

Next, we introduce some special vectors:
\begin{align}
\bm{Q}
&=\bm{b}^{\mathrm{fcc}}_{1}+\bm{b}^{\mathrm{fcc}}_{2}+\bm{b}^{\mathrm{fcc}}_{3}
\nonumber\\
&=\bm{b}^{(100)}_{3}+\bm{b}^{(100)}_{1}
=\bm{b}^{(010)}_{1}+\bm{b}^{(010)}_{2}
\nonumber\\
&=\bm{b}^{(001)}_{2}+\bm{b}^{(001)}_{3}
=\bm{b}^{(111)}_{3},
\\
\bm{\Delta} &=  \bm{a}_{1}+\bm{a}_{2}+\bm{a}_{3},
\label{eq:Delta}
\\
\tilde{\bm{Q}}^{(hkl)} &=
\begin{cases}
\bm{b}^{(hkl)}_{\gamma} &  \text{for $\mathrm{Sys}^{(100)\Vert(010)\Vert(001)}$}, \\
\bm{b}^{(111)}_{3} &  \text{for $\mathrm{Sys}^{(111)}$},
\end{cases}
\\
\tilde{\bm{\Delta}}^{(hkl)} &=
\begin{cases}
\bm{a}_{\beta} & \text{for $\mathrm{Sys}^{(100)\Vert(010)\Vert(001)}$}, \\
\bm{a}_{3} & \text{for $\mathrm{Sys}^{(111)}$}.
\end{cases}
\label{eq:Delta_tilde}
\end{align}
The above vectors satisfy the following relations:
\begin{align}
&\bm{Q}\cdot\bm{\Delta} = 3\pi, \quad
\tilde{\bm{Q}}^{(hkl)} \cdot \tilde{\bm{\Delta}}^{(hkl)} = \pi
.
\end{align}
For the lattice sites $\{\bm{r}\}$, an expression using $\bm{\Delta}$ is found:
\begin{align}
\bm{r} & =
\begin{cases}
\bm{R} & (\bm{r}\in A),
\\
\bm{R} + \bm{\Delta} & (\bm{r}\in B),
\end{cases}
\quad
\bm{R} = \sum_{\mu}n_{\mu}\bm{a}^{\mathrm{fcc}}_{\mu},
\end{align}
where $n_{\mu}\in\mathbb{Z}$.
Let the symbol $\tilde{\bm{r}}$ represent a position within each layer as follows:
\begin{align}
\tilde{\bm{r}}  &=
\begin{cases}
\tilde{\bm{R}}  & (\bm{r}\in A), \\
\tilde{\bm{R}} +\tilde{\bm{\Delta}}^{(hkl)} & (\bm{r}\in B),
\end{cases}
\\
\tilde{\bm{R}} &=
\begin{cases}
n_{\beta}\bm{a}^{(hkl)}_{\beta} +n_{\gamma}\bm{a}^{(hkl)}_{\gamma} & \text{for $\mathrm{Sys}^{(100)\Vert(010)\Vert(001)}$}, \\
n_{1}\bm{a}^{(111)}_{1} +n_{2}\bm{a}^{(111)}_{2} & \text{for $\mathrm{Sys}^{(111)}$}.
\end{cases}
\end{align}
For the lattice sites $\{\bm{r}\}$, another expression is found:
\begin{align}
\bm{r} &=\tilde{\bm{r}}
+
\begin{cases}
n_{\alpha}\bm{a}^{(hkl)}_{\alpha} & \text{for  $\mathrm{Sys}^{(100)\Vert(010)\Vert(001)}$},
\\
n_{3}\bm{a}^{(111)}_{3} & \text{for $\mathrm{Sys}^{(111)}$}.
\end{cases}
\end{align}
The above expression leads to the following relations:
\begin{align}
\left(-1\right)^{\bm{r}}
&= e^{i \bm{Q}\cdot\bm{r}}= e^{i \tilde{\bm{Q}}^{(hkl)} \cdot \bm{r}} = e^{i \tilde{\bm{Q}}^{(hkl)} \cdot \tilde{\bm{r}}}
.
\end{align}

The two-dimensional wavevector space ($\tilde{\bm{k}}$ space) of $\mathrm{Sys}^{(hkl)}$ is generated by the basis $\{\bm{b}^{(hkl)}_{\beta}, \bm{b}^{(hkl)}_{\gamma}\}$ for $\mathrm{Sys}^{(100)\Vert(010)\Vert(001)}$ and by the basis $\{\bm{b}^{(111}_{1}, \bm{b}^{(111)}_{2}\}$ for $\mathrm{Sys}^{(111)}$.
The projected Brillouin zone of $\mathrm{Sys}^{(hkl)}$ is obtained by projecting the $\tilde{\bm{k}}$ space to the $(hkl)$ plane.
However, we naively call the $\tilde{\bm{k}}$ space the projected Brillouin zone because of the one-to-one correspondence between them.

\section{\label{sec:Systems_with_No_Surfaces}Systems with No Surfaces}
\subsection{\label{sec:FT-Hamiltonian}Fourier-transformed Hamiltonian}
For systems periodic in all directions, we combine the Fourier-transformed spinors on each sublattice into a four-component spinor:
\begin{align}
& \hat{\psi}_{\bm{k}} =
\begin{pmatrix}
\hat{c}_{\mathrm{A}\bm{k}}\\
\hat{c}_{\mathrm{B}\bm{k}}
\end{pmatrix}
,
\quad
\hat{\psi}_{\bm{k}+\bm{G}} = \hat{\psi}_{\bm{k}},
\label{eq:psi}
\\
& \hat{c}_{\mathrm{A}\bm{k}} = \frac{1}{\sqrt{N}}\sum_{\bm{R}} e^{-i\bm{k}\cdot\bm{R}}\hat{c}_{\bm{R}}
,\quad
\hat{c}_{\mathrm{B}\bm{k}} = \frac{1}{\sqrt{N}}\sum_{\bm{R}} e^{-i\bm{k}\cdot\bm{R}}\hat{c}_{\bm{R}+\bm{\Delta}}
,
\end{align}
where $N$ is the number of Wigner-Seitz unit cells and $\bm{G}$ is a reciprocal lattice vector.
The above example adopts a gauge that respects the periodicity in the wavevector space, i.e., $\hat{\psi}_{\bm{k}+\bm{G}} = \hat{\psi}_{\bm{k}}$.
We can take another gauge, such as $\hat{c}_{\mathrm{B}\bm{k}} = \frac{1}{\sqrt{N}}\sum_{\bm{R}} e^{-i\bm{k}\cdot\left(\bm{R}+\bm{\Delta}\right)}\hat{c}_{\bm{R}+\bm{\Delta}}$, at the expense of the periodicity.
The latter gauge simplifies the Hamiltonian expression in the wavevector space.
We will discuss this issue later.
The Hamiltonian in terms of the operators in Eq.~(\ref{eq:psi}) is
\begin{align}
&\hat{H}  =
\sum_{\bm{k}\in\mathrm{BZ}}
\hat{\psi}^{\dagger}_{\bm{k}}
H_{\bm{k}}
\hat{\psi}_{\bm{k}}
,\quad
H_{\bm{k}} =
\begin{pmatrix}
m_{\bm{k}}\sigma_{0} & e^{-i\bm{k}\cdot\bm{\Delta}}T^{\dagger}_{\bm{k}} \\
e^{i\bm{k}\cdot\bm{\Delta}}T_{\bm{k}} & -m_{\bm{k}}\sigma_{0}
\end{pmatrix}
,
\label{eq:Hk}
\\
&m_{\bm{k}} =v_{s}+
\sum_{\mu<\nu}4t_{\mu\nu}\cos\left(\bm{k}\cdot\bm{a}_{\mu}\right)\cos\left(\bm{k}\cdot\bm{a}_{\nu}\right)
,
\label{eq:mk}
\\
&T_{\bm{k}} =-i\sum_{\mu}2t_{\mu}\sin\left(\bm{k}\cdot\bm{a}_{\mu}\right) U_{\mu}
.
\end{align}
The subscript $\bm{k}\in\mathrm{BZ}$ of the summation in Eq.~(\ref{eq:Hk}) means that the wavevector $\bm{k}$ runs over all of the first Brillouin zone of the reciprocal space of the FCC lattice. We can confirm the periodicity $H_{\bm{k}+\bm{G}} =H_{\bm{k}} $ for an arbitrary reciprocal lattice vector $\bm{G}$ via the relations $\bm{\Delta}=\bm{a}_{\mu}+\bm{a}^{\mathrm{fcc}}_{\mu}$ and $e^{2 i\bm{G}\cdot\bm{\Delta}}=1$.
The eigenvalues of $H_{\bm{k}} $ in Eq.~({\ref{eq:Hk}}) are easily obtained as $\pm\left\vert\epsilon_{\bm{k}}\right\vert$, where
\begin{align}
& \left\vert\epsilon_{\bm{k}}\right\vert=\sqrt{t^{2}_{\bm{k}}+m^{2}_{\bm{k}}},
\quad
t^{2}_{\bm{k}} = \sum_{\mu}\left\{2t_{\mu}\sin\left(\bm{k}\cdot\bm{a}_{\mu}\right)\right\}^{2}
.
\end{align}
The above result shows that the bulk band structure does not depend on the details of $U_{\mu}$.
For regions of $\bm{k}$ space where $m_{\bm{k}}$ is positive or negative definite, we can give continuous representations of eigenvectors with eigenvalues $\pm\mathrm{sgn}\left(m_{\bm{k}}\right) \left\vert\epsilon_{\bm{k}}\right\vert$ as follows:
\begin{align}
& u_{\bm{k}\chi+} =
\frac{1}{\sqrt{N_{\bm{k}} }}
\begin{pmatrix}
\left(\left\vert\epsilon_{\bm{k}}\right\vert+\left\vert m_{\bm{k}}\right\vert\right) \chi\\
e^{i\bm{k}\cdot\bm{\Delta}}\mathrm{sgn}\left(m_{\bm{k}}\right) T_{\bm{k}} \chi
\end{pmatrix}
,
\\
& u_{\bm{k}\chi-} =
\frac{1}{\sqrt{N_{\bm{k}} }}
\begin{pmatrix}
-e^{-i\bm{k}\cdot\bm{\Delta}}\mathrm{sgn}\left(m_{\bm{k}}\right)T^{\dagger}_{\bm{k}} \chi\\
\left(\left\vert\epsilon_{\bm{k}}\right\vert+\left\vert m_{\bm{k}}\right\vert\right) \chi
\end{pmatrix}
,
\\
& N_{\bm{k}} =2\left\vert\epsilon_{\bm{k}}\right\vert\left(\left\vert\epsilon_{\bm{k}}\right\vert+\left\vert m_{\bm{k}}\right\vert\right) ,
\end{align}
where $\chi$ is an arbitrary $c$-number spinor.
Here, we introduce the following representation for the above eigenvectors in the second quantized formalism:
\begin{align}
&\Ket{\bm{k}\chi\pm}
= \hat{\psi}^{\dagger}_{\bm{k}}u_{\bm{k}\chi\pm}\Ket{0}
,
\\
&
\hat{H}\Ket{\bm{k}\chi\pm}
=\pm\mathrm{sgn}\left(m_{\bm{k}}\right)
\left\vert\epsilon_{\bm{k}}\right\vert
\Ket{\bm{k}\chi\pm}
.
\end{align}
By definition, these states also have periodicity in momentum space, $\Ket{\left(\bm{k}+\bm{G}\right)\chi\pm}=\Ket{\bm{k}\chi\pm}$.
For regions where $m_{\bm{k}}$ can be zero, we take the following gauge to represent eigenvectors:
\begin{align}
&  \tilde{u}_{\bm{k}\chi+} =
\frac{1}{\sqrt{\tilde{N}_{\bm{k}} }}
\begin{pmatrix}
\left(\left\vert\epsilon_{\bm{k}}\right\vert+m_{\bm{k}}\right) \chi\\
e^{i\bm{k}\cdot\bm{\Delta}}T_{\bm{k}} \chi
\end{pmatrix}
,
\\
&  \tilde{u}_{\bm{k}\chi-} =
\frac{1}{\sqrt{\tilde{N}_{\bm{k}} }}
\begin{pmatrix}
-e^{-i\bm{k}\cdot\bm{\Delta}}T^{\dagger}_{\bm{k}} \chi\\
\left(\left\vert\epsilon_{\bm{k}}\right\vert+m_{\bm{k}}\right) \chi
\end{pmatrix}
,
\\
&  \tilde{N}_{\bm{k}} =2\left\vert\epsilon_{\bm{k}}\right\vert\left(\left\vert\epsilon_{\bm{k}}\right\vert+m_{\bm{k}}\right) .
\end{align}
The eigenvalues of $\tilde{u}_{\bm{k}\chi\pm}$ are $\pm\left\vert\epsilon_{\bm{k}}\right\vert$.
We can find the following representation for the above eigenvectors in the second quantized formalism:
\begin{align}
&\Ket{\bm{k}\chi\pm}
= \hat{\psi}^{\dagger}_{\bm{k}}\tilde{u}_{\bm{k}\chi\pm}\Ket{0}
,
\\
&
\hat{H}\Ket{\bm{k}\chi\pm}
=\pm\left\vert\epsilon_{\bm{k}}\right\vert
\Ket{\bm{k}\chi\pm}
.
\end{align}

Here, we check the parity of the eigenvectors with respect to $\hat{\mathcal{P}}_{\mathrm{m}}$ through the formula equivalent to Eq.~(\ref{eq:Pm}):
\begin{align}
&
\hat{\mathcal{P}}_{\mathrm{m}} \hat{\psi}_{\bm{k}} \hat{\mathcal{P}}^{-1}_{\mathrm{m}}
=
\begin{pmatrix}
1 & 0 \\
0 & -e^{2i\bm{k}\cdot\bm{\Delta}}
\end{pmatrix}
 \hat{\psi}_{-\bm{k}}.
\end{align}
As the price of periodicity in $\bm{k}$ space, the eigenstates transform in a somewhat complicated way to
\begin{align}
\hat{\mathcal{P}}_{\mathrm{m}}\Ket{\bm{k}\chi\pm}
&=
\begin{cases}
\qquad\qquad\;\Ket{\left(-\bm{k}\right)\chi+},\\
-e^{-2i\bm{k}\cdot\bm{\Delta}}\Ket{\left(-\bm{k}\right)\chi-}.
\end{cases}
\end{align}
At a time-reversal symmetric (TRS) point $\bm{k}_{\mathrm{TRS}}=\bm{G}/2$, $\ket{\bm{k}_{\mathrm{TRS}}\chi\pm}$ are eigenstates of $\hat{\mathcal{P}}_{\mathrm{m}}$.
The TRS points correspond to the eight high-symmetry points, i.e., $\Gamma$, $X_{1\sim3}$, and $L_{1\sim4}$,
\begin{align}
& X_{1\sim3}: \frac{1}{2}\left(\bm{b}^{\mathrm{fcc}}_{\mu}+\bm{b}^{\mathrm{fcc}}_{\nu}\right)_{\mu\neq\nu}
,\quad
 L_{1\sim4}: \frac{\bm{b}^{\mathrm{fcc}}_{\mu}}{2},\quad \frac{\bm{Q}}{2}
,
\end{align}
where the orders of the subscripts are $[\mu, \nu] = [2,3], [3,1], [1,2]$ for $ X_{1\sim3}$ and $\mu = 1,2,3$ for $L_{1\sim4}$.
The relation $e^{i\bm{b}^{\mathrm{fcc}}_{\mu}\cdot\bm{\Delta}}=-1$ leads to the following results:
\begin{align}
\hat{\mathcal{P}}_{\mathrm{m}}\Ket{\bm{k}_{\mathrm{TRS}}\chi\pm}
&=
\begin{cases}
\pm\Ket{\bm{k}_{\mathrm{TRS}}\chi\pm} & \text{for $\Gamma$, $X_{1\sim3}$}, \\
\quad\Ket{\bm{k}_{\mathrm{TRS}}\chi\pm} & \text{for $L_{1\sim4}$}.
\end{cases}
\end{align}

In the nonperiodic gauge with $\hat{c}_{\mathrm{B}\bm{k}} = \frac{1}{\sqrt{N}}\sum_{\bm{R}} e^{-i\bm{k}\cdot\left(\bm{R}+\bm{\Delta}\right)}\hat{c}_{\bm{R}+\bm{\Delta}}$, we can delete the factor $e^{i\bm{k}\cdot\bm{\Delta}}$ in all expressions above.
Then, $\hat{\mathcal{P}}_{\mathrm{m}}$ works on the eigenstates $\Ket{\bm{k}\chi\pm}_{\mathrm{np}}$ following the concise expression below:
\begin{align}
\hat{\mathcal{P}}_{\mathrm{m}}\Ket{\bm{k}\chi\pm}_{\mathrm{np}}
&= \pm\Ket{\left(-\bm{k}\right)\chi\pm}_{\mathrm{np}}.
\end{align}
Especially at $\bm{k}_{\mathrm{TRS}}$, we obtain
\begin{align}
\hat{\mathcal{P}}_{\mathrm{m}}\Ket{\bm{k}_{\mathrm{TRS}}\chi\pm}_{\mathrm{np}}
&=
\pm\Ket{\bm{k}_{\mathrm{TRS}}\chi\pm}_{\mathrm{np}}
.
\end{align}

\subsection{\label{sec:Z2-invariants}$\mathbb{Z}_{2}$ Topological Invariants}
The four $\mathbb{Z}_{2}$ topological invariants~\cite{Fu2007,Moore2007,Fu2007a,Roy2009} are constructed from the Pfaffian of the matrix:
\begin{align}
\left[w\left(\bm{k}\right)\right]_{mn}=\Bra{-\bm{k}m}\hat{\mathcal{T}}\Ket{\bm{k}n},
\label{eq:w-matrix}
\end{align}
where $m$ and $n$ represent quantum indices for the bulk valence bands.
The first step to obtaining the invariants is to derive the following indices at $\bm{k}_{\mathrm{TRS}}$s from $w\left(\bm{k}\right)$:
\begin{align}
\delta_{\bm{k}_{\mathrm{TRS}}}=\frac{\mathrm{Pf}\left[w\left(\bm{k}_{\mathrm{TRS}}\right)\right]}{\sqrt{\mathrm{det}\left[w\left(\bm{k}_{\mathrm{TRS}}\right)\right]}}
.
\end{align}
For the explicit calculation of $w\left(\bm{k}\right)$, we need to further specifically fix the gauge.
Our model Hamiltonian describes a four-band system with two valence bands in the half-filling case.
Here, we choose the following gauge:
\begin{align}
\left\{\Ket{\bm{k}1}, \Ket{\bm{k}2}\right\}
=
\begin{cases}
\left\{\Ket{\bm{k}\chi_{0}-} , \Ket{\bm{k}\left(\Theta\chi_{0}\right)-}\right\}& \mathrm{sgn}\left(m_{\bm{k}}\right) > 0,
\\
\left\{\Ket{\bm{k}\left(\Theta\chi_{0}\right)+}, \Ket{\bm{k}\chi_{0}+}\right\} & \mathrm{sgn}\left(m_{\bm{k}}\right) < 0,
\end{cases}
\end{align}
where $\chi_{0}$ is an arbitrary $c$-number spinor and $\Theta\chi_{0}$ is its time-reversal counterpart.
We obtain $\delta_{\bm{k}_{\mathrm{TRS}}}$ in this gauge as
\begin{align}
&\delta_{\bm{k}_{\mathrm{TRS}}}=-\mathrm{sgn}\left(m_{\bm{k}_{\mathrm{TRS}}}\right).
\label{eq:delta_kTRS}
\end{align}
This result is invariant under the change from the periodic gauge to the nonperiodic gauge by the replacement $\Ket{\bm{k}\chi\pm}\to\Ket{\bm{k}\chi\pm}_{\mathrm{np}}$ in Eq.~(\ref{eq:w-matrix}).

As in the discussion on systems with ordinary space-inversion symmetry~\cite{Fu2007a},
$\delta_{\bm{k}_{\mathrm{TRS}}}$ is related to the product of the modified parities of the doubly degenerate eigenstates of the valence bands at $\bm{k}_{\mathrm{TRS}}$.
Note that a given pair of such doubly degenerate states has a common parity.
We introduce the symbol $\xi_{\bm{k}_{\mathrm{TRS}}n}$ as the modified parity of the $n$-th pair of degenerate valence states at $\bm{k}_{\mathrm{TRS}}$ and define the matrix $v\left(\bm{k}\right)$ as
\begin{align}
\left[v\left(\bm{k}\right)\right]_{mn}=\Bra{-\bm{k}m}\hat{\mathcal{P}}_{\mathrm{m}}\hat{\mathcal{T}}\Ket{\bm{k}n}.
\end{align}
Then, we obtain the relation between $\delta_{\bm{k}_{\mathrm{TRS}}}$ and $\xi_{\bm{k}_{\mathrm{TRS}}n}$:
\begin{align}
\delta_{\bm{k}_{\mathrm{TRS}}}
=\frac{\mathrm{Pf}\left[v\left(\bm{k}_{\mathrm{TRS}}\right)\right]}{\sqrt{\mathrm{det}\left[v\left(\bm{k}_{\mathrm{TRS}}\right)\right]}}
\prod_{n}\xi_{\bm{k}_{\mathrm{TRS}}n}
.
\end{align}
The above relation suggests that we may extract the topological information of the set $\left\{\delta_{\bm{k}_{\mathrm{TRS}}}\right\}$ via the set $\left\{\xi_{\bm{k}_{\mathrm{TRS}}n}\right\}$ by selecting a gauge to make $v\left(\bm{k}_{\mathrm{TRS}}\right)$ independent of $\bm{k}_{\mathrm{TRS}}$.
For example, in the nonperiodic gauge, $\delta_{\bm{k}_{\mathrm{TRS}}}$ coincides with $\prod_{n}\xi_{\bm{k}_{\mathrm{TRS}}n}$ at all TRS points.
In contrast, in the periodic gauge, $\delta_{\bm{k}_{\mathrm{TRS}}}$ coincides with $\prod_{n}\xi_{\bm{k}_{\mathrm{TRS}}n}$ only at the symmetry points $\Gamma$ and $X_{1\sim3}$ and does not coincide at $L_{1\sim4}$ by a factor of $v\left(\bm{k}_{\mathrm{TRS}}\right)$.
However, as discussed later, $\{\delta_{\bm{k}_{\mathrm{TRS}}}\}$ at $L_{1\sim4}$ is of no importance to the topological information.
We can extract the topological information via the subset $\left\{\xi_{\bm{k}_{\mathrm{TRS}}n}\right\}$ at $\Gamma$ and $X_{1\sim3}$ in either gauge.

Equations~(\ref{eq:mk}) and (\ref{eq:delta_kTRS}) provide explicit expressions for $\{\delta_{\bm{k}_{\mathrm{TRS}}}\}$ with the parameters $t_{\mu\nu}$ and $v_{\mathrm{s}}$ as follows:
\begin{align}
\delta_{\Gamma} &= -\mathrm{sgn}\left(v_{\mathrm{s}}+4t_{12}+4t_{23}+4t_{31}\right)
,
\\
\delta_{\mathrm{X}_{1}} &= -\mathrm{sgn}\left(v_{\mathrm{s}}-4t_{12}+4t_{23}-4t_{31}\right)
,
\\
\delta_{\mathrm{X}_{2}} &= -\mathrm{sgn}\left(v_{\mathrm{s}}-4t_{12}-4t_{23}+4t_{31}\right)
,
\\
\delta_{\mathrm{X}_{3}} &= -\mathrm{sgn}\left(v_{\mathrm{s}}+4t_{12}-4t_{23}-4t_{31}\right)
,
\\
\delta_{\mathrm{L}_{1}} &= \delta_{\mathrm{L}_{2}} = \delta_{\mathrm{L}_{3}} =\delta_{\mathrm{L}_{4}} = -\mathrm{sgn}\left(v_{\mathrm{s}}\right)
\label{eq:delta_L}
.
\end{align}
We relabel the eight symmetry points using the following symbols to introduce the four $\mathbb{Z}_{2}$ invariants:
\begin{align}
&\Lambda_{(m_{1},m_{2},m_{3})}: \sum_{\mu}\frac{m_{\mu}}{2}\bm{b}^{\mathrm{fcc}}_{\mu}
,
\end{align}
where $m_{\mu}\in \{0,1\}$ ($\mu=1,2,3$).
The following products of $\{\delta_{\Lambda_{(m_{1},m_{2},m_{3})}}\}$ characterize the topological phases~\cite{Fu2007a}:
\begin{align}
\delta_{0} &= \prod_{m_{1},m_{2},m_{3}} \delta_{\Lambda_{(m_{1},m_{2},m_{3})}},
\label{eq:delta_0}
\\
\delta_{1} & = \prod_{m_{2},m_{3}} \delta_{\Lambda_{(1,m_{2},m_{3})}},
\\
\delta_{2} & = \prod_{m_{1},m_{3}} \delta_{\Lambda_{(m_{1},1,m_{3})}},
\\
\delta_{3} & = \prod_{m_{1},m_{2}} \delta_{\Lambda_{(m_{1},m_{2},1)}}.
\end{align}

\begin{figure}[h]
\vspace{5mm}
\begin{tabular}{ccc}
\begin{overpic}[scale=0.17]{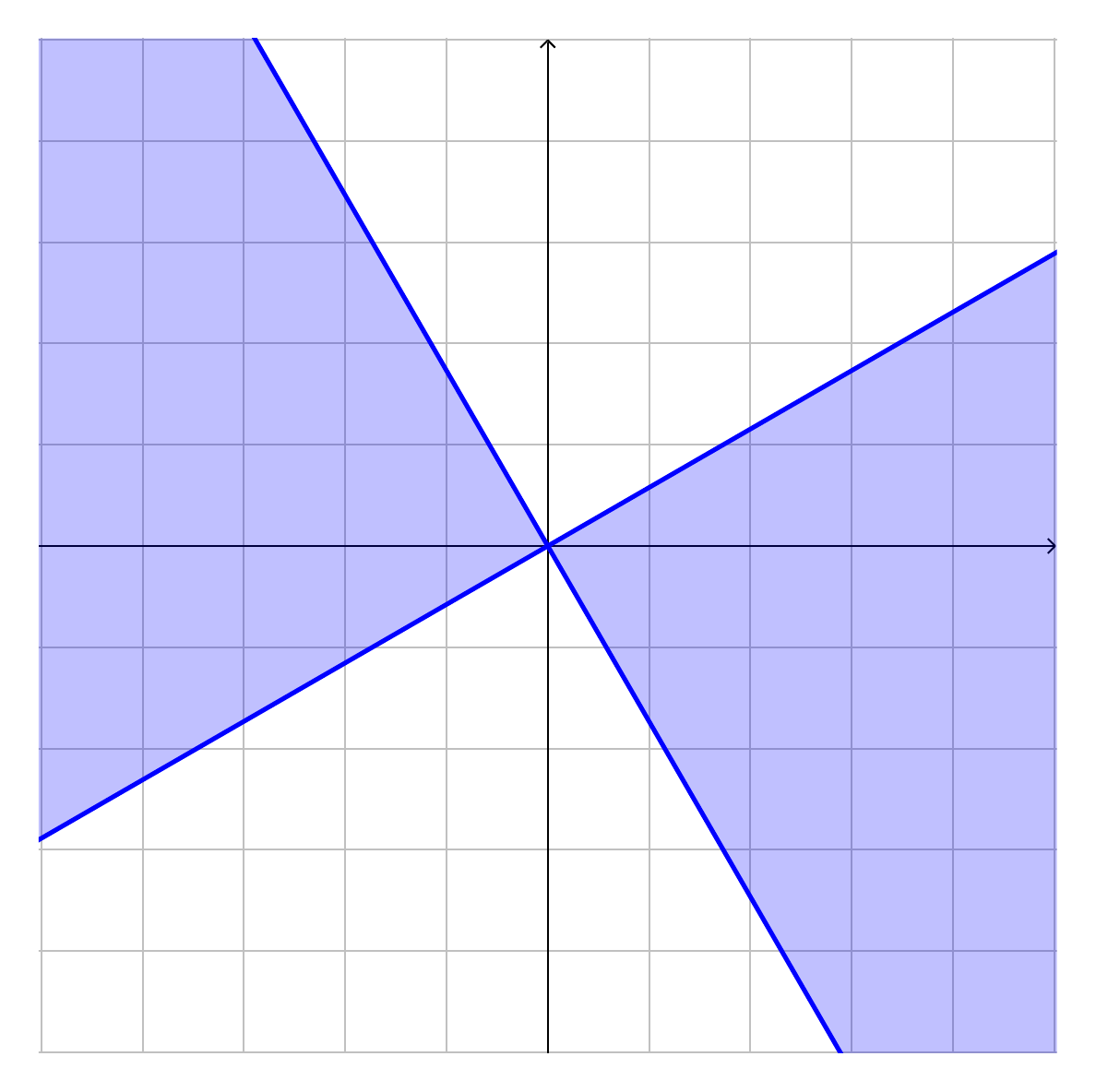}
\thicklines
\put( 2,  98){(a)}
\put( 95,  45){$4\sqrt{3}t_{\mathrm{nn}}$}
\put( 46,  96){$v_{\mathrm{s}}$}
\put( 56,  78){$v_{\mathrm{s}}=4t_{\mathrm{nn}}$}
\put( 17,  15){$v_{\mathrm{s}}=-12t_{\mathrm{nn}}$}
\end{overpic}
&
\hspace{8mm}
&
\begin{overpic}[scale=0.17]{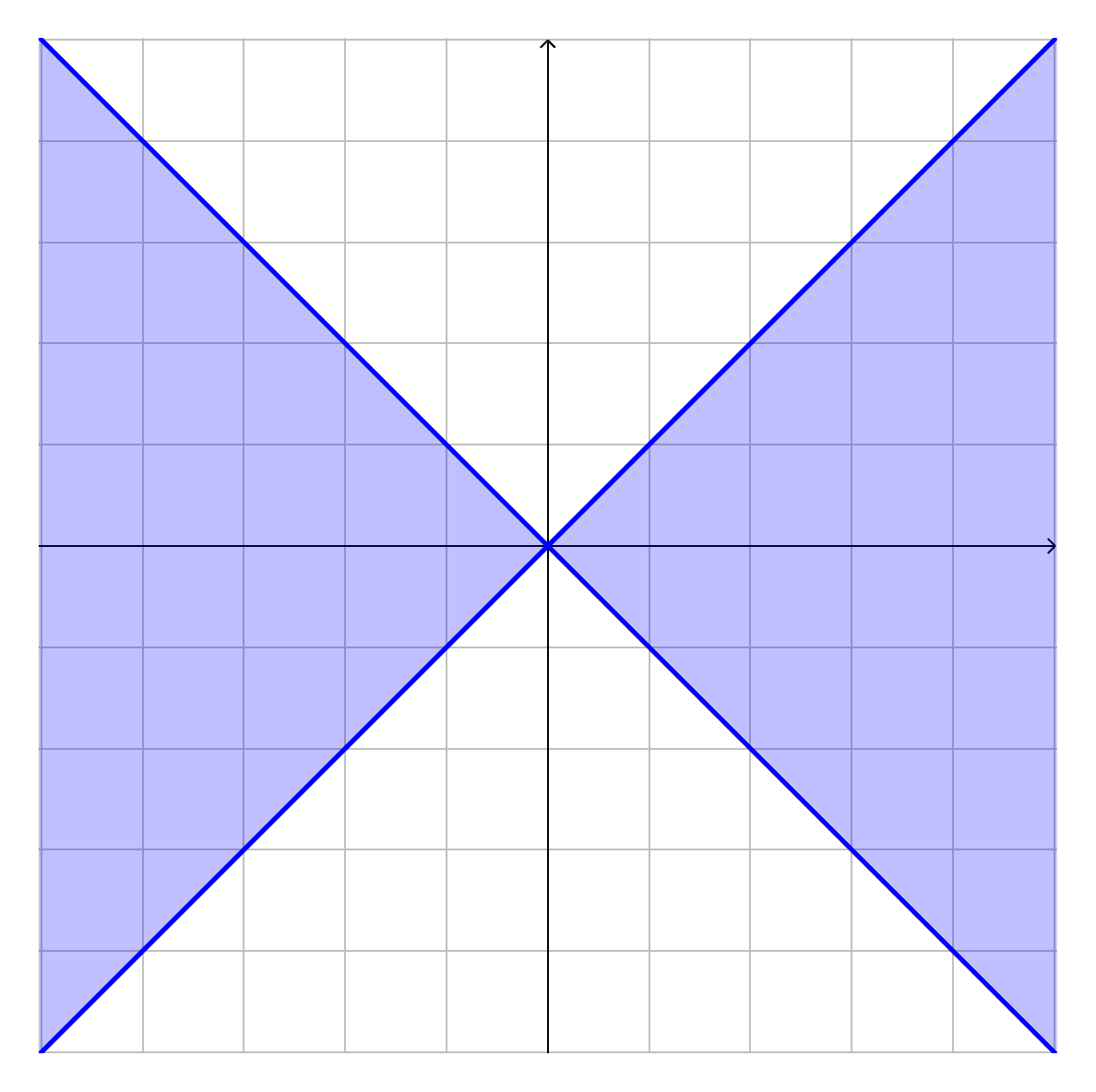}
\thicklines
\put( 2,  98){(b)}
\put( 95,  45){$4t_{\mu\nu}$}
\put( 46,  96){$v_{\mathrm{s}}$}
\put( 41,  78){$v_{\mathrm{s}}=4t_{\mu\nu}$}
\put( 36,  15){$v_{\mathrm{s}}=-4t_{\mu\nu}$}
\end{overpic}
\end{tabular}
\caption{(a) Phase diagram of the case with $t_{12}=t_{23}=t_{31}=t_{\mathrm{nn}}$.
(b) Phase diagram of the case with $t_{\mu\nu}\neq 0$ and $t_{\mu\nu}=t_{\nu\lambda}=0$, where $[\mu,\nu,\lambda]=[1,2,3]\Vert[2,3,1]\Vert[3,1,2]$.}
\label{fig:phase-diagrams}
\end{figure}

The index $\delta_{0}$ determines whether the system is a strong topological insulator.
Due to Eq.~(\ref{eq:delta_L}), Eq.~ (\ref{eq:delta_0}) yields
\begin{align}
\delta_{0} &= \delta_{\Gamma}\delta_{\mathrm{X}_{1}}\delta_{\mathrm{X}_{2}}\delta_{\mathrm{X}_{3}}
\delta_{\mathrm{L}_{1}}\delta_{\mathrm{L}_{2}}\delta_{\mathrm{L}_{3}}\delta_{\mathrm{L}_{4}}
= \delta_{\Gamma}\delta_{\mathrm{X}_{1}}\delta_{\mathrm{X}_{2}}\delta_{\mathrm{X}_{3}}
.
\end{align}
The system with $\delta_{0}=-1$ is in the strong topological insulator phase.
The indices $\delta_{\mu=1,2,3}$ classify phases more finely and define weak topological insulators.
Each of them is a product of $\delta_{\bm{k}_{\mathrm{TRS}}}$s at two $\mathrm{X}$ points and two $\mathrm{L}$ points.
Based on Eq.~(\ref{eq:delta_L}), we obtain the following:
\begin{align}
\delta_{1}
&= \delta_{\mathrm{L}_{1}} \delta_{\mathrm{X}_{2}} \delta_{\mathrm{X}_{3}} \delta_{\mathrm{L}_{4}}
= \delta_{\mathrm{X}_{2}} \delta_{\mathrm{X}_{3}}
,\\
\delta_{2}
&=\delta_{\mathrm{X}_{1}} \delta_{\mathrm{L}_{2}}  \delta_{\mathrm{X}_{3}}  \delta_{\mathrm{L}_{4}}
= \delta_{\mathrm{X}_{3}} \delta_{\mathrm{X}_{1}}
,\\
\delta_{3}
&= \delta_{\mathrm{X}_{1}} \delta_{\mathrm{X}_{2}} \delta_{\mathrm{L}_{3}}  \delta_{\mathrm{L}_{4}}
= \delta_{\mathrm{X}_{1}} \delta_{\mathrm{X}_{2}}
.
\end{align}
The topological phases are labeled by the four $\mathbb{Z}_{2}$ invariants $(\nu_{0};\nu_{1},\nu_{2},\nu_{3})$ defined below~\cite{Fu2007,Moore2007,Fu2007a,Roy2009}:
\begin{align}
&(-1)^{\nu_{\lambda}}=\delta_{\lambda} \quad (\lambda=0,1,2,3).
\end{align}
Here, we select two types of cases: (i) $t_{12}=t_{23}=t_{31}=t_{\mathrm{nn}}$ and
(ii) $t_{\mu\nu}\neq 0$, $t_{\nu\lambda}=t_{\lambda\mu}=0$,
where $[\mu,\nu,\lambda]=[1,2,3]\Vert[2,3,1]\Vert[3,1,2]$.
In case (i),
\begin{align}
\delta_{0}
&=\mathrm{sgn}\left[\left(v_{\mathrm{s}}+4t_{\mathrm{nn}}\right)^{2}-\left(8t_{\mathrm{nn}}\right)^{2}\right]
\nonumber\\
&=
\begin{cases}
+1 & \left(\left\vert v_{\mathrm{s}}+4t_{\mathrm{nn}}\right\vert >\left\vert 8t_{\mathrm{nn}}\right\vert \right), \\
-1 & \left(\left\vert v_{\mathrm{s}}+4t_{\mathrm{nn}}\right\vert <\left\vert 8t_{\mathrm{nn}}\right\vert \right),
\end{cases}
\\
\delta_{1}
&=\delta_{2} =\delta_{3} =\mathrm{sgn}\left[\left(v_{\mathrm{s}}-4t_{\mathrm{nn}}\right)^{2}\right]=+1
.
\end{align}
The system is in the strong topological insulator phase of type $(1;000)$ under the condition $\left\vert v_{\mathrm{s}}+4t_{\mathrm{nn}}\right\vert < \left\vert 8t_{\mathrm{nn}}\right\vert $ depicted by the shaded portion in Fig.~\ref{fig:phase-diagrams}(a).
In contrast, in case (ii),
\begin{align}
\delta_{0}
&=\mathrm{sgn}\left[\left(v_{\mathrm{s}}+4t_{\mu\nu}\right)^{2}\left(v_{\mathrm{s}}-4t_{\mu\nu}\right)^{2}\right]=+1
,\\
\delta_{\mu}
& = \delta_{\nu} =\mathrm{sgn}\left[v_{\mathrm{s}}^{2}-\left(4t_{\mu\nu}\right)^{2}\right]
=
\begin{cases}
+1 & \left(\left\vert v_{\mathrm{s}}\right\vert > \left\vert 4t_{\mu\nu}\right\vert \right), \\
-1 & \left(\left\vert v_{\mathrm{s}}\right\vert < \left\vert 4t_{\mu\nu}\right\vert \right),
\end{cases}
\\
 \delta_{\lambda}
&=\mathrm{sgn}\left[\left(v_{\mathrm{s}}-4t_{\mu\nu}\right)^{2}\right]=+1
.
\end{align}
The system of case (ii) with $[\mu,\nu,\lambda]=[1,2,3]\Vert[2,3,1]\Vert[3,1,2]$ is in the weak topological insulator phase of type $(0;110)\Vert(0;011)\Vert(0;101)$ under the condition $\left\vert v_{\mathrm{s}}\right\vert < \left\vert 4t_{\mu\nu}\right\vert $ depicted by the shaded portion in Fig.~\ref{fig:phase-diagrams}(b).

In concluding this section, we discuss the transition between the strong topological insulator phase of type $(1;000)$ in case (i) and the weak topological insulator phase of type $(0;110)\Vert(0;011)\Vert(0;101)$ in case (ii).
Let $\left\{t_{\mathrm{STI}}, v_{\mathrm{STI}}\right\}$ and $\left\{t_{\mathrm{WTI}}, v_{\mathrm{WTI}}\right\}$ satisfy the inequalities
$\left\vert v_{\mathrm{STI}}+4t_{\mathrm{STI}}\right\vert < \left\vert 8t_{\mathrm{STI}}\right\vert$,
$\left\vert v_{\mathrm{WTI}}\vert < \vert 4t_{\mathrm{WTI}}\right\vert$.
We consider the following trajectory in parameter space:
\begin{align}
\begin{cases}
t_{\mu\nu} = t_{\mathrm{STI}}\left(1-x\right) + t_{\mathrm{WTI}}x,
\\
t_{\nu\lambda} = t_{\lambda\mu} = t_{\mathrm{STI}}\left(1-x\right),
\\
v_{\mathrm{s}} = v_{\mathrm{STI}} \left(1-x\right) + v_{\mathrm{WTI}}x,
\end{cases}
\label{eq:trajectory}
\end{align}
where $0\leq x\leq 1$ and $[\mu,\nu,\lambda]=[1,2,3]\Vert[2,3,1]\Vert[3,1,2]$.
Then, we obtain the topological invariants as follows:
\begin{widetext}
\begin{align}
\delta_{0}
&=\mathrm{sgn}\left[\left\{\left(v_{\mathrm{STI}}+4t_{\mathrm{STI}}\right)\left(1-x\right)+\left(v_{\mathrm{WTI}}+4t_{\mathrm{WTI}}\right)x\right\}^{2}-\left\{8t_{\mathrm{STI}}\left(1-x\right)\right\}^{2}\right]
\nonumber\\
&=
\begin{cases}
+1 &  \left(\left\vert \left(v_{\mathrm{STI}}+4t_{\mathrm{STI}}\right)\left(1-x\right)+\left(v_{\mathrm{WTI}}+4t_{\mathrm{WTI}}\right)x\right\vert > \left\vert 8t_{\mathrm{STI}}\left(1-x\right)\right\vert \right), \\
-1 & \left(\left\vert \left(v_{\mathrm{STI}}+4t_{\mathrm{STI}}\right)\left(1-x\right)+\left(v_{\mathrm{WTI}}+4t_{\mathrm{WTI}}\right)x\right\vert < \left\vert 8t_{\mathrm{STI}}\left(1-x\right)\right\vert \right),
\end{cases}
\\
\delta_{\mu}
& = \delta_{\nu} 
=\mathrm{sgn}\left[\left\{\left(v_{\mathrm{STI}}-4t_{\mathrm{STI}}\right)\left(1-x\right)+v_{\mathrm{WTI}}x\right\}^{2}-\left(4t_{\mathrm{WTI}}x\right)^{2}\right]
\nonumber\\
&=
\begin{cases}
+1 &  \left(\left\vert\left(v_{\mathrm{STI}}-4t_{\mathrm{STI}}\right)\left(1-x\right)+v_{\mathrm{WTI}}x\right\vert > \left\vert4t_{\mathrm{WTI}}x\right\vert \right), \\
-1 & \left(\left\vert\left(v_{\mathrm{STI}}-4t_{\mathrm{STI}}\right)\left(1-x\right)+v_{\mathrm{WTI}}x\right\vert < \left\vert4t_{\mathrm{WTI}}x\right\vert \right), 
\end{cases}
\\
 \delta_{\lambda}
&=\mathrm{sgn}\left[\left\{\left(v_{\mathrm{STI}}-4t_{\mathrm{STI}}\right)\left(1-x\right)+\left(v_{\mathrm{WTI}}-4t_{\mathrm{WTI}}\right)x\right\}^{2}\right]=+1
.
\end{align}
\end{widetext}
The above results give the following transition points:
$x = x_{\mathrm{I}}$ as the solution of $\delta_{0}=0$ for $t_{\mathrm{STI}}t_{\mathrm{WTI}}>0$,
$x = x_{\mathrm{I\hspace{-1pt}IA}}$ and $x=x_{\mathrm{I\hspace{-1pt}IB}}$
as the solutions of $\delta_{0}=0$ and $\delta_{\mu} = \delta_{\nu}  =0$ for $t_{\mathrm{STI}}t_{\mathrm{WTI}}<0$,
where
\begin{align}
&x_{\mathrm{I}} = \frac{v_{\mathrm{STI}}-4t_{\mathrm{STI}}}{\left(v_{\mathrm{STI}}-4t_{\mathrm{STI}}\right) - \left(v_{\mathrm{WTI}}+4t_{\mathrm{WTI}}\right)}
,\\
&x_{\mathrm{I\hspace{-1pt}IA}} =  \frac{v_{\mathrm{STI}}+12t_{\mathrm{STI}}}{\left(v_{\mathrm{STI}}+12t_{\mathrm{STI}}\right) - \left(v_{\mathrm{WTI}}+4t_{\mathrm{WTI}}\right)}
,\\
&x_{\mathrm{I\hspace{-1pt}IB}}  = \frac{v_{\mathrm{STI}}-4t_{\mathrm{STI}}}{\left(v_{\mathrm{STI}}-4t_{\mathrm{STI}}\right) - \left(v_{\mathrm{WTI}}-4t_{\mathrm{WTI}}\right)}
.
\end{align}
Thus, we can find the following types of transitions:
\begin{widetext}
\begin{align}
\begin{cases}
(1;000) \overset{x_{\mathrm{I}}}{\longleftrightarrow} (0;110)\Vert(0;011)\Vert(0;101) 
& \left(t_{\mathrm{STI}}t_{\mathrm{WTI}} > 0\right),
\\
(1;000) \overset{x_{\mathrm{I\hspace{-1pt}IA}}}{\longleftrightarrow} (0;000) 
 \overset{x_{\mathrm{I\hspace{-1pt}IB}}}{\longleftrightarrow}  (0;110)\Vert(0;011)\Vert(0;101)
& \left(t_{\mathrm{STI}}t_{\mathrm{WTI}} < 0\right) \land \left(x_{\mathrm{I\hspace{-1pt}IA}} < x_{\mathrm{I\hspace{-1pt}IB}}\right),
\\
(1;000)  \overset{x_{\mathrm{I\hspace{-1pt}IB}}}{\longleftrightarrow} (1;110)\Vert(1;011)\Vert(1;101)
\overset{x_{\mathrm{I\hspace{-1pt}IA}}}{\longleftrightarrow} (0;110)\Vert(0;011)\Vert(0;101)
&\left(t_{\mathrm{STI}}t_{\mathrm{WTI}} < 0\right) \land \left(x_{\mathrm{I\hspace{-1pt}IA}} > x_{\mathrm{I\hspace{-1pt}IB}}\right).
\end{cases}
\label{eq:STI-WTI-transitions}
\end{align}
\end{widetext}
The transition of the first type also occurs when the condition $x_{\mathrm{I\hspace{-1pt}IA}} = x_{\mathrm{I\hspace{-1pt}IB}}$ accidentally holds.
In particular, the first and third types prominently demonstrate the unique feature of the series of lattice models.
%%%%%%%%%%%%%%%%%%%%%%%%%%%%%%%%%%%%%%%%%%%%%%%%%%
% Fig of Phase Diagrams
%%%%%%%%%%%%%%%%%%%%%%%%%%%%%%%%%%%%%%%%%%%%%%%%%%
%%%%%%%%%%%%%%%%%%%%%%%%%%%%%%%%%%%%%%%%%%%%%%%%%%

\section{\label{sec:Systems_with_Surfaces}Systems with ($hkl$) surfaces}
\subsection{\label{sec:PFT-Hamiltonian}Partially Fourier-transformed Hamiltonian}
This section considers systems with a pair of $(hkl)$ surfaces.
To prepare for this, we first define some constants:
\begin{align}
\tilde{N}
&:
\text{$\#$ of unit cells within each layer},
\nonumber\\
N_{\perp}
&:
\begin{cases}
\text{$\#$ of layers}
& \text{for $\mathrm{Sys}^{(100)\Vert(010)\Vert(001)}$},
\\
\text{$\#$ of pairs of layers}
& \text{for $\mathrm{Sys}^{(111)}$},
\end{cases}
\nonumber\\
\alpha &=
\begin{cases}
1 & \text{for $\mathrm{Sys}^{(100)}$},
\\
2 & \text{for $\mathrm{Sys}^{(010)}$},
\\
3 & \text{for $\mathrm{Sys}^{(001)}$ and $\mathrm{Sys}^{(111)}$}.
\end{cases}
\end{align}
Then, we partially Fourier transform the annihilation operators by imposing periodic boundary conditions in the directions parallel to the surfaces:
\begin{align}
& \hat{\psi}_{\tilde{\bm{k}}n} =
\begin{pmatrix}
\hat{c}_{\mathrm{A}\tilde{\bm{k}}n}\\
\hat{c}_{\mathrm{B}\tilde{\bm{k}}n}
\end{pmatrix}
,
\quad
\hat{\psi}_{(\tilde{\bm{k}}+\tilde{\bm{G}})n} = \hat{\psi}_{\tilde{\bm{k}}n}
,\\
& \hat{c}_{\mathrm{A}\tilde{\bm{k}}n} = \frac{1}{\sqrt{\tilde{N}}}\sum_{\tilde{\bm{R}}} e^{-i\tilde{\bm{k}}\cdot\tilde{\bm{R}}}\hat{c}_{\tilde{\bm{R}}+n\bm{a}^{(hkl)}_{\alpha}}
,\\
&
\hat{c}_{\mathrm{B}\tilde{\bm{k}}n} = \frac{1}{\sqrt{\tilde{N}}}\sum_{\tilde{\bm{R}}} e^{-i\tilde{\bm{k}}\cdot\tilde{\bm{R}}}\hat{c}_{\tilde{\bm{R}}+\tilde{\bm{\Delta}}^{(hkl)}+n\bm{a}^{(hkl)}_{\alpha}}
,
\end{align}
where $\tilde{\bm{G}}$ is a reciprocal lattice vector in the $\tilde{\bm{k}}$ space of $\mathrm{Sys}^{(hkl)}$.
The index $n\in\mathbb{Z}$ labels each layer of $\mathrm{Sys}^{(100)\Vert(010)\Vert(001)}$ and each pair of layers of $\mathrm{Sys}^{(111)}$ and takes values in the following range:
\begin{align}
\left\lfloor -\frac{N_{\perp}-1}{2} \right\rfloor \leq n \leq \left\lfloor \frac{N_{\perp}-1}{2} \right\rfloor
,
\end{align}
where $\left\lfloor x \right\rfloor$ is the floor function of $x$.

The partially Fourier-transformed Hamiltonian takes the following form,
\begin{align}
&\hat{H} =
\sum_{\tilde{\bm{k}}\in\mathrm{pBZ}}
\sum_{n}
\begin{pmatrix}
\hat{\psi}^{\dagger}_{\tilde{\bm{k}}n} & \hat{\psi}^{\dagger}_{\tilde{\bm{k}}(n+1)}
\end{pmatrix}
H^{(hkl)}_{\tilde{\bm{k}}}
\begin{pmatrix}
\hat{\psi}_{\tilde{\bm{k}}n} \\
\hat{\psi}_{\tilde{\bm{k}}(n+1)}
\end{pmatrix}
,\\
&H^{(hkl)}_{\tilde{\bm{k}}}
=
\begin{pmatrix}
D^{(hkl)}_{\tilde{\bm{k}}}  & F^{(hkl)\dagger}_{\tilde{\bm{k}}} \\
F^{(hkl)}_{\tilde{\bm{k}}}  & D^{(hkl)}_{\tilde{\bm{k}}}
\end{pmatrix}
,\\
&
D^{(hkl)}_{\tilde{\bm{k}}}  =
\begin{pmatrix}
m^{(hkl)}_{\tilde{\bm{k}}} \sigma_{0}& e^{-i\tilde{\bm{k}}\cdot\tilde{\bm{\Delta}}^{(hkl)}}\tilde{T}^{(hkl)\dagger}_{\tilde{\bm{k}}} \\
e^{i\tilde{\bm{k}}\cdot\tilde{\bm{\Delta}}^{(hkl)}}\tilde{T}^{(hkl)}_{\tilde{\bm{k}}}  & -m^{(hkl)}_{\tilde{\bm{k}}}\sigma_{0}
\end{pmatrix}
\label{eq:Dk}
,
\end{align}
\begin{widetext}
\begin{align}
&
F^{(hkl)}_{\tilde{\bm{k}}}  =
\begin{cases}
e^{i\tilde{\bm{k}}\cdot\bm{a}_{\beta}}
\begin{pmatrix}
\mu^{(hkl)}_{\tilde{\bm{k}}} \sigma_{0}& -e^{-i\tilde{\bm{k}}\cdot\tilde{\bm{\Delta}}^{(hkl)}}t_{\alpha} U^{\dagger}_{\alpha}\\
e^{i\tilde{\bm{k}}\cdot\tilde{\bm{\Delta}}^{(hkl)}} t_{\alpha} U_{\alpha} & -\mu^{(hkl)}_{\tilde{\bm{k}}}\sigma_{0}
\end{pmatrix}
& \text{for $\mathrm{Sys}^{(100)\Vert(010)\Vert(001)}$}
,\\
\begin{pmatrix}
\mu^{(111)}_{\tilde{\bm{k}}} \sigma_{0}& -e^{-i\tilde{\bm{k}}\cdot\tilde{\bm{\Delta}}^{(111)}}\tilde{T}^{(111)\dagger}_{-\tilde{\bm{k}}}\\
0 & -\mu^{(111)}_{\tilde{\bm{k}}}\sigma_{0} \\
\end{pmatrix}
& \text{for $\mathrm{Sys}^{(111)}$}
,
\end{cases}
\\
&
\tilde{T}^{(hkl)}_{\tilde{\bm{k}}} =
\begin{cases}
-i\left\{2t_{\beta}\sin\left(\tilde{\bm{k}}\cdot\bm{a}_{\beta}\right)U_{\beta}
+2t_{\gamma}\sin\left(\tilde{\bm{k}}\cdot\bm{a}_{\gamma}\right)U_{\gamma}
\right\}
& \text{for $\mathrm{Sys}^{(100)\Vert(010)\Vert(001)}$}
,\\
t_{1}e^{-i\tilde{\bm{k}}\cdot\bm{a}_{1}}U_{1}
+t_{2}e^{-i\tilde{\bm{k}}\cdot\bm{a}_{2}}U_{2}
+t_{3}e^{-i\tilde{\bm{k}}\cdot\bm{a}_{3}}U_{3}
& \text{for $\mathrm{Sys}^{(111)}$}
,
\end{cases}
\label{eq:Ttild(hkl)}
\\
&
m^{(hkl)}_{\tilde{\bm{k}}} =
\begin{cases}
v_{\mathrm{s}} + 4t_{\beta\gamma}\cos\left(\tilde{\bm{k}}\cdot\bm{a}_{\beta}\right)\cos\left(\tilde{\bm{k}}\cdot\bm{a}_{\gamma}\right)
& \text{for $\mathrm{Sys}^{(100)\Vert(010)\Vert(001)}$}
,\\
v_{\mathrm{s}} + \sum_{\mu<\nu}2t_{\mu\nu}\cos\left\{\tilde{\bm{k}}\cdot\left(\bm{a}_{\mu}-\bm{a}_{\nu}\right)\right\}
& \text{for $\mathrm{Sys}^{(111)}$}
,
\end{cases}
\label{eq:mdiag(hkl)}
\\
&
\mu^{(hkl)}_{\tilde{\bm{k}}} =
\begin{cases}
2t_{\alpha\beta}\cos\left(\tilde{\bm{k}}\cdot\bm{a}_{\beta}\right)+2t_{\gamma\alpha}\cos\left(\tilde{\bm{k}}\cdot\bm{a}_{\gamma}\right)
& \text{for $\mathrm{Sys}^{(100)\Vert(010)\Vert(001)}$}
,\\
t_{12}
+t_{23}e^{i\tilde{\bm{k}}\cdot\bm{a}^{(111)}_{1}}
+t_{31}e^{i\tilde{\bm{k}}\cdot\bm{a}^{(111)}_{2}}
& \text{for $\mathrm{Sys}^{(111)}$}
,
\end{cases}
\label{eq:moff(hkl)}
\end{align}
\end{widetext}
where the subscript $\tilde{\bm{k}}\in \mathrm{pBZ}$ means that the index $\tilde{\bm{k}}$ runs over the projected first Brillouin zone, and $\tilde{\bm{\Delta}}^{(hkl)}$ are given by Eq.~(\ref{eq:Delta_tilde}).

Next, we rearrange the indices of the sublattice and layer to express the Hamiltonian in a different form, which helps find the forms of surface-state eigenvectors.
\begin{widetext}
\begin{align}
&\hat{H}
=
\sum_{\tilde{\bm{k}}\in\mathrm{pBZ}}
\begin{pmatrix}
\hat{\Phi}^{\dagger}_{\mathrm{A}\tilde{\bm{k}}} & \hat{\Phi}^{\dagger}_{\mathrm{B}\tilde{\bm{k}}}
\end{pmatrix}
\mathcal{H}^{(hkl)}_{\tilde{\bm{k}}}
\begin{pmatrix}
\hat{\Phi}_{\mathrm{A}\tilde{\bm{k}}} \\
\hat{\Phi}_{\mathrm{B}\tilde{\bm{k}}}
\end{pmatrix}
,\quad
\mathcal{H}^{(hkl)}_{\tilde{\bm{k}}}
=
\begin{pmatrix}
\mathcal{M}^{(hkl)}_{\tilde{\bm{k}}}  &e^{-i\tilde{\bm{k}}\cdot\tilde{\bm{\Delta}}^{(hkl)}}\mathcal{T}^{(hkl)\dagger}_{\tilde{\bm{k}}} \\
e^{i\tilde{\bm{k}}\cdot\tilde{\bm{\Delta}}^{(hkl)}}\mathcal{T}^{(hkl)}_{\tilde{\bm{k}}}  & -\mathcal{M}^{(hkl)}_{\tilde{\bm{k}}}
\end{pmatrix}
,\\
&\hat{\Phi}_{X\tilde{\bm{k}}}
=
\begin{pmatrix}
\hat{c}_{X\tilde{\bm{k}}\left\lfloor -\frac{N_{\perp}-1}{2} \right\rfloor}
& \cdots
& \hat{c}_{X\tilde{\bm{k}}\left\lfloor \frac{N_{\perp}-1}{2} \right\rfloor}
\end{pmatrix}
^{\mathrm{T}}
\quad \left(X = \mathrm{A}, \mathrm{B}\right)
,\\
&\left[\mathcal{M}^{(hkl)}_{\tilde{\bm{k}}}\right]_{mn}
=
\begin{cases}
\left(
m^{(hkl)}_{\tilde{\bm{k}}} \delta_{mn}
+e^{i\tilde{\bm{k}}\cdot\bm{a}_{\beta}} \mu^{(hkl)}_{\tilde{\bm{k}}} \delta_{m(n+1)}
+e^{-i\tilde{\bm{k}}\cdot\bm{a}_{\beta}} \mu^{(hkl)}_{\tilde{\bm{k}}} \delta_{(m+1)n}
\right)\sigma_{0}
& \text{for $\mathrm{Sys}^{(100)\Vert(010)\Vert(001)}$}
,\\
 \left(
m^{(111)}_{\tilde{\bm{k}}} \delta_{mn}
+\mu^{(111)}_{\tilde{\bm{k}}} \delta_{m(n+1)}
+\overline{\mu}^{(111)}_{\tilde{\bm{k}}} \delta_{(m+1)n}
\right)\sigma_{0}
& \text{for $\mathrm{Sys}^{(111)}$}
,
\end{cases}
\\
&\left[\mathcal{T}^{(hkl)}_{\tilde{\bm{k}}}\right]_{mn}
=
\begin{cases}
\tilde{T}^{(hkl)}_{\tilde{\bm{k}}}\delta_{mn}
+
e^{i\tilde{\bm{k}}\cdot\bm{a}_{\beta}} t_{\alpha}U_{\alpha}\delta_{m(n+1)}
-e^{-i\tilde{\bm{k}}\cdot\bm{a}_{\beta}} t_{\alpha}U_{\alpha}\delta_{(m+1)n}
& \text{for $\mathrm{Sys}^{(100)\Vert(010)\Vert(001)}$}
,\\
\tilde{T}^{(111)}_{\tilde{\bm{k}}}\delta_{mn}
-\tilde{T}^{(111)}_{-\tilde{\bm{k}}}
\delta_{(m+1)n}
& \text{for $\mathrm{Sys}^{(111)}$}
,
\end{cases}
\end{align}
\end{widetext}
where $\tilde{T}^{(hkl)}_{\tilde{\bm{k}}}$, $m^{(hkl)}_{\tilde{\bm{k}}}$, and $\mu^{(hkl)}_{\tilde{\bm{k}}}$
are given by Eqs.~(\ref{eq:Ttild(hkl)}), (\ref{eq:mdiag(hkl)}), and (\ref{eq:moff(hkl)}), respectively.
$\overline{\mu}^{(111)}_{\tilde{\bm{k}}}$ is the complex conjugate of $\mu^{(111)}_{\tilde{\bm{k}}}$.

\subsection{\label{sec:Ansatz}Ansatz for Surface-state Eigenvectors}
We deduce the ansatz for the structures of surface-state eigenvectors by assuming that each surface-state eigenvector layer-dependent component is decoupled from its spin- and sublattice-dependent components as follows:
\begin{align}
u^{(hkl)}_{Y\tilde{\bm{k}}\pm}
&=
v^{(hkl)}_{Y\tilde{\bm{k}}\pm}
\otimes w^{(hkl)}_{Y\tilde{\bm{k}}\pm}
\quad \left(Y=\mathrm{U}, \mathrm{L}\right)
.
\end{align}
$u^{(hkl)}_{\mathrm{U}\tilde{\bm{k}}\pm}$ and $u^{(hkl)}_{\mathrm{L}\tilde{\bm{k}}\pm}$ represent the eigenvectors of surface states localized around the upper and lower $(hkl)$ surfaces, respectively.
$\{v^{(hkl)}_{Y\tilde{\bm{k}}\pm}\}_{Y=\mathrm{U}, \mathrm{L}}$ are their spin- and sublattice-dependent components, and $\{w^{(hkl)}_{Y\tilde{\bm{k}}\pm}\}_{Y=\mathrm{U},\mathrm{L}}$ are their layer-dependent components.

Let $\chi^{(hkl)}_{\tilde{\bm{k}}\pm}$ be $c$-number spinors.
The ansatz for $\{v^{(100)\Vert(010)\Vert(001)}_{Y\tilde{\bm{k}}\pm}\}_{Y=\mathrm{U}, \mathrm{L}}$ is expressed as
\begin{widetext}
\begin{align}
&v^{(hkl)}_{\mathrm{U}\tilde{\bm{k}}\pm}
\propto
\begin{pmatrix}
\chi^{(hkl)}_{\tilde{\bm{k}}\pm}\\
-e^{i\tilde{\bm{k}}\cdot\tilde{\bm{\Delta}}^{(hkl)}}
\mathrm{sgn}\left(t_{\alpha}\mu^{(hkl)}_{\tilde{\bm{k}}} \right)
U_{\alpha}\chi^{(hkl)}_{\tilde{\bm{k}}\pm}
\end{pmatrix}
,\quad
v^{(hkl)}_{\mathrm{L}\tilde{\bm{k}}\pm}
\propto
\begin{pmatrix}
\chi^{(hkl)}_{\tilde{\bm{k}}\pm} \\
e^{i\tilde{\bm{k}}\cdot\tilde{\bm{\Delta}}^{(hkl)}}
\mathrm{sgn}\left(t_{\alpha}\mu^{(hkl)}_{\tilde{\bm{k}}} \right)
U_{\alpha}\chi^{(hkl)}_{\tilde{\bm{k}}\pm}
\end{pmatrix}
,\\
&T^{\parallel(hkl)}_{\tilde{\bm{k}}}\chi^{(hkl)}_{\tilde{\bm{k}}\pm}
=\pm\mathrm{sgn}\left(t_{\alpha}\mu^{(hkl)}_{\tilde{\bm{k}}} \right)
\left\vert\bm{t}^{\parallel(hkl)}_{\tilde{\bm{k}}} \right\vert
\chi^{(hkl)}_{\tilde{\bm{k}}\pm}
,\quad
T^{\parallel(hkl)}_{\tilde{\bm{k}}}
=-U^{\dagger}_{\alpha}\tilde{T}^{(hkl)}_{\tilde{\bm{k}}}
=\bm{t}^{\parallel(hkl)}_{\tilde{\bm{k}}}\cdot\bm{\sigma}
\label{eq:Tpara}
,
\\
&\bm{t}^{\parallel(hkl)}_{\tilde{\bm{k}}}
=\sum_{\mu=\beta,\gamma}
2t_{\mu}\sin\left(\tilde{\bm{k}}\cdot\bm{a}_{\mu}\right)
\left\{
\sin\phi_{\alpha}\cos\phi_{\mu}\bm{n}_{\alpha}
-\cos\phi_{\alpha}\sin\phi_{\mu}\bm{n}_{\mu}
-\sin\phi_{\alpha}\sin\phi_{\mu}\left(\bm{n}_{\alpha}\times\bm{n}_{\mu}\right)
\right\}
\label{eq:tpara}
,
\\
&\left\vert\bm{t}^{\parallel(hkl)}_{\tilde{\bm{k}}}\right\vert
=2\sqrt{t^{2}_{\beta}\sin^{2}\left(\tilde{\bm{k}}\cdot\bm{a}_{\beta}\right)+t^{2}_{\gamma}\sin^{2}\left(\tilde{\bm{k}}\cdot\bm{a}_{\gamma}\right)}
\label{eq:tpara_abs}
,
\end{align}
\end{widetext}
where $\tilde{\bm{\Delta}}^{(hkl)}$, $\tilde{T}^{(hkl)}_{\tilde{\bm{k}}}$, and $\mu^{(hkl)}_{\tilde{\bm{k}}}$ are given by Eqs.~(\ref{eq:Delta_tilde}), (\ref{eq:Ttild(hkl)}), and (\ref{eq:moff(hkl)}).
The right part of Eq.~(\ref{eq:Tpara}) is obtained from Eq.~(\ref{eq:UdU}).

With the help of Eqs.~(\ref{eq:nn-cond}) and (\ref{eq:UdU}), the ansatz for $\{v^{(111)}_{Y\tilde{\bm{k}}\pm}\}_{Y=\mathrm{U}, \mathrm{L}}$ is expressed as
\begin{widetext}
\begin{align}
&v^{(111)}_{\mathrm{U}\tilde{\bm{k}}\pm}
\propto
\begin{pmatrix}
-\gamma_{\tilde{\bm{k}}\pm}
\chi^{(111)}_{\tilde{\bm{k}}\pm}
\\
\frac{\mu^{(111)}_{\tilde{\bm{k}}}}{\left\vert \mu^{(111)}_{\tilde{\bm{k}}} \right\vert}
\frac{e^{i\tilde{\bm{k}}\cdot\tilde{\bm{\Delta}}^{(111)}}}{\left\vert \epsilon_{\tilde{\bm{k}}\pm}\right\vert}
\tilde{T}^{(111)}_{-\tilde{\bm{k}}}
\chi^{(111)}_{\tilde{\bm{k}}\pm}
\end{pmatrix}
,\quad
v^{(111)}_{\mathrm{L}\tilde{\bm{k}}\pm}
\propto
\begin{pmatrix}
\chi^{(111)}_{\tilde{\bm{k}}\pm} \\
\gamma_{\tilde{\bm{k}}\pm}
\frac{\mu^{(111)}_{\tilde{\bm{k}}}}{\left\vert \mu^{(111)}_{\tilde{\bm{k}}} \right\vert}
\frac{e^{i\tilde{\bm{k}}\cdot\tilde{\bm{\Delta}}^{(111)}}}{\left\vert \epsilon_{\tilde{\bm{k}}\pm}\right\vert}
\tilde{T}^{(111)}_{-\tilde{\bm{k}}}\chi^{(111)}_{\tilde{\bm{k}}\pm}
\end{pmatrix}
,\\
&\tilde{T}^{(111)\dagger}_{-\tilde{\bm{k}}}\tilde{T}^{(111)}_{-\tilde{\bm{k}}}\chi^{(111)}_{\tilde{\bm{k}}\pm}
=\left\vert \epsilon_{\tilde{\bm{k}}\pm}\right\vert^{2} \chi^{(111)}_{\tilde{\bm{k}}\pm}
,
\quad
\tilde{T}^{(111)\dagger}_{\tilde{\bm{k}}}\tilde{T}^{(111)}_{\tilde{\bm{k}}}
=t^{2}_{1}+t^{2}_{2}+t^{2}_{3}-\bm{W}_{\tilde{\bm{k}}}\cdot\bm{\sigma}
,
\\
&\bm{W}_{\tilde{\bm{k}}}
=\sum_{\mu<\nu}2t_{\mu}t_{\nu}\sin\left\{\tilde{\bm{k}}\cdot\left(\bm{a}_{\mu}-\bm{a}_{\nu}  \right)\right\}
\left\{
-\sin\phi_{\mu}\cos\phi_{\nu}\bm{n}_{\mu}
+\cos\phi_{\mu}\sin\phi_{\nu}\bm{n}_{\nu}
+\sin\phi_{\mu}\sin\phi_{\nu}\left(\bm{n}_{\mu}\times\bm{n}_{\nu}\right)
\right\}
,\\
&
\left\vert \epsilon_{\tilde{\bm{k}}\pm}\right\vert
=\sqrt{t^{2}_{1}+t^{2}_{2}+t^{2}_{3}\pm\left\vert\bm{W}_{\tilde{\bm{k}}}\right\vert}
\label{eq:epsilon(111)}
,\quad
\left\vert\bm{W}_{\tilde{\bm{k}}}\right\vert
=2\sqrt{\sum_{\mu<\nu}t^{2}_{\mu}t^{2}_{\nu}\sin^{2}\left\{\tilde{\bm{k}}\cdot\left(\bm{a}_{\mu}-\bm{a}_{\nu}  \right)\right\}}
,\\
&\gamma_{\tilde{\bm{k}}\pm}
=
\sqrt{
1+
\left\vert\frac{\epsilon_{\tilde{\bm{k}}\pm}}{2\mu^{(111)}_{\tilde{\bm{k}}}}
\right\vert^{2}
}
-\left\vert
\frac{\epsilon_{\tilde{\bm{k}}\pm}}{2\mu^{(111)}_{\tilde{\bm{k}}}}
\right\vert
\quad
 \left(0\le\gamma_{\tilde{\bm{k}}\pm}\le1\right)
\label{eq:gamma(111)}
,
\end{align}
\end{widetext}
where $\tilde{\bm{\Delta}}^{(111)}$, $\tilde{T}^{(111)}_{\tilde{\bm{k}}}$, and $\mu^{(111)}_{\tilde{\bm{k}}}$ are given by Eqs.~(\ref{eq:Delta_tilde}), (\ref{eq:Ttild(hkl)}), and (\ref{eq:moff(hkl)}), respectively.
We can find the condition in which $\{u^{(hkl)}_{Y\tilde{\bm{k}}\pm}\}_{Y=\mathrm{U},\mathrm{L}}$ become eigenvectors by operating $\mathcal{H}^{(hkl)}_{\tilde{\bm{k}}}$ on $\{u^{(hkl)}_{Y\tilde{\bm{k}}\pm}\}_{Y=\mathrm{U},\mathrm{L}}$.
The scalar-valued parameters $\gamma_{\tilde{\bm{k}}\pm}$ are determined by preliminary adjustments that simplify the action of $\mathcal{H}^{(111)}_{\tilde{\bm{k}}}$ on $\{u^{(111)}_{Y\tilde{\bm{k}}\pm}\}_{Y=\mathrm{U},\mathrm{L}}$.
This adjustment is interpreted as the requirement for a kind of coherency in the action of off-diagonal elements of $\mathcal{H}^{(111)}_{\tilde{\bm{k}}}$ on $\{u^{(111)}_{Y\tilde{\bm{k}}\pm}\}_{Y=\mathrm{U},\mathrm{L}}$.
This process is technical and intricate.
We comment that the above results make the form of action of $\mathcal{H}^{(111)}_{\tilde{\bm{k}}}$ on $\{u^{(111)}_{Y\tilde{\bm{k}}\pm}\}_{Y=\mathrm{U},\mathrm{L}}$ the same as that of other $\mathcal{H}^{(hkl)}_{\tilde{\bm{k}}}$ on $\{u^{(hkl)}_{Y\tilde{\bm{k}}\pm}\}_{Y=\mathrm{U},\mathrm{L}}$.

Then, for every $\mathrm{Sys}^{(hkl)}$ treated in this paper, the action of $\mathcal{H}^{(hkl)}_{\tilde{\bm{k}}}$ on $\{u^{(hkl)}_{Y\tilde{\bm{k}}\pm}\}_{Y=\mathrm{U},\mathrm{L}}$ takes the form
\begin{widetext}
\begin{align}
&\mathcal{H}^{(hkl)}_{\tilde{\bm{k}}} u^{(hkl)}_{Y\tilde{\bm{k}}\pm}
=\left(E^{(hkl)}_{Y\tilde{\bm{k}}\pm}
+\Gamma^{(hkl)}_{Y\tilde{\bm{k}}\pm}
\otimes\Omega^{(hkl)}_{Y\tilde{\bm{k}}\pm}\right) u^{(hkl)}_{Y\tilde{\bm{k}}\pm}
\quad
\left(Y=\mathrm{U}, \mathrm{L}\right)
\label{eq:Hhkl-eigeneq}
,\\
&
E^{(hkl)}_{\mathrm{U}\tilde{\bm{k}}\pm}
=-E^{(hkl)}_{\mathrm{L}\tilde{\bm{k}}\pm}
,\quad
\Gamma^{(hkl)}_{\mathrm{U}\tilde{\bm{k}}\pm}
=-\tau_{2}^{\dagger}\Gamma^{(hkl)}_{\mathrm{L}\tilde{\bm{k}}\pm}\tau_{2}
,\quad
\Omega^{(hkl)}_{\mathrm{U}\tilde{\bm{k}}\pm}=\Omega^{(hkl)\dagger}_{\mathrm{L}\tilde{\bm{k}}\pm}
,\\
&E^{(hkl)}_{\mathrm{L}\tilde{\bm{k}}\pm}
=
\begin{cases}
\mp\left\vert\bm{t}^{\parallel(hkl)}_{\tilde{\bm{k}}}\right\vert
=
\mp2\sqrt{t^{2}_{\beta}\sin^{2}\left(\tilde{\bm{k}}\cdot\bm{a}_{\beta}\right)+t^{2}_{\gamma}\sin^{2}\left(\tilde{\bm{k}}\cdot\bm{a}_{\gamma}\right)}
& \text{for $\mathrm{Sys}^{(100)\Vert(010)\Vert(001)}$}
,\\
 \frac{1}{\sqrt{\left\vert\epsilon_{\tilde{\bm{k}}\pm}\right\vert^{2}+\left\vert 2\mu^{(111)}_{\tilde{\bm{k}}}\right\vert^{2}}}
\left\{
\left\vert\epsilon_{\tilde{\bm{k}}\pm}\right\vert m^{(111)}_{\tilde{\bm{k}}}
+\frac{1}{\left\vert\epsilon_{\tilde{\bm{k}}\pm}\right\vert}
\operatorname{Re}\left(
2\mu^{(111)}_{\tilde{\bm{k}}}
\det\tilde{T}^{(111)}_{-\tilde{\bm{k}}}
\right)
\right\}
& \text{for $\mathrm{Sys}^{(111)}$}
,
\end{cases}
\label{eq:E(hkl)}
\\
&\Gamma^{(hkl)}_{\mathrm{L}\tilde{\bm{k}}\pm}
=
\begin{cases}
\tau_{3}
& \text{for $\mathrm{Sys}^{(100)\Vert(010)\Vert(001)}$}
,\\
-\left\vert\frac{\epsilon_{\tilde{\bm{k}}\pm}}{2\mu^{(111)}_{\tilde{\bm{k}}}}
\right\vert\tau_{0}
+\sqrt{
1+
\left\vert\frac{\epsilon_{\tilde{\bm{k}}\pm}}{2\mu^{(111)}_{\tilde{\bm{k}}}}
\right\vert^2
}\tau_{3}
& \text{for $\mathrm{Sys}^{(111)}$}
,
\end{cases}
\label{eq:Gamma(hkl)}
\\
&\left[\Omega^{(hkl)}_{\mathrm{L}\tilde{\bm{k}}\pm}\right]_{mn}
=
\begin{cases}
m^{(hkl)}_{\tilde{\bm{k}}} \delta_{mn}
+
e^{i\tilde{\bm{k}}\cdot\bm{a}_{\beta}}
\left(\left\vert \mu^{(hkl)}_{\tilde{\bm{k}}}\right\vert-\left\vert t_{\alpha}\right\vert
\right)\delta_{m(n+1)}
+
e^{-i\tilde{\bm{k}}\cdot\bm{a}_{\beta}}
\left(\left\vert \mu^{(hkl)}_{\tilde{\bm{k}}}\right\vert+\left\vert t_{\alpha}\right\vert
\right)\delta_{(m+1)n}
& \text{for $\mathrm{Sys}^{(100)\Vert(010)\Vert(001)}$}
,\\
m^{\mathrm{mod}}_{\tilde{\bm{k}}\pm} \delta_{mn}
+
\left\vert\gamma_{\tilde{\bm{k}}\pm}\right\vert \mu^{(111)}_{\tilde{\bm{k}}}
\delta_{m(n+1)}
+
\frac{\overline{\mu}^{(111)}_{\tilde{\bm{k}}}}{\left\vert\gamma_{\tilde{\bm{k}}\pm}\right\vert}
\delta_{(m+1)n}
& \text{for $\mathrm{Sys}^{(111)}$}
,
\end{cases}
\label{eq:Omega(hkl)}
\\
&m^{\mathrm{mod}}_{\tilde{\bm{k}}\pm}
=
\frac{\left\vert 2\mu^{(111)}_{\tilde{\bm{k}}}\right\vert}{\sqrt{\left\vert\epsilon_{\tilde{\bm{k}}\pm}\right\vert^{2}+\left\vert 2\mu^{(111)}_{\tilde{\bm{k}}}\right\vert^{2}}}
\Biggl[
m^{(111)}_{\tilde{\bm{k}}}
+\frac{1}{\left\vert 2\mu^{(111)}_{\tilde{\bm{k}}}\right\vert\left\vert\epsilon_{\tilde{\bm{k}}\pm}\right\vert}
\biggl\{
-\left\vert\frac{\epsilon_{\tilde{\bm{k}}\pm}}{2\mu^{(111)}_{\tilde{\bm{k}}}}\right\vert
\operatorname{Re}\left(
2\mu^{(111)}_{\tilde{\bm{k}}}
\det\tilde{T}^{(111)}_{-\tilde{\bm{k}}}
\right)
\nonumber
\\
&
\hspace{8.6cm}
+i
\sqrt{
1+
\left\vert\frac{\epsilon_{\tilde{\bm{k}}\pm}}{2\mu^{(111)}_{\tilde{\bm{k}}}}
\right\vert^2
}
\operatorname{Im}\left(
2\mu^{(111)}_{\tilde{\bm{k}}}
\det\tilde{T}^{(111)}_{-\tilde{\bm{k}}}
\right)
\biggr\}
\Biggr]
,\label{eq:mmodL}
\end{align}
\end{widetext}
where $\tilde{T}^{(hkl)}_{\tilde{\bm{k}}}$, $m^{(hkl)}_{\tilde{\bm{k}}}$, $\mu^{(hkl)}_{\tilde{\bm{k}}}$, $\bm{t}^{\parallel(hkl)}_{\tilde{\bm{k}}}$, $\left\vert\epsilon_{\tilde{\bm{k}}\pm}\right\vert$ and $\gamma_{\tilde{\bm{k}}\pm}$ are given by Eqs.~(\ref{eq:Ttild(hkl)}), (\ref{eq:mdiag(hkl)}), (\ref{eq:moff(hkl)}), (\ref{eq:tpara}), (\ref{eq:epsilon(111)}), and (\ref{eq:gamma(111)}), respectively.
$\{E^{(hkl)}_{Y\tilde{\bm{k}}\pm}\}_{Y=\mathrm{U},\mathrm{L}}$ are the eigenvalue candidates.
$\{\Gamma^{(hkl)}_{Y\tilde{\bm{k}}\pm}\}_{Y=\mathrm{U},\mathrm{L}}$ and $\{\Omega^{(hkl)}_{Y\tilde{\bm{k}}\pm}\}_{Y=\mathrm{U},\mathrm{L}}$ represent $2\times 2$ matrices with sublattice indices and $N_{\perp}\times N_{\perp}$ matrices with layer indices, respectively.
$\{\tau_{\mu}\}_{\mu=1,2,3}$ is another set of Pauli matrices for the sublattice degrees of freedom, and $\tau_{0}$ is another $2\times 2$ identity matrix.
Equation~(\ref{eq:Hhkl-eigeneq}) suggests that $\{u^{(hkl)}_{Y\tilde{\bm{k}}\pm}\}_{Y=\mathrm{U},\mathrm{L}}$ can be the eigenvectors of $\mathcal{H}^{(hkl)}_{\tilde{\bm{k}}}$ with eigenvalues $\{E^{(hkl)}_{Y\tilde{\bm{k}}\pm}\}_{Y=\mathrm{U},\mathrm{L}}$ when the layer-dependent components $\{w^{(hkl)}_{Y\tilde{\bm{k}}\pm}\}_{Y=\mathrm{U},\mathrm{L}}$ satisfy equations $\Omega^{(hkl)}_{Y\tilde{\bm{k}}\pm} w^{(hkl)}_{Y\tilde{\bm{k}}\pm} =0$.
The matrices $\{\Omega^{(100)\Vert(010)\Vert(001)}_{Y\tilde{\bm{k}}S}\}_{Y=\mathrm{U},\mathrm{L}}$ are independent of the subscript $S=\pm$.
Thus, this is also the case for $\{w^{(100)\Vert(010)\Vert(001)}_{Y\tilde{\bm{k}}S}\}_{Y=\mathrm{U},\mathrm{L}}$.
In contrast, $\{\Omega^{(111)}_{Y\tilde{\bm{k}}S}\}_{Y=\mathrm{U},\mathrm{L}}$ and $\{w^{(111)}_{Y\tilde{\bm{k}}S}\}_{Y=\mathrm{U},\mathrm{L}}$ depend on $S=\pm$.

To conclude this subsection, we examine the dependence of $\{E^{(hkl)}_{Y\tilde{\bm{k}}\pm}\}_{Y=\mathrm{U},\mathrm{L}}$ and $\{\Omega^{(hkl)}_{Y\tilde{\bm{k}}\pm}\}_{Y=\mathrm{U},\mathrm{L}}$ on $\{U_{\mu}\}$ parametrized by $\bm{n}_{\mu}$ and $\phi_{\mu}$.
Equations~(\ref{eq:E(hkl)}) and (\ref{eq:Omega(hkl)}) suggest that only $\{E^{(111)}_{Y\tilde{\bm{k}}\pm}\}_{Y=\mathrm{U},\mathrm{L}}$ and $\{\Omega^{111)}_{Y\tilde{\bm{k}}\pm}\}_{Y=\mathrm{U},\mathrm{L}}$ can depend on $\{U_{\mu}\}$ through $\det\tilde{T}^{(111)}_{\tilde{\bm{k}}}$.
Note that $m^{\mathrm{mod}}_{\tilde{\bm{k}}\pm}$ in $\{\Omega^{111)}_{Y\tilde{\bm{k}}\pm}\}_{Y=\mathrm{U},\mathrm{L}}$ contain $\det\tilde{T}^{(111)}_{\tilde{\bm{k}}}$.
With the parametrization of $U_{\mu}$ in Eq.~(\ref{eq:U}), the definition of $\tilde{T}^{(111)}_{\tilde{\bm{k}}}$ in Eq.~(\ref{eq:Ttild(hkl)}) yields the expressions
\begin{align}
\left[\tilde{T}^{(111)}_{\tilde{\bm{k}}}\right]_{11}
&=\sum_{\mu}t_{\mu}e^{-i\tilde{\bm{k}}\cdot\bm{a}_{\mu}}
\left(
\cos\phi_{\mu}+in_{\mu,3}\sin\phi_{\mu}
\right)
,\\
\left[\tilde{T}^{(111)}_{\tilde{\bm{k}}}\right]_{12}
&=\sum_{\mu}t_{\mu}e^{-i\tilde{\bm{k}}\cdot\bm{a}_{\mu}}
i\sin\phi_{\mu}
\left(
n_{\mu,1}-in_{\mu,2}
\right)
,\\
\left[\tilde{T}^{(111)}_{\tilde{\bm{k}}}\right]_{21}
&=\sum_{\mu}t_{\mu}e^{-i\tilde{\bm{k}}\cdot\bm{a}_{\mu}}
i\sin\phi_{\mu}
\left(
n_{\mu,1}+in_{\mu,2}
\right)
,\\
\left[\tilde{T}^{(111)}_{\tilde{\bm{k}}}\right]_{22}
&=\sum_{\mu}t_{\mu}e^{-i\tilde{\bm{k}}\cdot\bm{a}_{\mu}}
\left(
\cos\phi_{\mu}-in_{\mu,3}\sin\phi_{\mu}
\right)
.
\end{align}
These expressions with the condition Eq.~(\ref{eq:nn-cond}) for $\bm{n}_{\mu}$ and $\phi_{\mu}$ prove that $\det \tilde{T}^{(111)}_{\tilde{\bm{k}}}$ is independent of $\bm{n}_{\mu}$ and $\phi_{\mu}$.
\begin{align}
\det \tilde{T}^{(111)}_{\tilde{\bm{k}}}
=&\sum_{\mu,\nu}t_{\mu}t_{\nu}e^{-i\tilde{\bm{k}}\cdot\left(\bm{a}_{\mu}+\bm{a}_{\nu}\right)}
\nonumber\\
&\times
\left(
\cos\phi_{\mu}\cos\phi_{\nu}+\sin\phi_{\mu}\sin\phi_{\nu} \bm{n}_{\mu}\cdot\bm{n}_{\nu}
\right)
\nonumber\\
=&\sum_{\mu}t^{2}_{\mu}e^{-2i\tilde{\bm{k}}\cdot\bm{a}_{\mu}}
.
\end{align}
Thus, this is also the case for $\{E^{(111)}_{Y\tilde{\bm{k}}\pm}\}_{Y=\mathrm{U},\mathrm{L}}$ and $\{\Omega^{(111)}_{Y\tilde{\bm{k}}\pm}\}_{Y=\mathrm{U},\mathrm{L}}$.

This fact ensures that the following discussion holds for arbitrary $\{U_{\mu}\}$ satisfying the condition Eq.~(\ref{eq:U-cond-Solv}) equivalent to the condition Eq.~(\ref{eq:nn-cond}).

\subsection{\label{sec:Analytical_Solutions}Analytical Solutions of Surface States}
$\{u^{(hkl)}_{Y\tilde{\bm{k}}\pm}\}_{Y=\mathrm{U},\mathrm{L}}$ can be the eigenvectors of $\mathcal{H}^{(hkl)}_{\tilde{\bm{k}}}$ when $\{w^{(hkl)}_{Y\tilde{\bm{k}}\pm}\}_{Y=\mathrm{U},\mathrm{L}}$ are the eigenvectors of $\{\Omega^{(hkl)}_{Y\tilde{\bm{k}}\pm}\}_{Y=\mathrm{U},\mathrm{L}}$ with zero eigenvalues.
Since the matrices $\{\Omega^{(hkl)}_{Y\tilde{\bm{k}}\pm}\}_{Y=\mathrm{U},\mathrm{L}}$ are tridiagonal for all cases addressed here, the eigenvalue problems of $\{\Omega^{(hkl)}_{Y\tilde{\bm{k}}\pm}\}_{Y=\mathrm{U},\mathrm{L}}$ are tractable and formally solvable except for the boundary condition issue.
We adopt the strategy of taking the semi-infinite limit $N_{\perp}\to\infty$ instead of strictly treating the boundary conditions of a finite-thickness system.
More precisely, we assume that the thickness of a given system is sufficiently thicker than any penetration depth of its surface states.
We introduce, for convenience, other layer indices $\{n_{Y}\}_{Y=\mathrm{U},\mathrm{L}}$ related to the original index $n$ ($\left\lfloor -\frac{N_{\perp}-1}{2} \right\rfloor\leq n \leq \left\lfloor \frac{N_{\perp}-1}{2} \right\rfloor$) as
\begin{align}
&n_{\mathrm{U}} =  \left\lfloor\frac{N_{\perp}+1}{2}\right\rfloor -n
\quad(1\le n_{\mathrm{U}} \le N_{\perp})
,
\\
&n_{\mathrm{L}} = \left\lceil\frac{N_{\perp}+1}{2}\right\rceil +n
\quad(1\le n_{\mathrm{L}} \le N_{\perp})
,
\end{align}
where $\left\lceil x \right\rceil$ is the ceiling function of $x$.

With the boundary conditions immediately outside the surfaces,
$\left[w^{(hkl)}_{Y\tilde{\bm{k}}\pm}\right]_{n} = 0$ at $n_{Y}=0$ ($Y=\mathrm{U}, \mathrm{L}$),
the solutions are constructed as follows:
\begin{align}
\left[w^{(hkl)}_{\mathrm{U}\tilde{\bm{k}}S}\right]_{n}
&\propto
\frac{1}{2i}\left(
\overline{\varphi}^{(hkl)n_{\mathrm{U}}}_{\tilde{\bm{k}}S+}-\overline{\varphi}^{(hkl)n_{\mathrm{U}}}_{\tilde{\bm{k}}S-}
\right)
,\\
\left[w^{(hkl)}_{\mathrm{L}\tilde{\bm{k}}S}\right]_{n}
&\propto
\frac{1}{2i}\left(
\varphi^{(hkl)n_{\mathrm{L}}}_{\tilde{\bm{k}}S+}-\varphi^{(hkl)n_{\mathrm{L}}}_{\tilde{\bm{k}}S-}
\right)
,
\end{align}
\begin{widetext}
\begin{align}
\varphi^{(hkl)}_{\tilde{\bm{k}}S\pm}
&=
\begin{cases}
e^{i\tilde{\bm{k}}\cdot\bm{a}_{\beta}}
\left\{
-\frac{\mathrm{sgn}\left(\mu^{(hkl)}_{\tilde{\bm{k}}}\right)m^{(hkl)}_{\tilde{\bm{k}}}}{2\left(\left\vert \mu^{(hkl)}_{\tilde{\bm{k}}}\right\vert+\left\vert t_{\alpha}\right\vert\right)}
\pm
\sqrt{
\left(
\frac{m^{(hkl)}_{\tilde{\bm{k}}}}{2\left(\left\vert \mu^{(hkl)}_{\tilde{\bm{k}}}\right\vert+\left\vert t_{\alpha}\right\vert\right)}
\right)^{2}
-\frac{\left\vert \mu^{(hkl)}_{\tilde{\bm{k}}}\right\vert-\left\vert t_{\alpha}\right\vert}{\left\vert \mu^{(hkl)}_{\tilde{\bm{k}}}\right\vert+\left\vert t_{\alpha}\right\vert}
}
\right\}
& \text{for $\mathrm{Sys}^{(100)\Vert(010)\Vert(001)}$}
,\\
\left\vert\gamma_{\tilde{\bm{k}}S}\right\vert
\left\{
-\frac{m^{\mathrm{mod}}_{\tilde{\bm{k}}S}}{2\overline{\mu}^{(111)}_{\tilde{\bm{k}}}}
\pm
\sqrt{
\left(
\frac{m^{\mathrm{mod}}_{\tilde{\bm{k}}S}}{2\overline{\mu}^{(111)}_{\tilde{\bm{k}}}}
\right)^{2}
-\frac{\mu^{(111)}_{\tilde{\bm{k}}}}{\overline{\mu}^{(111)}_{\tilde{\bm{k}}}}
}
\right\}
& \text{for $\mathrm{Sys}^{(111)}$}
,
\end{cases}
\end{align}
\end{widetext}
where $m^{(hkl)}_{\tilde{\bm{k}}}$, $\mu^{(hkl)}_{\tilde{\bm{k}}}$, and $m^{\mathrm{mod}}_{\tilde{\bm{k}}S}$ are given by Eqs.~(\ref{eq:mdiag(hkl)}), (\ref{eq:moff(hkl)}), and (\ref{eq:mmodL}), respectively.
We can see that $\varphi^{(100)\Vert(010)\Vert(001)}_{\tilde{\bm{k}}S\pm}$ do not depend on the subscript $S=\pm$, while $\varphi^{(111)}_{\tilde{\bm{k}}S\pm}$ depend on $S=\pm$ through $\left\vert\gamma_{\tilde{\bm{k}}S}\right\vert$ and $m^{\mathrm{mod}}_{\tilde{\bm{k}}S}$.
Thus, this is also the case for $\{w^{(hkl)}_{Y\tilde{\bm{k}}S}\}_{Y=\mathrm{U},\mathrm{L}}$.

To conclude this subsection, we discuss the conditions under which the above solutions can be surface states.
We derive the conditions in which the vectors $\{w^{(hkl)}_{Y\tilde{\bm{k}}\pm}\}_{Y=\mathrm{U},\mathrm{L}}$ have suitable properties as layer-dependent components of surface-state eigenvectors.
In general, the boundary conditions on the decaying side of surface states make the problem complicated.
However, for semi-infinite systems, requiring that $\left\vert w^{(hkl)}_{\mathrm{U}\tilde{\bm{k}}\pm}\right\vert \Vert \left\vert w^{(hkl)}_{\mathrm{L}\tilde{\bm{k}}\pm}\right\vert$ decay as the layer index $n$ decreases $\Vert$ increases is sufficient for $u^{(hkl)}_{\mathrm{U}\tilde{\bm{k}}\pm}\Vert u^{(hkl)}_{\mathrm{L}\tilde{\bm{k}}\pm}$ to be surface-state eigenvectors localized around the upper$\Vert$lower surface.

The attenuation condition for $\mathrm{Sys}^{(100)\Vert(010)\Vert(001)}$ is given by
\begin{widetext}
\begin{align}
&
\sqrt{\left\vert
\frac{\left\vert \mu^{(hkl)}_{\tilde{\bm{k}}}\right\vert-\left\vert t_{\alpha}\right\vert}{\left\vert \mu^{(hkl)}_{\tilde{\bm{k}}}\right\vert+\left\vert t_{\alpha}\right\vert}
\right\vert}
<\min
\left[
\left\vert
-\frac{m^{(hkl)}_{\tilde{\bm{k}}}}{2\sqrt{\left\vert \mu^{(hkl)}_{\tilde{\bm{k}}}\right\vert^{2}-\left\vert t_{\alpha}\right\vert^{2}}}
\pm
\sqrt{
\frac{\left\vert m^{(hkl)}_{\tilde{\bm{k}}}\right\vert^{2}}{4\left(\left\vert \mu^{(hkl)}_{\tilde{\bm{k}}}\right\vert^{2}-\left\vert t_{\alpha}\right\vert^{2}\right)}
-1
}
\right\vert
\right]
,
\label{eq:condition-hkl)}
\end{align}
\end{widetext}
where $m^{(hkl)}_{\tilde{\bm{k}}}$ and $\mu^{(hkl)}_{\tilde{\bm{k}}}$ are defined by Eqs.~(\ref{eq:mdiag(hkl)}) and (\ref{eq:moff(hkl)}), respectively.
In contrast, the attenuation condition for $\mathrm{Sys}^{(111)}$ is given by
\begin{align}
&
\left\vert
\gamma_{\tilde{\bm{k}}S}
\right\vert
<\min
\left[
\left\vert
-\frac{m^{\mathrm{mod}}_{\tilde{\bm{k}}S}}{\left\vert2\mu^{(111)}_{\tilde{\bm{k}}}\right\vert}
\pm
\sqrt{
\left(
\frac{m^{\mathrm{mod}}_{\tilde{\bm{k}}S}}{\left\vert2\mu^{(111)}_{\tilde{\bm{k}}}\right\vert}
\right)^{2}
-1
}
\right\vert
\right]
,
\label{eq:condition-111)}
\end{align}
where $\mu^{(111)}_{\tilde{\bm{k}}}$, $\gamma_{\tilde{\bm{k}}\pm}$ and $m^{\mathrm{mod}}_{\tilde{\bm{k}}}$ are defined by Eqs.~(\ref{eq:moff(hkl)}), (\ref{eq:gamma(111)}), and (\ref{eq:mmodL}), respectively.

\section{\label{sec:Symmetry_Points}Symmetry Points in Brillouin Zones}
This section explains the labeling of symmetry points appearing in the band diagrams of the main article and this Supplemental Material.
We name some symmetry points of the conventional Brillouin zone of the FCC lattice as follows:
\begin{align}
X:
& \frac{1}{2}\left(\bm{b}^{\mathrm{fcc}}_{2}+\bm{b}^{\mathrm{fcc}}_{3}\right),
& X':
& \frac{1}{2}\left(\bm{b}^{\mathrm{fcc}}_{1}+\bm{b}^{\mathrm{fcc}}_{2}\right),
\nonumber
\\
W:
& \frac{1}{4}\left(\bm{b}^{\mathrm{fcc}}_{1}+3\bm{b}^{\mathrm{fcc}}_{2}+2\bm{b}^{\mathrm{fcc}}_{3}\right),
& W':
& \frac{1}{4}\left(2\bm{b}^{\mathrm{fcc}}_{1}+3\bm{b}^{\mathrm{fcc}}_{2}+\bm{b}^{\mathrm{fcc}}_{3}\right),
&&
\nonumber
\\
U:
& \frac{1}{8}\left(5\bm{b}^{\mathrm{fcc}}_{2}+3\bm{b}^{\mathrm{fcc}}_{3}\right),
& U':
& \frac{1}{8}\left(3\bm{b}^{\mathrm{fcc}}_{1}+5\bm{b}^{\mathrm{fcc}}_{2}\right),
\nonumber
\\
L:
& \frac{1}{2}\bm{b}^{\mathrm{fcc}}_{2},
& K:
& \frac{3}{8}\left(\bm{b}^{\mathrm{fcc}}_{1}+2\bm{b}^{\mathrm{fcc}}_{2}+\bm{b}^{\mathrm{fcc}}_{3}\right),
\end{align}
where $\{\bm{b}^{\mathrm{fcc}}_{\mu}\}$ is the reciprocal lattice basis of Eq.~(\ref{eq:bfcc}).
These symmetry points appear in panel (a) of each set of band diagrams in this Supplemental Material.
%	   FCC basis    Cartesian
%	X1: ( 0  1/2 1/2)  ( 1    0   0 )
%	U1: ( 0  5/8 3/8)  ( 1  -1/4 1/4)
%	L : ( 0  1/2  0 )  (1/2 -1/2 1/2)
%	G : ( 0   0   0 )  ( 0    0   0 )
%	X1: ( 0  1/2 1/2)  ( 1    0   0 )
%	W1: (1/4 3/4 1/2)  ( 1    0  1/2)
%	K : (3/8 3/4 3/8)  (3/4   0  3/4)
%	W3: (1/2 3/4 1/4)  (1/2   0   1 )
%	X3: (1/2 1/2  0 )  ( 0    0   1 )
%	G : ( 0   0   0 )  ( 0    0   0 )
%	L : ( 0  1/2  0 )  (1/2 -1/2 1/2)
%	U3: (3/8 5/8  0 )  (1/4 -1/4  1 )
%	X3: (1/2 1/2  0 )  ( 0    0   1 )

The symmetry points of the surface Brillouin zone of $\mathrm{Sys}^{(111)}$ are the projections of the following reciprocal vectors to the $(111)$ surface and are labeled by the symbols $K$, $K'$, $M_{1}$, $M_{2}$, and $M_{3}$ in panel (a) of each set of band diagrams in the main article and panel (b) of each set of band diagrams in this Supplemental Material:
\begin{align}
K:
&\frac{1}{3}\left(2\bm{b}^{(111)}_{1}+\bm{b}^{(111)}_{2}\right),
&K':
&\frac{1}{3}\left(\bm{b}^{(111)}_{1}+2\bm{b}^{(111)}_{2}\right),
\nonumber
\\
M_{1}:
&\frac{1}{2}\bm{b}^{(111)}_{1},
&M_{2}:
&\frac{1}{2}\left(\bm{b}^{(111)}_{1}+\bm{b}^{(111)}_{2}\right),
\nonumber
\\
M_{3}:
&\frac{1}{2}\bm{b}^{(111)}_{2},
&&
\end{align}
where $\{\bm{b}^{(111)}_{\mu}\}$ is the reciprocal lattice basis of Eq.~(\ref{eq:bhkl}) with $(hkl)=(111)$.
%     Definition in the program of 20210523 version
%	      (111)-basis    Cartesian
%	G : (  0    0    0 )  (  0    0    0 )
%	M2: ( 1/2  1/2   0 )  (  0    0   -1 )
%	M3: (  0   1/2   0 )  (-1/2  1/2 -1/2)
%	K': ( 1/3  2/3   0 )  (-1/3  1/3  -1 )
%	G : (  0    0    0 )  (  0    0    0 )
%	K : ( 2/3  1/3   0 )  ( 1/3 -1/3  -1 )
%	M2: ( 1/2  1/2   0 )  (  0    0   -1 )
%	M1: ( 1/2   0    0 )  ( 1/2 -1/2 -1/2)
%	G : (  0    0    0 )  (  0    0    0 )

The symmetry points of the surface Brillouin zone of $\mathrm{Sys}^{(100)\Vert(001)}$ are the projections of the following reciprocal vectors to the ${(100)\Vert(001)}$ surface and are labeled by the symbols $X$, $X'$, $M$, and $M'$
in panel (b) of each set of band diagrams in the main article and
in panels (c) and (d) of each set of band diagrams in this Supplemental Material:
\begin{align}
X:
& -\frac{1}{2}\bm{b}^{(hkl)}_{\beta},
& X':
&\frac{1}{2}\bm{b}^{(hkl)}_{\gamma},
\nonumber
\\
M:
&\frac{1}{2}\left(\bm{b}^{(hkl)}_{\beta}+\bm{b}^{(hkl)}_{\gamma}\right),
& M':
&\frac{1}{2}\left(-\bm{b}^{(hkl)}_{\beta}+\bm{b}^{(hkl)}_{\gamma}\right),
\end{align}
where $\{\bm{b}^{(hkl)}_{\mu}\}$ is the reciprocal lattice basis of Eq.~(\ref{eq:bhkl}) with $(hkl)=(100)\Vert(001)$
and $[\beta,\gamma]=[2,3]\Vert[1,2]$ for $\mathrm{Sys}^{(100)\Vert(001)}$.
%     Definition in the program of 20210523 version
%	(001)-basis         Cartesian
%	G: (  0    0   0 )  (  0    0    0 )
%	X':(  0   1/2  0 )  ( 1/2  1/2 -1/2)
%	-X:(-1/2   0   0 )  (-1/2  1/2  1/2)
%	M':(-1/2  1/2  0 )  (  0    1    0 )
%	G: (  0    0   0 )  (  0    0    0 )
%	M: ( 1/2  1/2  0 )  (  1    0   -1 )
%	X':(  0   1/2  0 )  ( 1/2  1/2 -1/2)	
%	X: (1/2   0   0 )  ( 1/2 -1/2 -1/2)
%	G: (  0    0   0 )  (  0    0    0 )
%	-----------
%	   (100)-basis       Cartesian
%	G: (  0    0    0 )  (  0    0    0 )
%	X':(  0    0   1/2)  (-1/2  1/2  1/2)
%	-X:(  0  -1/2   0 )  ( 1/2 -1/2  1/2)
%	M':(  0  -1/2  1/2)  (  0    0    1 )
%	G: (  0    0    0 )  (  0    0    0 )
%	M: (  0   1/2  1/2)  ( -1    1    0 )
%	X':(  0    0   1/2)  (-1/2  1/2  1/2)
%	X: (  0   1/2   0 )  (-1/2  1/2 -1/2)
%	G: (  0    0    0 )  (  0    0    0 )
%
%	NOTE: M'-point is equivalent to M-point, while GM'-line is not equivalent to GM-line.
Although $X$ and its time reversal represent the same symmetry point, we introduce $\overline{X}$ as the time reversal of $X$.
The same is true for $M$ and $M'$.
These redundancies are purposely introduced to specify the drawing path of the projected band diagram of $\mathrm{Sys}^{(100)\Vert(001)}$.
For example, we distinguish between two different symmetry lines $X'\overline{X}$ and $X' X$ and between two different symmetry lines $\Gamma M$ and $\Gamma M'$.

Finally, we supplement the discussion of the band structures of the surface-state analytical solutions and the corresponding numerical solutions for the sets of parameters listed in Table~\ref{table:Parameter_Sets}, where $t_{\mathrm{n}}$ is an energy unit.
These are special cases of Example I\hspace{-1pt}I expressed by Eq.~(\ref{eq:example-II}) with isotropic nearest-neighbor hopping $t_{\mu}=t_{\mathrm{n}}$ ($\mu=1,2,3$). In the main article, only TI-$t_{\mathrm{nn}}$ and TI-$t_{12}$ are selected.
The parameter sets TI-$t_{\mathrm{nn}}$, TI-$t_{12}$, and TI-$t_{23}$ correspond to an isotropic strong topological insulator and two anisotropic weak topological insulators.
The two anisotropic weak topological insulators are essentially the same but face different directions concerning a pair of surfaces under consideration.
All numerical results are obtained with a thickness of 60 pairs of layers for $\mathrm{Sys}^{(111)}$ and a thickness of 60 layers for $\mathrm{Sys}^{(100)\Vert(001)}$. Each layer contains $360\times 360$ unit cells.

\begin{table}[h]
 \caption{Parameter sets}
 \label{table:Parameter_Sets}
 \centering
  \begin{tabular}{l c rrr c r c l}
    \hline \hline
    Set name&& $ t_{12}/t_{\mathrm{n}}$ & $t_{23}/t_{\mathrm{n}}$ & $t_{31}/t_{\mathrm{n}}$ && $v_{\mathrm{s}}/t_{\mathrm{n}}$&  &Model system\\
    \hline
    TI-$t_{\mathrm{nn}}$       && $-0.150$ & $-0.150$ & $-0.150$ && $ 0.800$ && Strong TI\\
    TI-$t_{12}$      && $-0.300$ & $ 0.000$ & $ 0.000$ && $ 0.200$ && Weak TI\\
    TI-$t_{23}$     && $ 0.000$ & $-0.300$ & $ 0.000$ && $ 0.200$ && Weak TI\\
    \hline \hline
  \end{tabular}
\end{table}

\begin{figure}[htb]
\vspace{5mm}
\begin{tabular}{ccc}
\begin{overpic}[scale=0.25, clip]{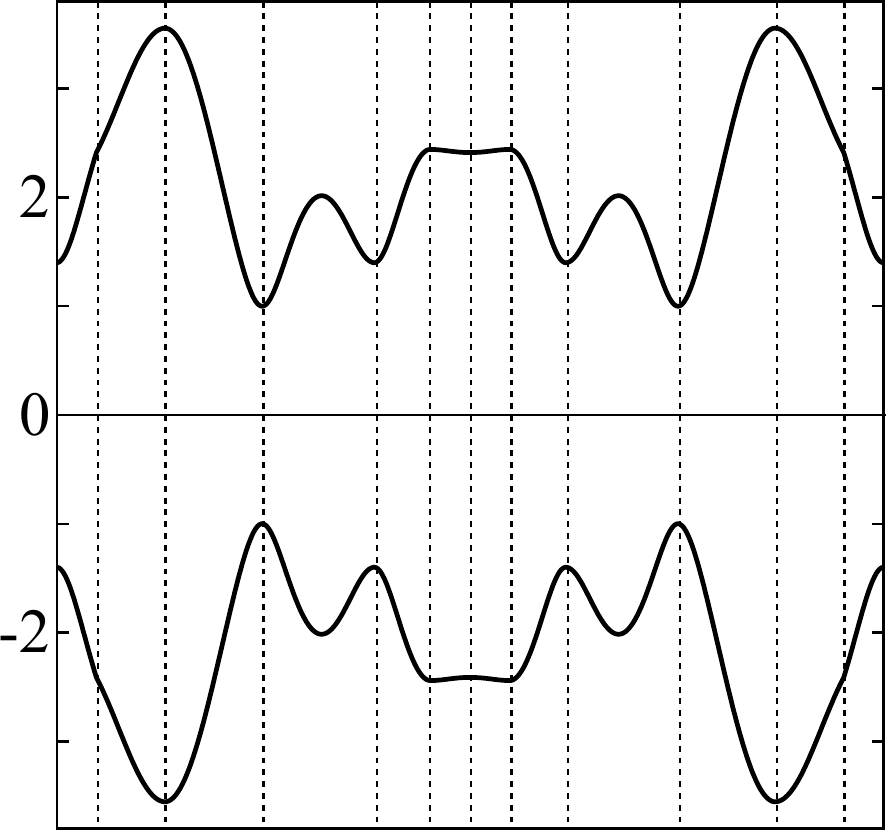}
\thicklines
\put( -6, 102){(a)}
\put( -8, 43){\rotatebox{90}{$\epsilon/t_{\mathrm{n}}$}}
\put(    3,    -8){$\mathrm{X}$}
\put(   9, 102){$\mathrm{U}$}
\put(  17,    -8){$\mathrm{L}$}
\put(  29,    -8){$\Gamma$}
\put(  42,    -8){$\mathrm{X}$}
\put(  47, 102){$\mathrm{W}$}
\put(  53,    -8){$\mathrm{K}$}
\put(  57, 102){$\mathrm{W}'$}
\put(  65,    -8){$\mathrm{X}'$}
\put(  79,    -8){$\Gamma$}
\put(  90,    -8){$\mathrm{L}$}
\put(  98, 102){$\mathrm{U}'$}
\put(103,    -8){$\mathrm{X}'$}
\end{overpic}
&
\hspace{2mm}
&
\begin{overpic}[scale=0.25, clip]{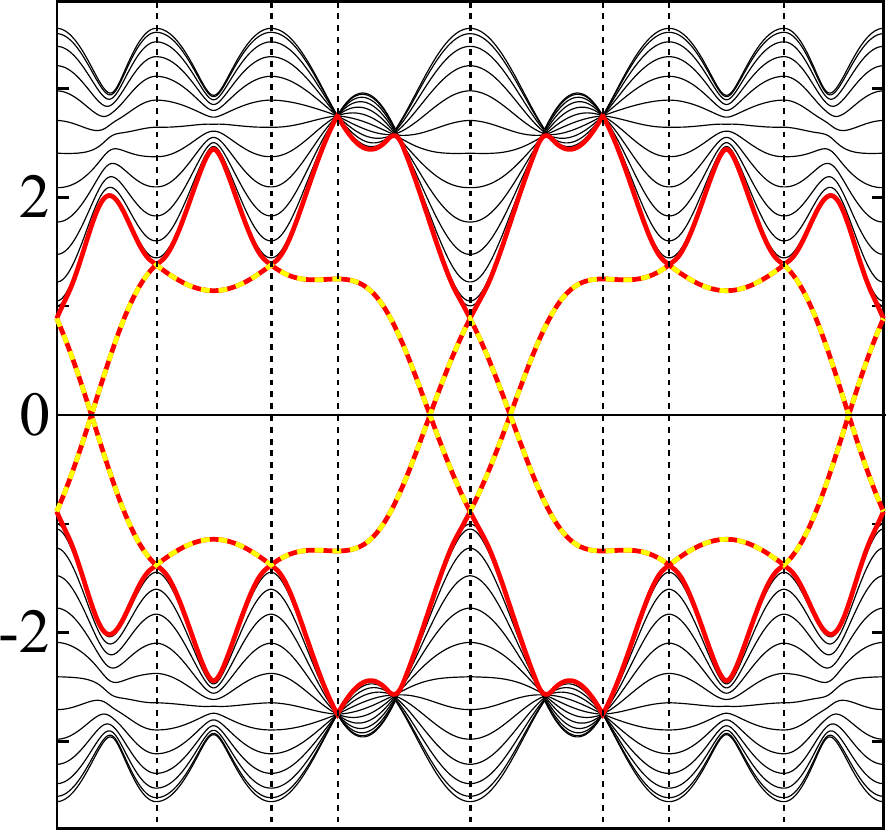}
\thicklines
\put( -6, 102){(b)}
\put(   4,    -8){$\Gamma$}
\put(  14, 102){$\mathrm{M}_{2}$}
\put(  28, 102){$\mathrm{M}_{3}$}
\put(  37,    -8){$\mathrm{K}'$}
\put(  54,    -8){$\Gamma$}
\put( 69,    -8){$\mathrm{K}$}
\put( 76, 102){$\mathrm{M}_{2}$}
\put( 90, 102){$\mathrm{M}_{1}$}
\put(103,   -8){$\Gamma$}
\end{overpic}
\\
\vspace{5mm}
\\
\begin{overpic}[scale=0.25, clip]{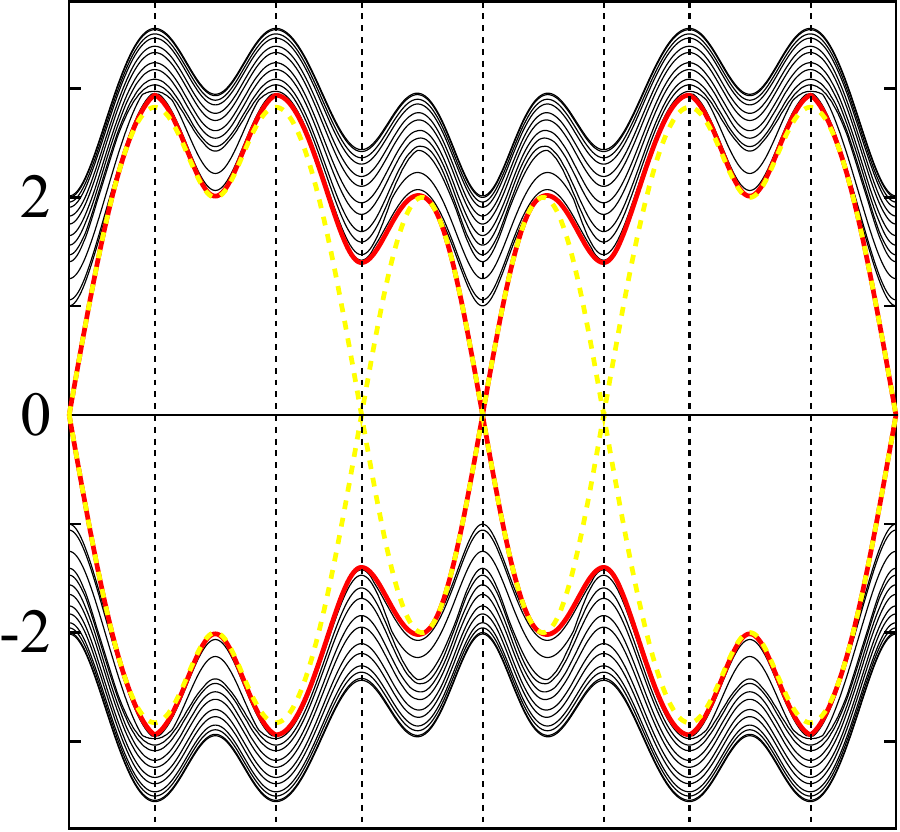}
\thicklines
\put( -4, 102){(c)}
\put( -8, 43){\rotatebox{90}{$\epsilon/t_{\mathrm{n}}$}}
\put(    5,  -8){$\Gamma$}
\put(  15, 102){$\mathrm{X}'$}
\put(  30, 102){$\overline{\mathrm{X}}$}
\put(  39,  -8){$\mathrm{M}'$}
\put(  55,  -8){$\Gamma$}
\put(  68,  -8){$\mathrm{M}$}
\put(  80, 102){$\mathrm{X}'$}
\put(  95, 102){$\mathrm{X}$}
\put(105,  -8){$\Gamma$}
\end{overpic}
&
\hspace{2mm}
&
\begin{overpic}[scale=0.25, clip]{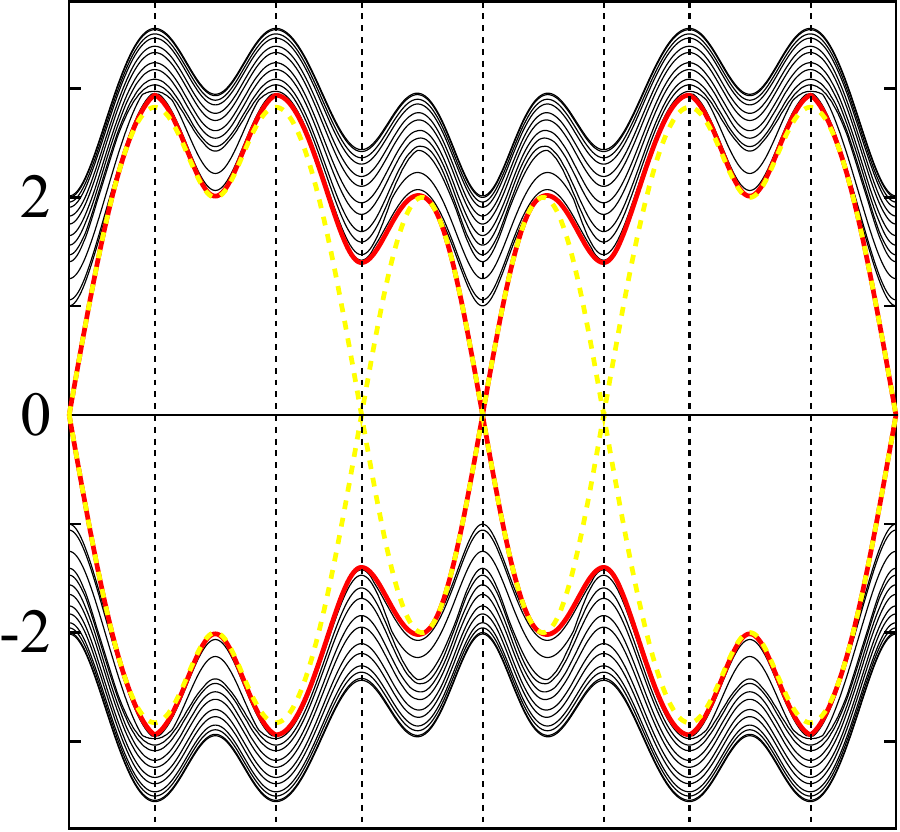}
\thicklines
\put( -5, 102){(d)}
\put(    5,  -8){$\Gamma$}
\put(  15, 102){$\mathrm{X}'$}
\put(  30, 102){$\overline{\mathrm{X}}$}
\put(  39,  -8){$\mathrm{M}'$}
\put(  55,  -8){$\Gamma$}
\put(  68,  -8){$\mathrm{M}$}
\put(  80, 102){$\mathrm{X}'$}
\put(  95, 102){$\mathrm{X}$}
\put(105,  -8){$\Gamma$}
\end{overpic}
\end{tabular}
\vspace{0mm}
\caption{
(a) Band diagram and (b)-(d) projected band diagrams of a strong topological insulator ($t_{\mathrm{nn}}/t_{\mathrm{n}}=-0.15, v_{\mathrm{s}}/t_{\mathrm{n}}=0.8$) in (a) a 3D torus shape without a surface and in a slab shape with (b) $(111)$ surfaces, (c) $(100)$ surfaces, and (d) $(001)$ surfaces.
}
\label{fig:band-diagrams_strong-TI_tnn}
\end{figure}

\begin{figure}[t]
\vspace{5mm}
\begin{tabular}{ccc}
\begin{overpic}[scale=0.25, clip]{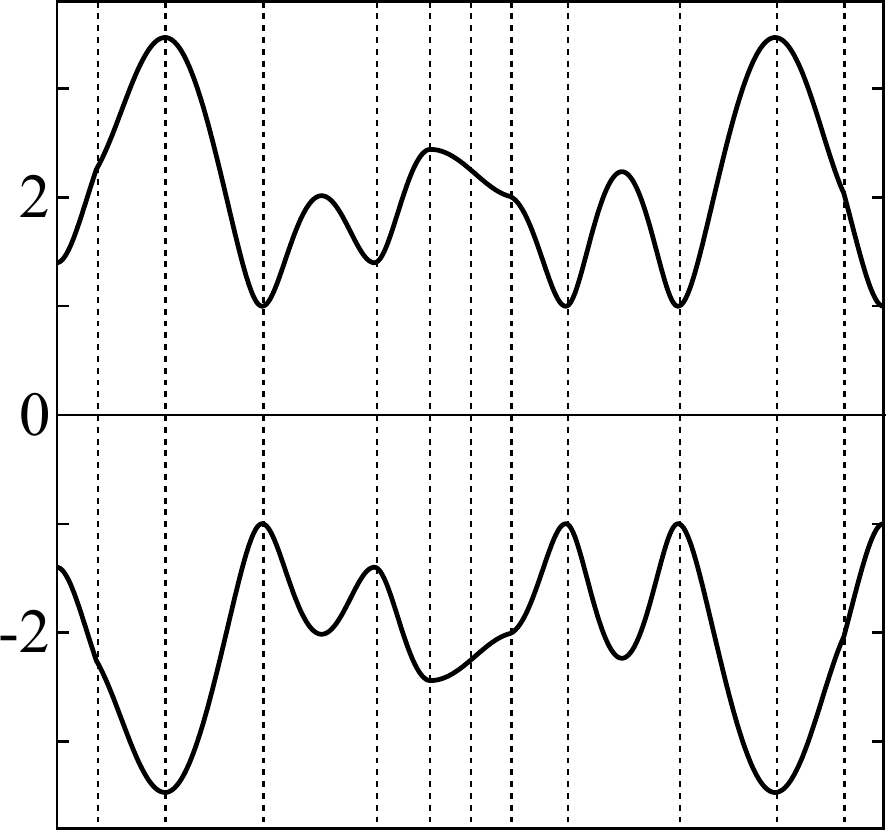}
\thicklines
\put( -6, 102){(a)}
\put( -8, 43){\rotatebox{90}{$\epsilon/t_{\mathrm{n}}$}}
\put(    3,    -8){$\mathrm{X}$}
\put(   9, 102){$\mathrm{U}$}
\put(  17,    -8){$\mathrm{L}$}
\put(  29,    -8){$\Gamma$}
\put(  42,    -8){$\mathrm{X}$}
\put(  47, 102){$\mathrm{W}$}
\put(  53,    -8){$\mathrm{K}$}
\put(  57, 102){$\mathrm{W}'$}
\put(  65,    -8){$\mathrm{X}'$}
\put(  79,    -8){$\Gamma$}
\put(  90,    -8){$\mathrm{L}$}
\put(  98, 102){$\mathrm{U}'$}
\put(103,    -8){$\mathrm{X}'$}
\end{overpic}
&
\hspace{2mm}
&
\begin{overpic}[scale=0.25, clip]{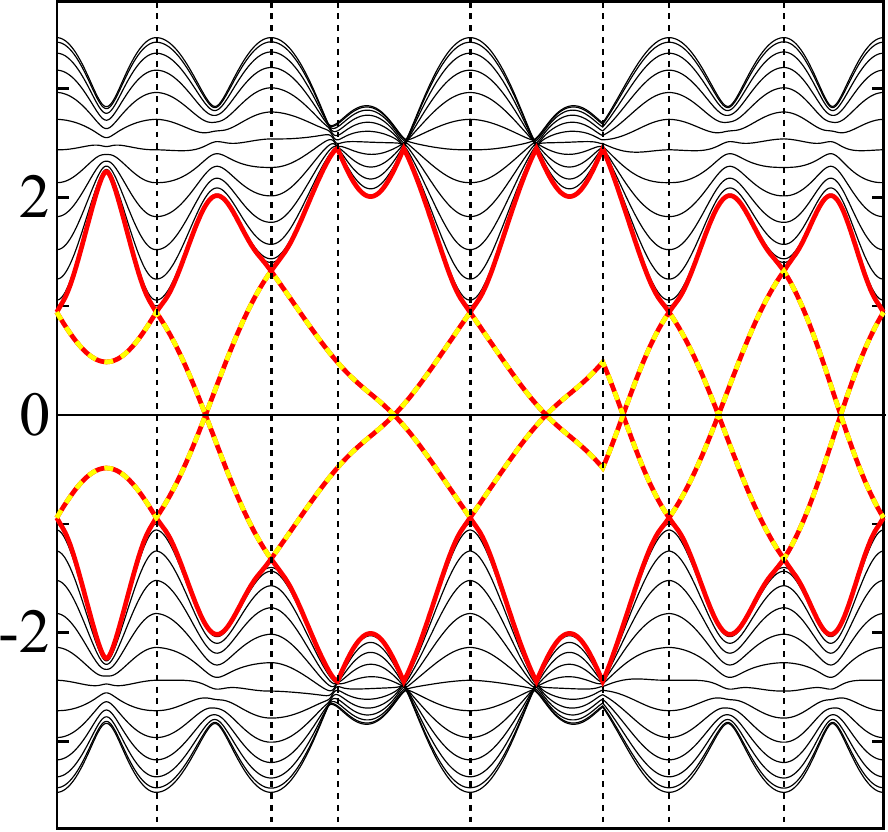}
\thicklines
\put( -6, 102){(b)}
%\put( -4, 104){(b)}
\put(   4,    -8){$\Gamma$}
\put(  14, 102){$\mathrm{M}_{2}$}
\put(  28, 102){$\mathrm{M}_{3}$}
\put(  37,    -8){$\mathrm{K}'$}
\put(  54,    -8){$\Gamma$}
\put( 69,    -8){$\mathrm{K}$}
\put( 76, 102){$\mathrm{M}_{2}$}
\put( 90, 102){$\mathrm{M}_{1}$}
\put(103,   -8){$\Gamma$}
\end{overpic}
\\
\vspace{5mm}
\\
\begin{overpic}[scale=0.25, clip]{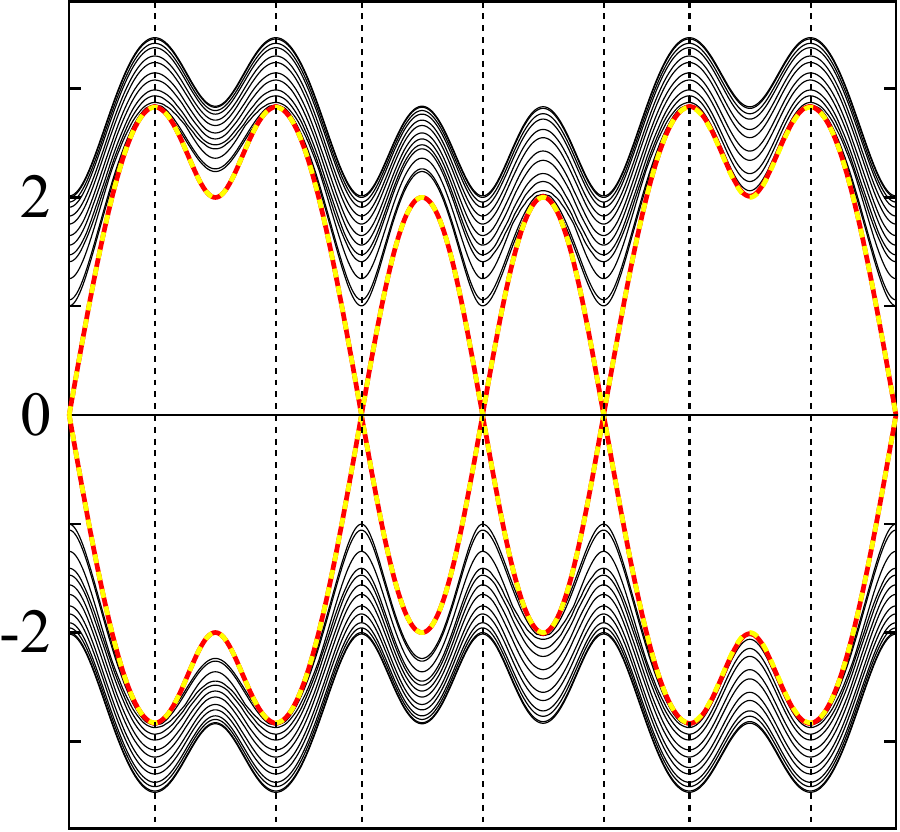}
\thicklines
\put( -4, 102){(c)}
%\put( -2, 104){(c)}
\put( -8, 43){\rotatebox{90}{$\epsilon/t_{\mathrm{n}}$}}
\put(    5,  -8){$\Gamma$}
\put(  15, 102){$\mathrm{X}'$}
\put(  30, 102){$\overline{\mathrm{X}}$}
\put(  39,  -8){$\mathrm{M}'$}
\put(  55,  -8){$\Gamma$}
\put(  68,  -8){$\mathrm{M}$}
\put(  80, 102){$\mathrm{X}'$}
\put(  95, 102){$\mathrm{X}$}
\put(105,  -8){$\Gamma$}
\end{overpic}
&
\hspace{2mm}
&
\begin{overpic}[scale=0.25, clip]{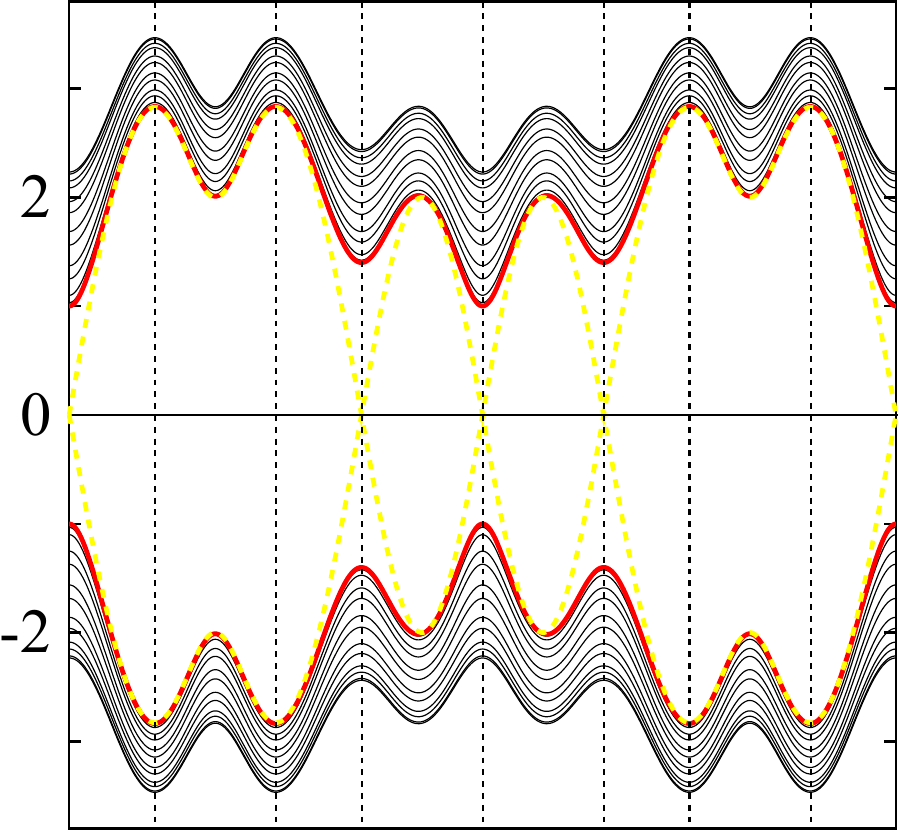}
\thicklines
\put( -5, 102){(d)}
%\put( -3, 104){(d)}
\put(    5,  -8){$\Gamma$}
\put(  15, 102){$\mathrm{X}'$}
\put(  30, 102){$\overline{\mathrm{X}}$}
\put(  39,  -8){$\mathrm{M}'$}
\put(  55,  -8){$\Gamma$}
\put(  68,  -8){$\mathrm{M}$}
\put(  80, 102){$\mathrm{X}'$}
\put(  95, 102){$\mathrm{X}$}
\put(105,  -8){$\Gamma$}
\end{overpic}
\end{tabular}
\vspace{0mm}
\caption{
(a) Band diagram and (b)-(d) projected band diagrams of a weak topological insulator ($t_{12}/t_{\mathrm{n}}=-0.3, v_{\mathrm{s}}/t_{\mathrm{n}}=0.2$) in (a) a 3D torus shape without a surface and in a slab shape with (b) $(111)$ surfaces, (c) $(100)$ surfaces, and (d) $(001)$ surfaces.
}
\label{fig:band-diagrams_weak-TI_t12}
\end{figure}

\begin{figure}[htb]
\vspace{5mm}
\begin{tabular}{ccc}
\begin{overpic}[scale=0.25, clip]{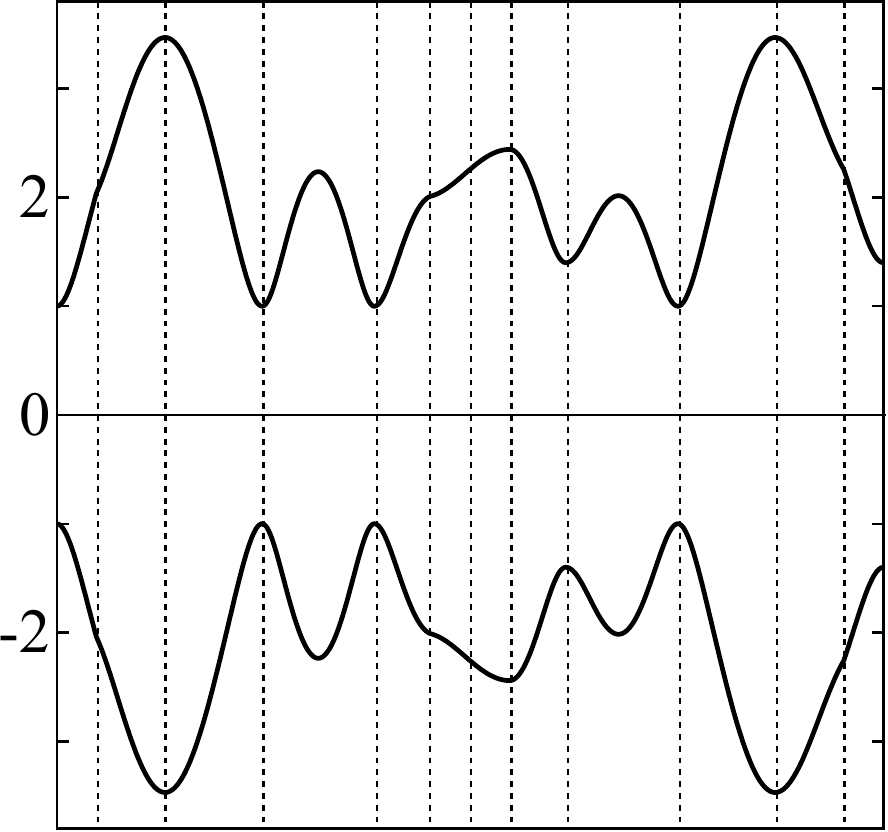}
\thicklines
\put( -6, 102){(a)}
\put( -8, 43){\rotatebox{90}{$\epsilon/t_{\mathrm{n}}$}}
\put(    3,    -8){$\mathrm{X}$}
\put(   9, 102){$\mathrm{U}$}
\put(  17,    -8){$\mathrm{L}$}
\put(  29,    -8){$\Gamma$}
\put(  42,    -8){$\mathrm{X}$}
\put(  47, 102){$\mathrm{W}$}
\put(  53,    -8){$\mathrm{K}$}
\put(  57, 102){$\mathrm{W}'$}
\put(  65,    -8){$\mathrm{X}'$}
\put(  79,    -8){$\Gamma$}
\put(  90,    -8){$\mathrm{L}$}
\put(  98, 102){$\mathrm{U}'$}
\put(103,    -8){$\mathrm{X}'$}
\end{overpic}
&
\hspace{2mm}
&
\begin{overpic}[scale=0.25, clip]{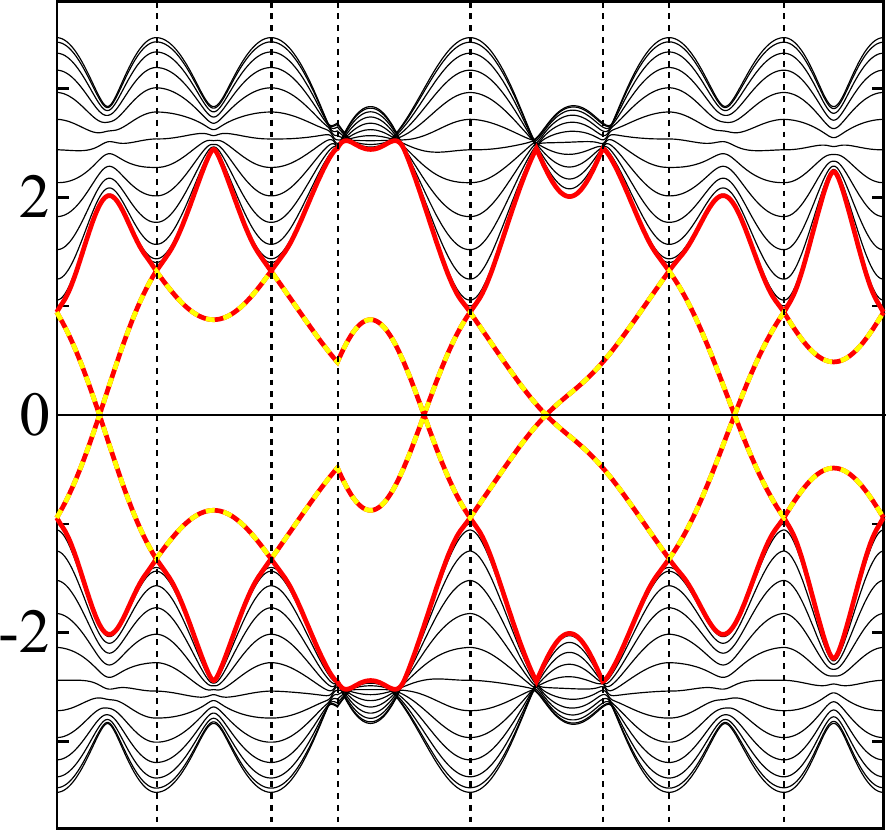}
\thicklines
\put( -6, 102){(b)}
%\put( -4, 104){(b)}
\put(   4,    -8){$\Gamma$}
\put(  14, 102){$\mathrm{M}_{2}$}
\put(  28, 102){$\mathrm{M}_{3}$}
\put(  37,    -8){$\mathrm{K}'$}
\put(  54,    -8){$\Gamma$}
\put( 69,    -8){$\mathrm{K}$}
\put( 76, 102){$\mathrm{M}_{2}$}
\put( 90, 102){$\mathrm{M}_{1}$}
\put(103,   -8){$\Gamma$}
\end{overpic}
\\
\vspace{5mm}
\\
\begin{overpic}[scale=0.25, clip]{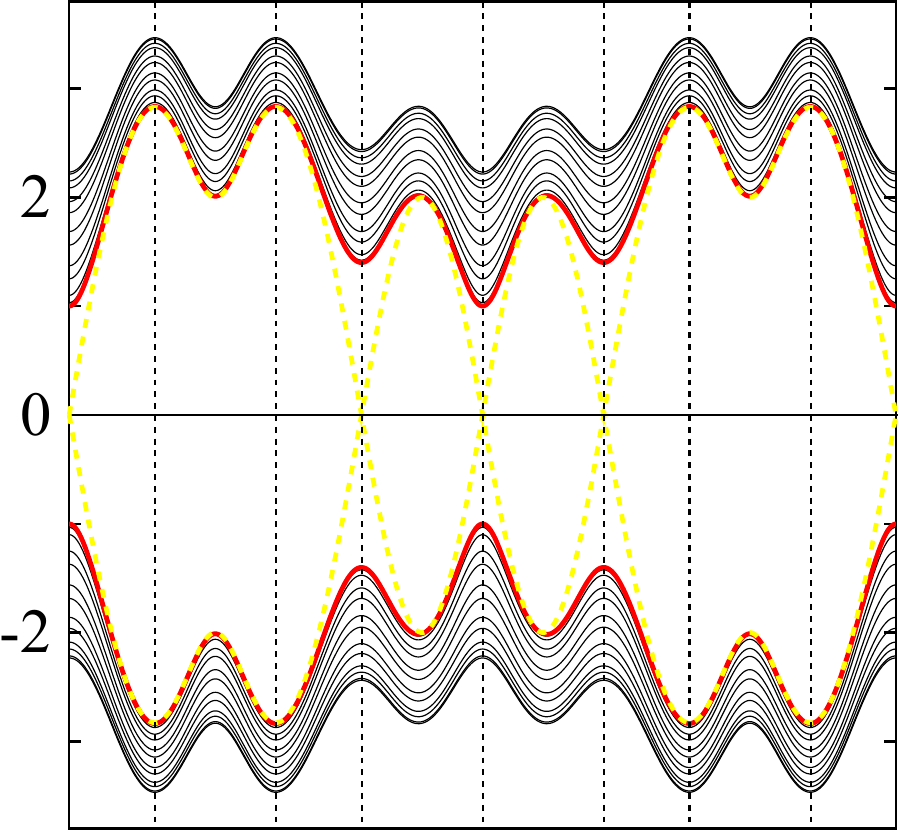}
\thicklines
\put( -4, 102){(c)}
\put( -8, 43){\rotatebox{90}{$\epsilon/t_{\mathrm{n}}$}}
\put(    5,  -8){$\Gamma$}
\put(  15, 102){$\mathrm{X}'$}
\put(  30, 102){$\overline{\mathrm{X}}$}
\put(  39,  -8){$\mathrm{M}'$}
\put(  55,  -8){$\Gamma$}
\put(  68,  -8){$\mathrm{M}$}
\put(  80, 102){$\mathrm{X}'$}
\put(  95, 102){$\mathrm{X}$}
\put(105,  -8){$\Gamma$}
\end{overpic}
&
\hspace{2mm}
&
\begin{overpic}[scale=0.25, clip]{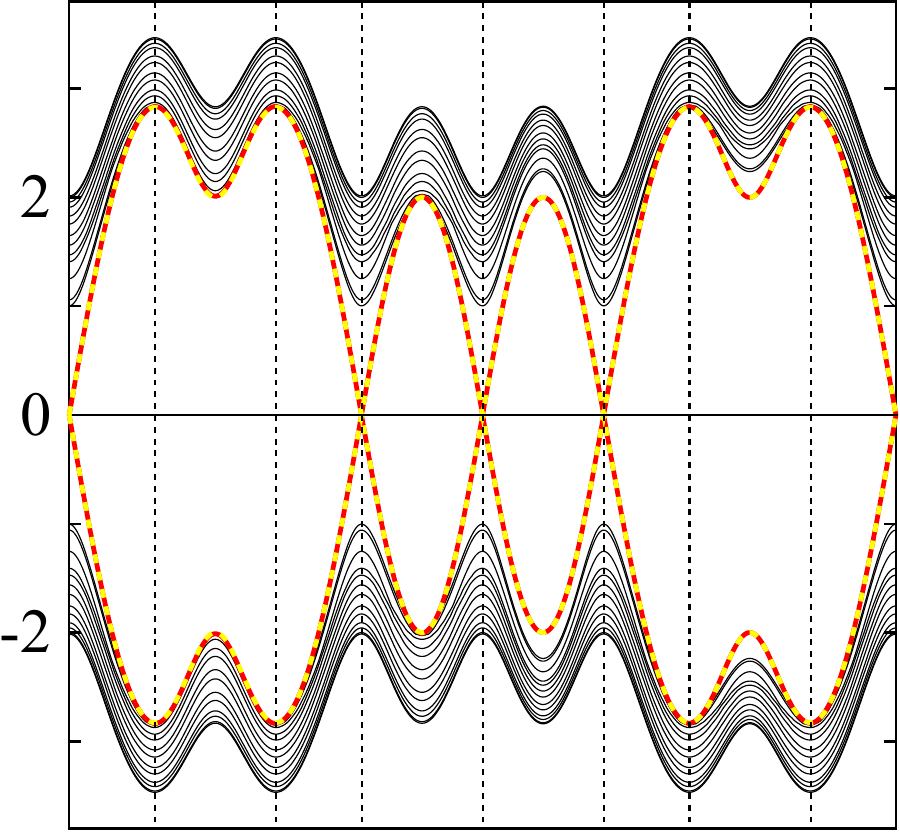}
\thicklines
\put( -6, 102){(d)}
\put(    5,  -8){$\Gamma$}
\put(  15, 102){$\mathrm{X}'$}
\put(  30, 102){$\overline{\mathrm{X}}$}
\put(  39,  -8){$\mathrm{M}'$}
\put(  55,  -8){$\Gamma$}
\put(  68,  -8){$\mathrm{M}$}
\put(  80, 102){$\mathrm{X}'$}
\put(  95, 102){$\mathrm{X}$}
\put(105,  -8){$\Gamma$}
\end{overpic}
\end{tabular}
\vspace{0mm}
\caption{
(a) Band diagram and (b)-(d) projected band diagrams of a weak topological insulator ($t_{23}/t_{\mathrm{n}}=-0.3, v_{\mathrm{s}}/t_{\mathrm{n}}=0.2$)
in (a) a 3D torus shape without a surface and in a slab shape with (b) $(111)$ surfaces, (c) $(100)$ surfaces, and (d) $(001)$ surfaces.
}
\label{fig:band-diagrams_weak-TI_t23}
\end{figure}

\begin{figure}[htb]
\vspace{5mm}
\begin{tabular}{ccc}
\begin{overpic}[scale=0.25, clip]{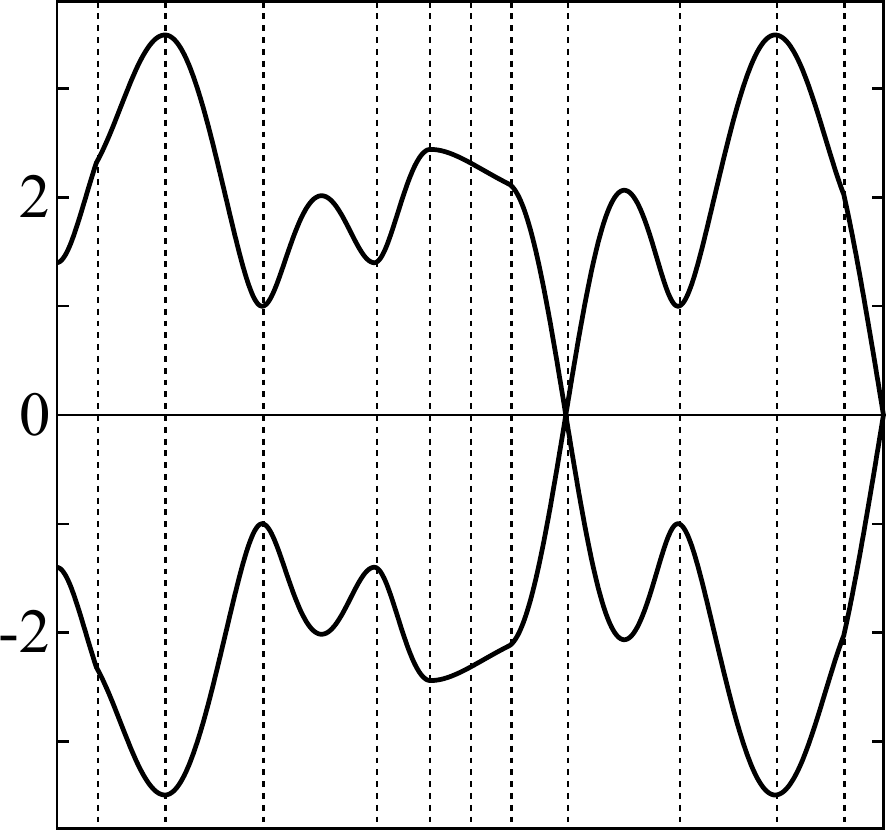}
\thicklines
\put( -6, 102){(a)}
\put( -8, 43){\rotatebox{90}{$\epsilon/t_{\mathrm{n}}$}}
\put(    3,    -8){$\mathrm{X}$}
\put(   9, 102){$\mathrm{U}$}
\put(  17,    -8){$\mathrm{L}$}
\put(  29,    -8){$\Gamma$}
\put(  42,    -8){$\mathrm{X}$}
\put(  47, 102){$\mathrm{W}$}
\put(  53,    -8){$\mathrm{K}$}
\put(  57, 102){$\mathrm{W}'$}
\put(  65,    -8){$\mathrm{X}'$}
\put(  79,    -8){$\Gamma$}
\put(  90,    -8){$\mathrm{L}$}
\put(  98, 102){$\mathrm{U}'$}
\put(103,    -8){$\mathrm{X}'$}
\end{overpic}
&
\hspace{2mm}
&
\begin{overpic}[scale=0.25, clip]{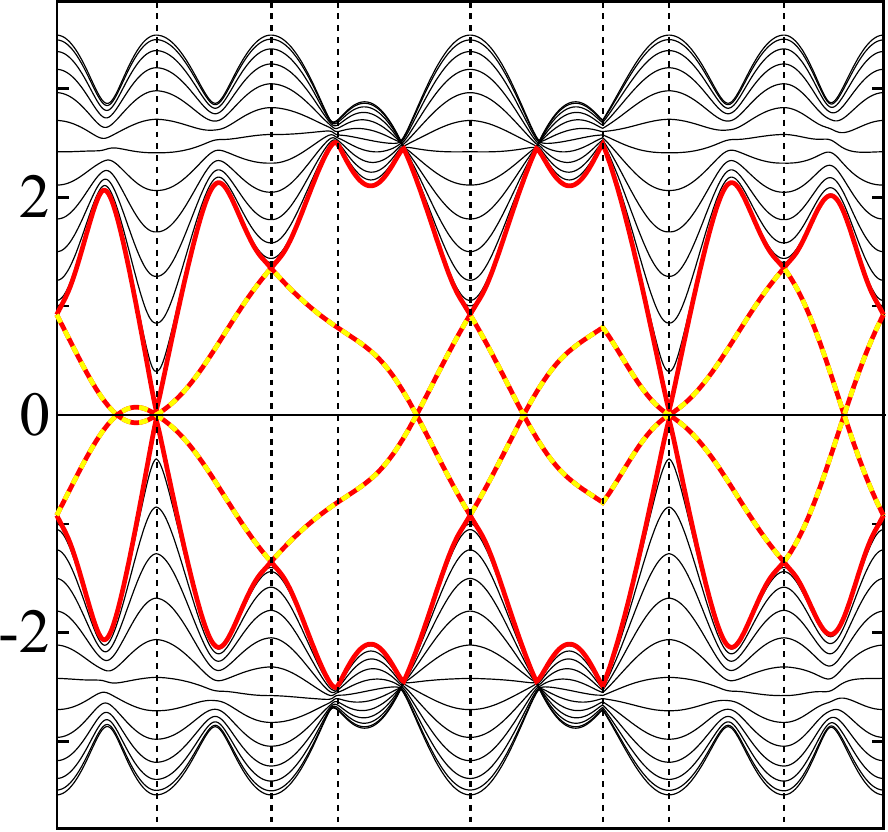}
\thicklines
\put( -6, 102){(b)}
\put(   4,    -8){$\Gamma$}
\put(  14, 102){$\mathrm{M}_{2}$}
\put(  28, 102){$\mathrm{M}_{3}$}
\put(  37,    -8){$\mathrm{K}'$}
\put(  54,    -8){$\Gamma$}
\put( 69,    -8){$\mathrm{K}$}
\put( 76, 102){$\mathrm{M}_{2}$}
\put( 90, 102){$\mathrm{M}_{1}$}
\put(103,   -8){$\Gamma$}
\end{overpic}
\\
\vspace{5mm}
\\
\begin{overpic}[scale=0.25, clip]{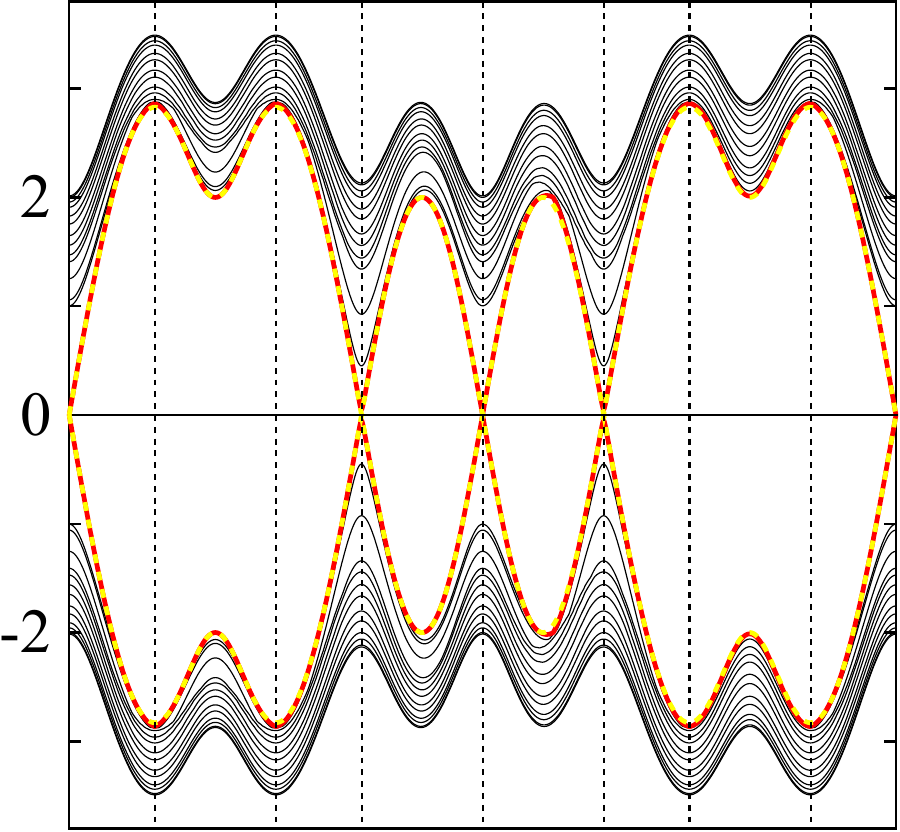}
\thicklines
\put( -4, 102){(c)}
\put( -8, 43){\rotatebox{90}{$\epsilon/t_{\mathrm{n}}$}}
\put(    5,  -8){$\Gamma$}
\put(  15, 102){$\mathrm{X}'$}
\put(  30, 102){$\overline{\mathrm{X}}$}
\put(  39,  -8){$\mathrm{M}'$}
\put(  55,  -8){$\Gamma$}
\put(  68,  -8){$\mathrm{M}$}
\put(  80, 102){$\mathrm{X}'$}
\put(  95, 102){$\mathrm{X}$}
\put(105,  -8){$\Gamma$}
\end{overpic}
&
\hspace{2mm}
&
\begin{overpic}[scale=0.25, clip]{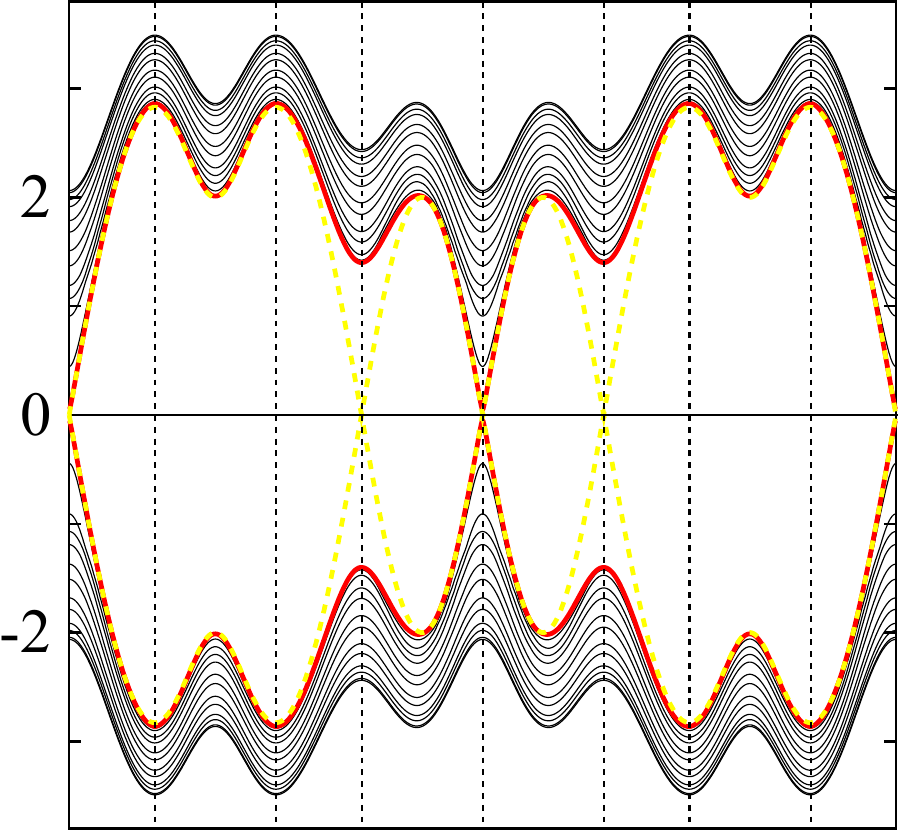}
\thicklines
\put( -6, 102){(d)}
\put(    5,  -8){$\Gamma$}
\put(  15, 102){$\mathrm{X}'$}
\put(  30, 102){$\overline{\mathrm{X}}$}
\put(  39,  -8){$\mathrm{M}'$}
\put(  55,  -8){$\Gamma$}
\put(  68,  -8){$\mathrm{M}$}
\put(  80, 102){$\mathrm{X}'$}
\put(  95, 102){$\mathrm{X}$}
\put(105,  -8){$\Gamma$}
\end{overpic}
\end{tabular}
\vspace{0mm}
\caption{
(a) Band diagram and (b)-(d) projected band diagrams of a semimetal at the transition point ($t_{12}/t_{\mathrm{n}}=-0.2375, t_{23}/t_{\mathrm{n}}=t_{31}/t_{\mathrm{n}}=-0.0625, v_{\mathrm{s}}/t_{\mathrm{n}}=0.45$)
in (a) a 3D torus shape without a surface and in a slab shape with (b) $(111)$ surfaces, (c) $(100)$ surfaces, and (d) $(001)$ surfaces.
}
\label{fig:band-diagrams_semimetal}
\end{figure}

Figures~\ref{fig:band-diagrams_strong-TI_tnn}-\ref{fig:band-diagrams_weak-TI_t23} show the sets of band diagrams of TI-$t_{\mathrm{nn}}$, TI-$t_{12}$, and TI-$t_{23}$. On the other hand, Fig.~\ref{fig:band-diagrams_semimetal} shows the sets of band diagrams of a semimental at the transition point $x_{\mathrm{I}}$ on the trajectory Eq.~(\ref{eq:trajectory}) connecting TI-$t_{\mathrm{nn}}$ and TI-$t_{12}$.
Panel (a) of each figure shows the band diagram of the case with no surfaces, where periodic boundary conditions are imposed in all directions.
A complete bulk bandgap is apparent in every panel (a) of Figs.~\ref{fig:band-diagrams_strong-TI_tnn}-\ref{fig:band-diagrams_weak-TI_t23}, while the bulk bandgap is closing in Fig.~\ref{fig:band-diagrams_semimetal}(a).
Panels (b), (c), and (d) of each figure show the projected band diagrams for $\mathrm{Sys}^{(111)}$, $\mathrm{Sys}^{(100)}$, and $\mathrm{Sys}^{(001)}$, respectively.
Periodic boundary conditions are imposed in the directions parallel to each given pair of surfaces.
The band dispersions in panels (b)-(d) are plotted at every ten indices from the top and bottom with thin black lines, and the middle four bands are plotted with thick red lines.
Dashed yellow lines in panels (b)-(d) represent the candidates of energy eigenvalues of surface states in Eq.~(\ref{eq:E(hkl)}), while only the middle two of the four appear in panel (b).

%%%%%%%%%%%%%%%%%%%%%%%%%%%%%%%%%%%%%%%%%%%%%%%%%%
% Figs of Band Diagrams of topological insulators
%%%%%%%%%%%%%%%%%%%%%%%%%%%%%%%%%%%%%%%%%%%%%%%%%%

%%%%%%%%%%%%%%%%%%%%%%%%%%%%%%%%%%%%%%%%%%%%%%%%%%

We first check $\mathrm{Sys}^{(111)}$ of TI-$t_{\mathrm{nn}}$, TI-$t_{12}$ and TI-$t_{23}$ all together.
In any of them, the lower half of the massless Dirac-type surface bands at the upper $(111)$ surface and the upper half of the massless Dirac-type surface bands at the lower $(111)$ surface satisfy the attenuation condition for $\mathrm{Sys}^{(111)}$, Eq.~(\ref{eq:condition-111)}), along almost the entire symmetry lines of the projected Brillouin zone.
The analytical and numerical dispersions almost entirely overlap, as shown in every panel (b) of Figs.~\ref{fig:band-diagrams_strong-TI_tnn}-\ref{fig:band-diagrams_weak-TI_t23}. They form vivid two-tone-colored dashed lines.
In contrast to the above situation of the $(111)$ surface states, the situation of the $(100) \Vert (001)$ surface states shows various aspects depending on the phases of topological insulators.

For TI-$t_{\mathrm{nn}}$, the attenuation condition for $\mathrm{Sys}^{(100) \Vert (001)}$, Eq.~(\ref{eq:condition-hkl)}), only holds over some portion of massless Dirac-type surface bands around $\Gamma$ at both the upper and lower $(100) \Vert (001)$ surfaces.
The yellow dashed lines and the red lines partially overlap in Figs.~\ref{fig:band-diagrams_strong-TI_tnn}(c) and \ref{fig:band-diagrams_strong-TI_tnn}(d).
The analytical and numerical dispersions appear to overlap well along $X'\overline{X}$ and $X' X$, while the attenuation condition does not hold there.
This type of overlap occurs in the proximity of bulk dispersions.
Analytical solutions of such cases no longer describe any surface states. 

For TI-$t_{12}$, the attenuation condition for $\mathrm{Sys}^{(100)}$ holds over a large portion of the entire symmetry lines except for around the midpoint of $\Gamma M$ and on $X' X$. 
Fig.~\ref{fig:band-diagrams_weak-TI_t12}(c) shows that the dashed yellow lines and the red lines overlap along almost the entire symmetry lines.
In contrast, the attenuation condition for $\mathrm{Sys}^{(001)}$ does not hold anywhere.
Fig.~\ref{fig:band-diagrams_weak-TI_t12}(d) indicates that TI-$t_{12}$ has no $(001)$ surface states.
These results are consistent with the expected features of weak topological insulators and the anisotropy pattern of each case.

For TI-$t_{23}$, the attenuation condition for $\mathrm{Sys}^{(100)}$ does not hold anywhere. 
Fig.~\ref{fig:band-diagrams_weak-TI_t23}(c) indicates that TI-$t_{23}$ has no $(100)$ surface states.
In contrast, the attenuation condition for $\mathrm{Sys}^{(001)}$ holds over a large portion of the entire symmetry lines except for around the midpoint of $M' \Gamma$ and on $X'\overline{X}$.
Fig.~\ref{fig:band-diagrams_weak-TI_t23}(d) shows that the dashed yellow lines and the red lines overlap along almost the entire symmetry lines.
These results are again consistent with the expected features of weak topological insulators and the anisotropy pattern of each case.

As discussed in Sec.~\ref{sec:Z2-invariants}, the transitions between the phases $(1;000)$ and $(0;110)\Vert(0;011)\Vert(0;101)$ highlight the unique features of the series of lattice models. For example, between TI-$t_{\mathrm{nn}}$ and TI-$t_{12}$, there emerges a direct transition of the first type in Eq.~(\ref{eq:STI-WTI-transitions}), $(1;000) \overset{x_{\mathrm{I}}}{\longleftrightarrow} (0;110)$, at $x_{\mathrm{I}}=\frac{7}{12}$ ($t_{12}/t_{\mathrm{n}}=-0.2375, t_{23}/t_{\mathrm{n}}=t_{31}/t_{\mathrm{n}}=-0.0625, v_{\mathrm{s}}/t_{\mathrm{n}}=0.45$).
Fig.~\ref{fig:band-diagrams_semimetal} shows the band diagrams of a semimetal at this transition point and suggests that a discussion similar to the above also holds even for semimetals at transition points. 

In fact, we have confirmed the validity of our analytical solutions of surface states at various types of transition points.
Since, apart from minor differences, the prominent characteristics of the above specific example are commonly found in other examples, we discuss only the details of Fig.~\ref{fig:band-diagrams_semimetal} below.
First, Fig.~\ref{fig:band-diagrams_semimetal} shows that the bulk bandgap is closing at $X'$ which corresponds to $M_{2}$ for $\mathrm{Sys}^{(111)}$, $M (M')$ for $\mathrm{Sys}^{(100)}$, and $\Gamma$ for $\mathrm{Sys}^{(001)}$.

The lower half of the massless Dirac-type surface bands at the upper $(111)$ surface and the upper half of the massless Dirac-type surface bands at the lower $(111)$ surface satisfy the attenuation condition for $\mathrm{Sys}^{(111)}$ along almost the entire symmetry lines of the projected Brillouin zone except for on $\mathrm{M}_{2}$.
The analytical and numerical dispersions almost entirely overlap, as shown in Fig.~\ref{fig:band-diagrams_semimetal}(b). 

The attenuation condition for $\mathrm{Sys}^{(100)}$ holds over a large portion of the entire symmetry lines except for around $X'$, $X$, the midpoint of $\Gamma M$ and on $X' X$.
Interestingly, the attenuation condition holds even in some domains where surface bands are close to bulk bands, e.g., around $M (M')$ and the midpoint of $X'\overline{X}$.
(Note that $\overline{X}$ and $X$ represent the same point, while $X'\overline{X}$ and $X' X$ are different symmetry lines.)
Fig.~\ref{fig:band-diagrams_semimetal}(c) shows that the dashed yellow lines and the red lines overlap along almost the entire symmetry lines.
In contrast, the attenuation condition for $\mathrm{Sys}^{(001)}$ holds only around $\Gamma$.
However, as in the previous example, the attenuation condition holds even in the proximity of massless Dirac-type bulk bands.
The yellow dashed lines and the red lines partially overlap in Fig.~\ref{fig:band-diagrams_semimetal}(d).

%\nocite{*}

%